\documentclass[iop,apj]{emulateapj}

\usepackage{amssymb,amsmath}
\pdfoutput=1

\def\Msun{{M_{\odot}}}
\def\hst{{\it HST}}

\newcommand{\kms}{\>{\rm km}\,{\rm s}^{-1}}
\newcommand{\masyr}{\>{\rm mas}\,{\rm yr}^{-1}}
\newcommand{\kpc}{\>{\rm kpc}}
\newcommand{\Gyr}{\>{\rm Gyr}}

\shorttitle{The Space Motion of Leo~I}
\shortauthors{Sohn et al.}

\begin{document}

\title{The Space Motion of Leo~I:\\
Hubble Space Telescope Proper Motion and Implied Orbit}

\author{
Sangmo Tony Sohn\altaffilmark{1}, 
Gurtina Besla\altaffilmark{2,3},
Roeland P. van der Marel\altaffilmark{1}, 
Michael Boylan-Kolchin\altaffilmark{4}, 
Steven R. Majewski\altaffilmark{5}
and
James S. Bullock\altaffilmark{4}, 
}
\altaffiltext{1}{Space Telescope Science Institute, 
                 3700 San Martin Drive, Baltimore, MD 21218, USA}
\altaffiltext{2}{Department of Astronomy, Columbia University, 
                 New York, NY 10027, USA}
\altaffiltext{3}{Hubble Fellow}
\altaffiltext{4}{Department of Physics and Astronomy, 
                 Center for Cosmology, University of California, 
                 4129 Reines Hall, Irvine, CA 92697, USA}
\altaffiltext{5}{Department of Astronomy, University of Virginia, 
                 Charlottesville, VA 22904-4325, USA}
\email{tsohn@stsci.edu}


\begin{abstract}
We present the first absolute proper motion measurement of Leo~I,
based on two epochs of {\it Hubble Space Telescope} ({\it HST})
ACS/WFC images separated by $\sim 5$ years in time. The average shift
of Leo~I stars with respect to $\sim 100$ background galaxies implies
a proper motion of $(\mu_{W}, \mu_{N}) = (0.1140 \pm 0.0295, -0.1256
\pm 0.0293) {\rm mas\ yr}^{-1}$. The implied Galactocentric velocity
vector, corrected for the reflex motion of the Sun, has radial and
tangential components $V_{\rm rad} = 167.9 \pm 2.8 \kms$ and $V_{\rm
  tan} = 101.0 \pm 34.4 \kms$, respectively. We study the detailed
orbital history of Leo~I by solving its equations of motion backward
in time for a range of plausible mass models for the Milky Way and its
surrounding galaxies. Leo~I entered the Milky Way virial radius $2.33
\pm 0.21$ Gyr ago, most likely on its first infall.  It had a
pericentric approach $1.05 \pm 0.09$ Gyr ago at a Galactocentric
distance of $91 \pm 36$ kpc.  We associate these time scales with
characteristic time scales in Leo~I's star formation history, which
shows an enhanced star formation activity $\sim 2$ Gyr ago and
quenching $\sim 1$ Gyr ago.  There is no indication from our
calculations that other galaxies have significantly influenced Leo~I's
orbit, although there is a small probability that it may have
interacted with either Ursa Minor or Leo~II within the last $\sim 1$
Gyr. For most plausible Milky Way masses, the observed velocity
implies that Leo~I is bound to the Milky Way. However, it may not be
appropriate to include it in models of the Milky Way satellite
population that assume dynamical equilibrium, given its recent infall.
Solution of the complete (non-radial) timing equations for the Leo~I
orbit implies a Milky Way mass $M_{\rm {MW,vir}} =
3.15_{-1.36}^{+1.58} \times 10^{12} \Msun$, with the large uncertainty
dominated by cosmic scatter. In a companion paper, we compare the new
observations to the properties of Leo~I subhalo analogs extracted from
cosmological simulations.
\end{abstract}


\keywords{Astrometry ---
galaxies: kinematics and dynamics --- 
galaxies: individual (Leo~I) ---
Galaxy: kinematics and dynamics ---
Galaxy: halo ---    
Local Group}


\section{Introduction}
\label{sec:Intro}

Structures in the Universe cluster on various scales. The Milky Way
(MW) is no exception, as evidenced by its system of satellites. Most 
of these satellites are dwarf spheroidal galaxies (dSphs). These are 
the most dark matter dominated stellar systems currently known, with
mass-to-light ($M/L$) ratios of up to a few thousand in units of
$\Msun/L_{\odot}$ \citep[e.g.,][]{wol10}

In the current paradigm for galaxy formation, dark halos of galaxies
form through the accumulation of smaller subunits. The MW satellite
system is one of the best objects to study these hierarchical
evolutionary processes in action, due to its proximity. In the last
decade, many wide-field ground-based surveys have led to discoveries
in this area.  For example, the The Two Micron All Sky Survey (2MASS) 
unveiled the ongoing disruption of the Sagittarius dSph that has 
produced a giant stream of stars wrapping around the entire MW at 
least a single time \citep[e.g.,][]{maj03}, and the SDSS has revealed 
many other such streams that once belonged to either dwarf galaxies 
or globular clusters \citep[][and references therein]{gri09}. 
In addition, there is evidence for recent accretion and build up of 
the MW satellite system: \hst\ proper motion measurements of the two 
most-massive MW satellites, the Large and Small Magellanic Clouds 
(LMC and SMC) \citep{kal06a,kal06b,pia08}, suggest that these galaxies 
were not born as MW satellites but instead may be falling into the 
Local Group for the first time \citep{bes07,kal13}.

As tracer objects, MW satellites are valuable tools for studying the
size and mass of the MW halo because their orbits contain important
information about the host potential. Distant satellite galaxies are
of particular interest because: (1) they probe the dark halo at the
largest radii; and (2) their kinematics may not have been fully
virialized yet. Measuring the space motions of distant satellites
with respect to the MW is therefore crucial for gaining insights into
the MW virial mass and the mass assembly at late epochs.

So far, there are only three known objects thought to be associated with
the MW at a distance beyond 200 kpc from the Galactic center: the dSphs 
Leo~I, Leo~II, and Canes Venatici I. Leo~I, unlike the others, has an 
unusually large Galactocentric radial velocity at its extreme distance
\citep{mat98,mat08}. Because of this, Leo~I has played an important
role in our interpretation of the MW satellite system. One reason for
this is that Leo~I disproportionately affects MW mass estimates based
on the assumption of equilibrium kinematics: including or excluding it
from the MW satellite population sample produces very different
estimates \citep[e.g.,][]{zar89,kul92,koc96,wil99,wat10}.

Another issue on which Leo~I has generated much interest and debate is
on the topic of its specific orbit. Its large radial velocity led
\citet{byr94} to suggest that Leo~I once belonged to M31 and is now on
a hyperbolic orbit flying past the MW. However, in an earlier study,
\citet{zar89} argued against the possibility that Leo~I is unbound to
the MW. More recently, \citet{soh07} and \citet{mat08} carried out 
orbital analyses combined with high-precision radial velocities of 
individual Leo~I member stars to study the orbit in detail. The former 
study suggested that Leo~I was tidally disrupted on one or two 
perigalactic passages about a massive Local Group member (most likely 
the MW), whereas the latter study proposed involvement of a third body 
that may have injected Leo~I into its present orbit a few Gyr before 
its last perigalacticon. The orbit of Leo~I can also shed light on 
studies using satellites as test particles in a cosmological context
\citep[e.g.,][]{li08,roc12}. Unfortunately, many orbital scenarios 
remain possible as long as only one component of Leo~I's velocity 
(along the line of sight) is known. To make progress, it is necessary 
to know also the proper motion of Leo~I to yield the full 
three-dimensional Galactocentric velocity.

Due to its distance, it has not previously been possible to measure
the proper motion of Leo~I. The most distant MW satellite with a
measured proper motion so far is Leo~II, for which a measurement was
obtained with the second-generation {\hst} instrument WFPC2
\citep{lep11}. However, the large uncertainty of this proper motion
measurement with an accuracy of $0.132 \masyr$, corresponding to 
$144 \kms$ at the distance of Leo~II, limits its usefulness in 
constraining models. Leo~I is located at a distance of $\sim 260$
kpc, which is even $\sim 40$ kpc farther away than Leo~II.

However, we have recently pioneered a method to measure the proper 
motions of galaxies as far away as M31 \citep{soh12} using the 
third- and/or fourth-generation {\hst} instruments ACS and WFC3. 
This involves sophisticated data analysis techniques to measure from 
deep images taken years apart the relative shifts of thousands of 
stars in the galaxy with respect to hundreds of distant compact 
galaxies in the background. We were able to achieve a final proper 
motion accuracy of $0.012 \masyr$, which yields a velocity 
uncertainty of $44 \kms$ at the distance of M31, using 18 independent 
measurements on three different fields. Leo~I is at a distance 
three times closer than M31, so the same level of velocity 
uncertainty is well within the reach of 
our proven techniques. Because this would
certainly yield useful new constraints on the orbit of Leo~I, 
we designed an observational program to measure the absolute proper 
motion of Leo~I for the first time. We report here on the results 
and implications of our program.

This paper is organized as follows. In Section~\ref{sec:proper motion}
we describe the data and analysis steps, and we present the inferred
Leo~I proper motion. In Section~\ref{sec:Orbit}, we correct the
measured proper motion for the solar motion to derive the
Galactocentric space motion. We then explore the implications for the
past orbit of Leo~I under a variety of assumptions for the mass and
mass distribution of the MW and other Local Group galaxies. In
Section~\ref{sec:Conclusions}, we discuss the implications of the
results for our understanding of both Leo~I and the MW, and summarize
the main results of the paper.

This is the first paper in a series of two. In Paper~II \citep{boy13},
we compare the new observations of Leo~I to the properties of Leo~I
subhalo analogs extracted from state-of-the-art cosmological
simulations. We use this comparison to place additional constraints on
the mass of the MW, the properties of its satellite system, and the
past history of Leo~I.


\section{The Proper Motion of Leo~I}
\label{sec:proper motion}


\subsection{Hubble Space Telescope Data}
\label{sec:Data}

\begin{figure*}
\epsscale{0.7}
\plotone{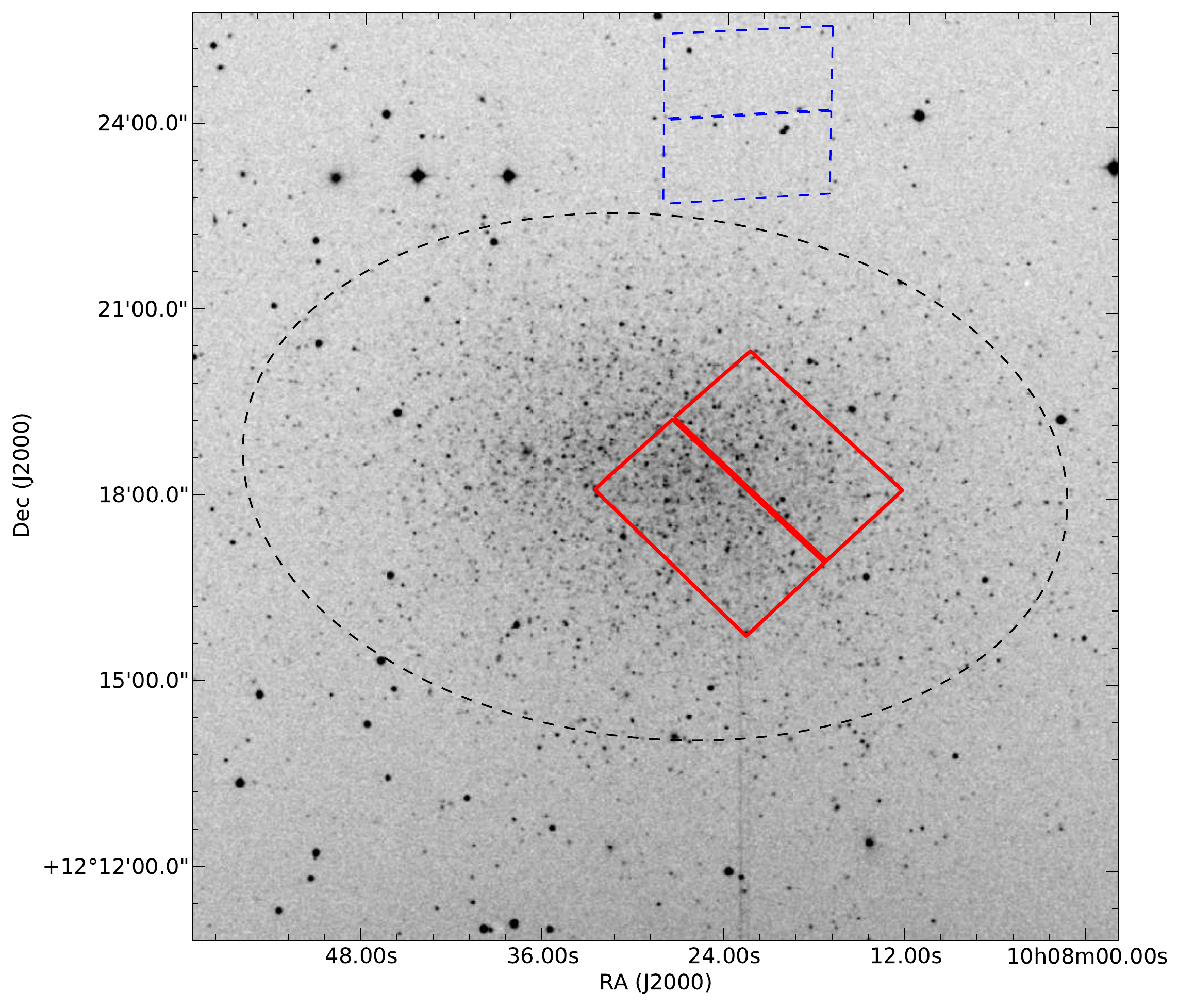}
\caption{Field location of our ACS/WFC observations plotted in red
  over a 15\arcmin$\times$15\arcmin\ section of the sky centered on
  Leo~I from the STScI Digital Sky Survey. The line that bisects the 
  ACS/WFC field is a small gap between two CCDs; the CCD readout 
  direction is roughly perpendicular to this. The dashed ellipse
  represents the King limiting boundary as derived by \citet{soh07}. 
  The parallel WFC3/UVIS field is plotted in blue dashed lines.
\label{fig:ACSPointing}}
\end{figure*}

The data used in this study to measure the proper motion of Leo~I
consist of images taken with \hst\ in two different epochs separated
by $\sim5$ years in time. For the first-epoch data, we used the images
taken in February 2006 for the science program GO-10520 (PI:
T. Smecker-Hane) to study the star formation history of Leo~I.  A
field slightly offset from the center of Leo~I was imaged with the
ACS/WFC using the F435W and F814W filters. The ACS/WFC covers a 
field-of-view of $\sim 200" \times 200"$ at a scale of 0.05
arcsec/pixel. Figure~\ref{fig:ACSPointing} shows the field location
overlaid on a STScI Digital Sky Survey image centered on Leo~I. Sets
of 7 and 6 images with exposure times of 1,700 sec each were obtained
in F435W and F814W, respectively, and three additional 440 sec F814W
images were obtained as well.

The second-epoch data were obtained in January 2011 for our science
program GO-12270 (PI: S. Sohn). We pre-analyzed the first-epoch data
to enable optimal design of the second-epoch observations. This
analysis indicated that the F814W filter provides a slightly better
astrometric handle on extended objects than the F435W filter 
\citep[see also][for a discussion of the wavelength dependence of the
astrometric accuracy]{mah08}. We therefore took second-epoch 
observations only with F814W. We obtained 12 images with individual 
exposure times ranging from 1,267 to 1,364 sec. The resulting total 
exposure time for the second epoch was $\sim 16$ ksec. Individual 
exposures were dithered using a pattern designed to optimize the 
sampling of the point-spread functions (PSFs) for stars that fall on 
different parts of the detector. This ``pixel-phase coverage'' is 
crucial for creating a high-resolution stacked image from a limited 
number of individual exposures.

We matched the orientation and field center of the second-epoch
observations as closely as possible to those of the first-epoch
observations. However, due to unavailability of the same guide stars
used for the first-epoch observations, we had to use an orientation
that differed by $\sim1$\arcdeg. We also obtained parallel
observations with WFC3/UVIS in the second epoch for an off-target
field. This field, also shown in Figure~\ref{fig:ACSPointing}, was
imaged in F438W and F814W to allow a study of the stellar population
in the outer halo of Leo~I. However, these observations are not
discussed further in the present paper.


\subsection{Measurement Technique}
\label{sec:MeasuringPM}

To measure the proper motion of Leo~I, we compare the two epochs of 
\hst\ F814W imaging data and determine the average shift of Leo~I 
stars relative to distant background galaxies.\footnote{The 
first-epoch F435W data were used only to extract color information 
on the sources in the field.} This requires a method that
allows accurate positions to be measured for both stars and compact
galaxies.  \citet{mah08} presented a method that accomplishes this by
constructing and fitting an individual template for each source in an
image. \citet{soh12} implemented, expanded and applied this method to
measure the proper motion of the galaxy M31 using \hst\ ACS/WFC and
WFC3/UVIS images of three distinct fields imaged over a 5--7 year 
time baseline. We adopt their method here to also analyze the new 
Leo~I data. We discuss only the main outline and results of the proper
motion derivation, and refer the reader to \citet{soh10} and
\citet{soh12} for more details about the methodology.

All the science {\tt \_flt.fits} images for the first and second
epochs were downloaded from the archive. To each image we first
applied the Charge Transfer Efficiency (CTE) correction routine 
developed by \citet{and10}. We then used the {\tt img2xym\_WFC.09x10} 
program \citep{and06} to determine a position and a flux for each star 
in each exposure. The positions were subsequently corrected for the 
known ACS/WFC geometric distortions. Separate distortion solutions 
were used for the first- and second-epoch data, to account for a 
difference between pre- and post-SM4 ACS/WFC data \citep[see Section 
3.3 of][]{soh12}. We then adopt the first exposure of the second epoch 
({\tt jbjm01kkq}) as the frame of reference. We cross-identify stars in 
this exposure and the same stars in other exposures. We use the 
distortion-corrected positions of the cross-identified stars to 
construct a six-parameter linear transformation between the two frames. 
These transformations are then used in a program that constructs a 
stacked image, cleaned of cosmic rays and detector artifacts, of the 
different exposures in each filter+epoch combination. The stacked 
images were super-sampled by a factor of 2 relative to the native 
ACS/WFC pixel scale for better sampling.

Stars and galaxies were identified from the stacked second-epoch F814W
image, which provides our deepest view of the field. First, a list of
point sources was constructed from the sources detected by {\tt
img2xym\_WFC.09x10}. The selection of Leo~I stars from this list is
relatively straightforward, since our target field is numerically and
spatially dominated by Leo~I stars (Leo~I is located at high Galactic
latitude $b = 49\degr$, so few foreground stars are expected). To
select only bona fide and well-measured Leo~I stars, we require: (1)
small RMS scatter between the 12 independent position measurements;
(2) consistent position in the color-magnitude diagram (obtained from
combination with the first-epoch F435W results) with the expected
Leo~I stellar evolutionary features; and (3) consistent proper motion
with the other Leo~I stars. This yielded a list of 36,000 stars
suitable for proper motion analysis. For the selection of background
galaxies in the target field, we started with a candidate list
generated by running {\tt SExtractor} \citep{ber96} on the stacked
image. From this list we then carefully identified 116 compact
background galaxies by eye.

\begin{figure}
\epsscale{1.0}
\plotone{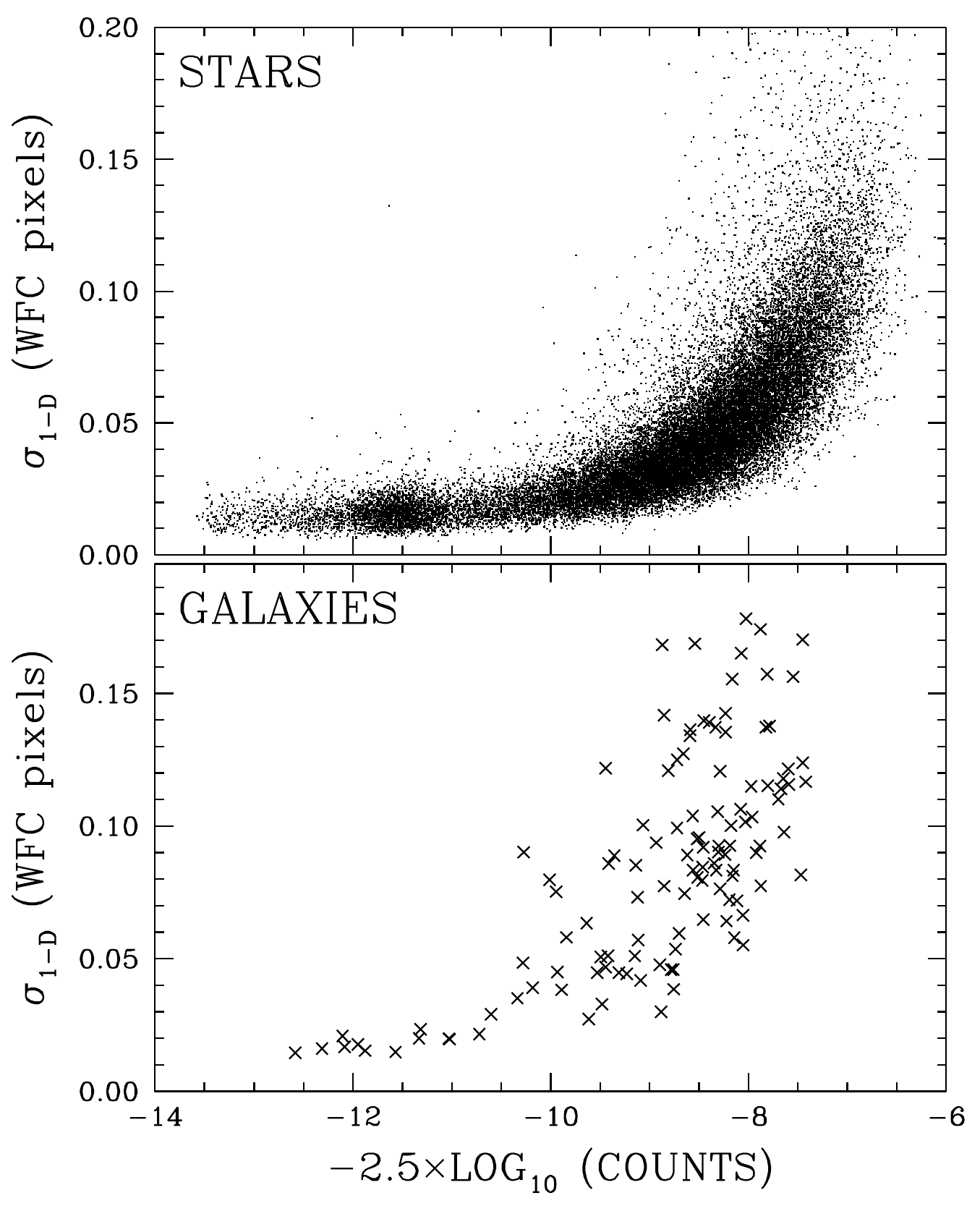}
\caption{One-dimensional positional uncertainty per exposure as a 
  function of instrumental magnitude, for Leo~I stars 
  ({\it upper panels}) and background galaxies ({\it lower panels}). 
  The error is defined as 
  $\sigma_{1-D} = \sqrt{\frac{1}{2}({\sigma_{x}}^{2}+{\sigma_{y}}^{2})}$. 
  Here $\sigma_{x}$ and $\sigma_{y}$ are the per-coordinate RMS 
  residuals with respect to the average for the multiple second-epoch 
  measurements. The errors for the first-epoch measurements are 
  slightly larger, but have a similar dependence on instrumental 
  magnitude. All random errors are propagated into our final proper 
  motion measurement for Leo I as described in the text.
\label{fig:SigmaXY}}
\end{figure}

For each star/galaxy in each F814W exposure of each epoch we then 
measured in consistent fashion a position using the template-fitting 
method. Templates were constructed from the second-epoch stacked image 
via interpolation. The template-fitting for the first-epoch data 
included an additional $7\times7$-pixel convolution kernel, to allow 
for PSF differences between epochs. This kernel was determined from the 
data for bright and isolated Leo~I stars, without any assumed
field-dependence. This is similar to what was done in our analysis of
M31 \citep{soh12}, except that in that case the role of the epochs was
reversed (since for M31 our first-epoch data were the deepest).

The template-fitted positions were corrected as before for the ACS/WFC
geometric distortion. Again, star positions in individual exposures
were used to determine six-parameter linear transformations with
respect to the first exposure ({\tt jbjm01kkq}, adopted as the frame
of reference) but now using only the selected Leo~I stars. These
linear transformations were then used to transform the measured
positions of all selected stars and background galaxies in all
exposures into the reference frame.

The individual exposures lead to multiple determinations for the
position of each star or background galaxy in each epoch. We average
these determinations to obtain the average position of each source in
each epoch. The RMS scatter between measurements in a given epoch 
quantifies the random positional uncertainty in a single measurement. 
The error in the mean is smaller by $\sqrt{N}$, where $N$ is the 
number of exposures in the epoch. 

Figure~\ref{fig:SigmaXY} shows the one-dimensional positional errors
per second-epoch exposure for Leo~I stars (upper panels) and
background galaxies (lower panels), as a function of instrumental
magnitude (defined as $m_{\rm instr} = -2.5 \log[{\rm
    counts}]$). Brighter objects produce a higher signal-to-noise
ratio than fainter objects, and therefore have more accurately
determined positions. Stars are more compact than background galaxies,
and therefore generally have more accurately determined positions.
However, the brightest galaxies have positional uncertainties
comparable to those of the brightest stars.

The proper motion of a source is the difference between its average
position in the two epochs, divided by the time baseline (4.93
years). By construction, our method aligns the star fields between the
two epochs. So with this convention, Leo~I stars have zero motion on
average, and the background galaxies move over time. Of course, in
reality the background galaxies are stationary and the Leo~I stars
move in the foreground. So the {\it actual} average proper motion of
Leo~I stars is obtained as $-1$ times the {\it measured} average
proper motion of the background galaxies.

\begin{figure*}
\epsscale{1.0}
\plotone{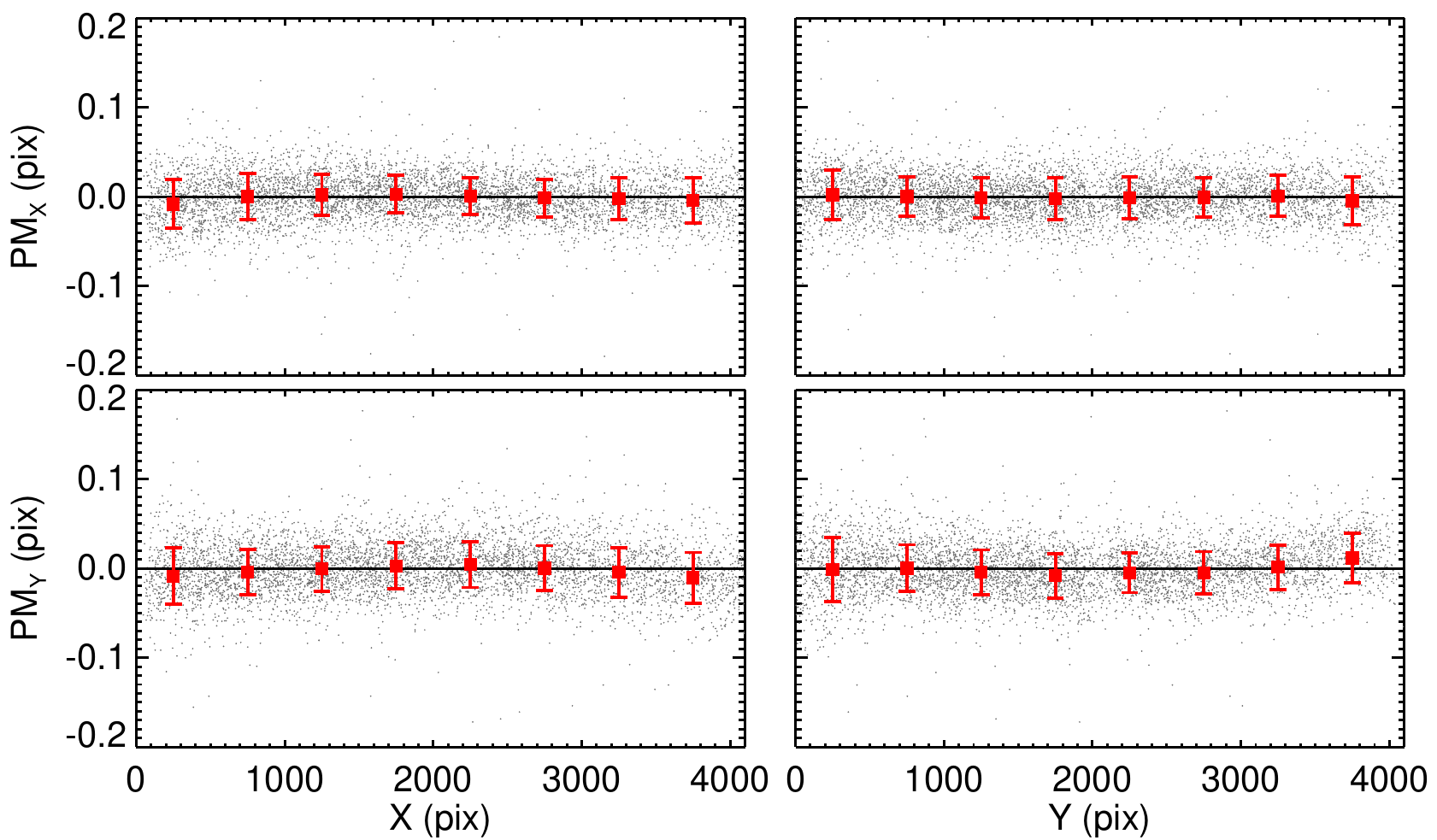}
\caption{Displacements of individual stars brighter than instrumental 
  magnitude of $-10$ (dark gray dots) versus detector location 
  between one of the 1,700 sec first-epoch images ({\tt j9gz04tsq}) 
  and the average of the second-epoch images, plotted separately for 
  $X$ and $Y$ positions. The average and RMS displacements of stars 
  in every 500-pixel bin are shown in red. The displacements average 
  to zero by construction. The scatter is a measure of the 
  per-exposure positional accuracy for a star in this brightness 
  range. Low-level trends are indicative of residual detector effects. 
  The units are in native ACS/WFC pixels, and $X$ and $Y$ positions 
  are in the reference frame.
\label{fig:starpm1}}
\end{figure*}

Figures~\ref{fig:starpm1} shows the measured proper motion of each 
star brighter than instrumental magnitude $-10$ in $X$ and $Y$ as a 
function of detector coordinates, for one of the 1,700 sec 
first-epoch images. The motions are zero on average by construction, 
but there remain small residual trends with position on the detector 
at levels $\lesssim 0.01$ pixel. This could be due, e.g., to 
limitations in the adopted geometric distortion corrections. These
trends are corrected by measuring the displacement of each background
galaxy with respect to only those Leo~I stars that lie in the vicinity
of the galaxy. This ``local correction'' removes any remaining
systematic proper motion residuals associated with the detector
position. Each local correction was constructed using stars of similar
brightness ($\pm 1$ mag) and within a 200 pixel region centered on the
given background galaxy. The 200 pixel size was chosen to provide a 
good compromise between having a sufficient number of stars (typically 
in the range 25-250), and not including too many distant sources.

\begin{figure*}
\epsscale{1.0}
\plotone{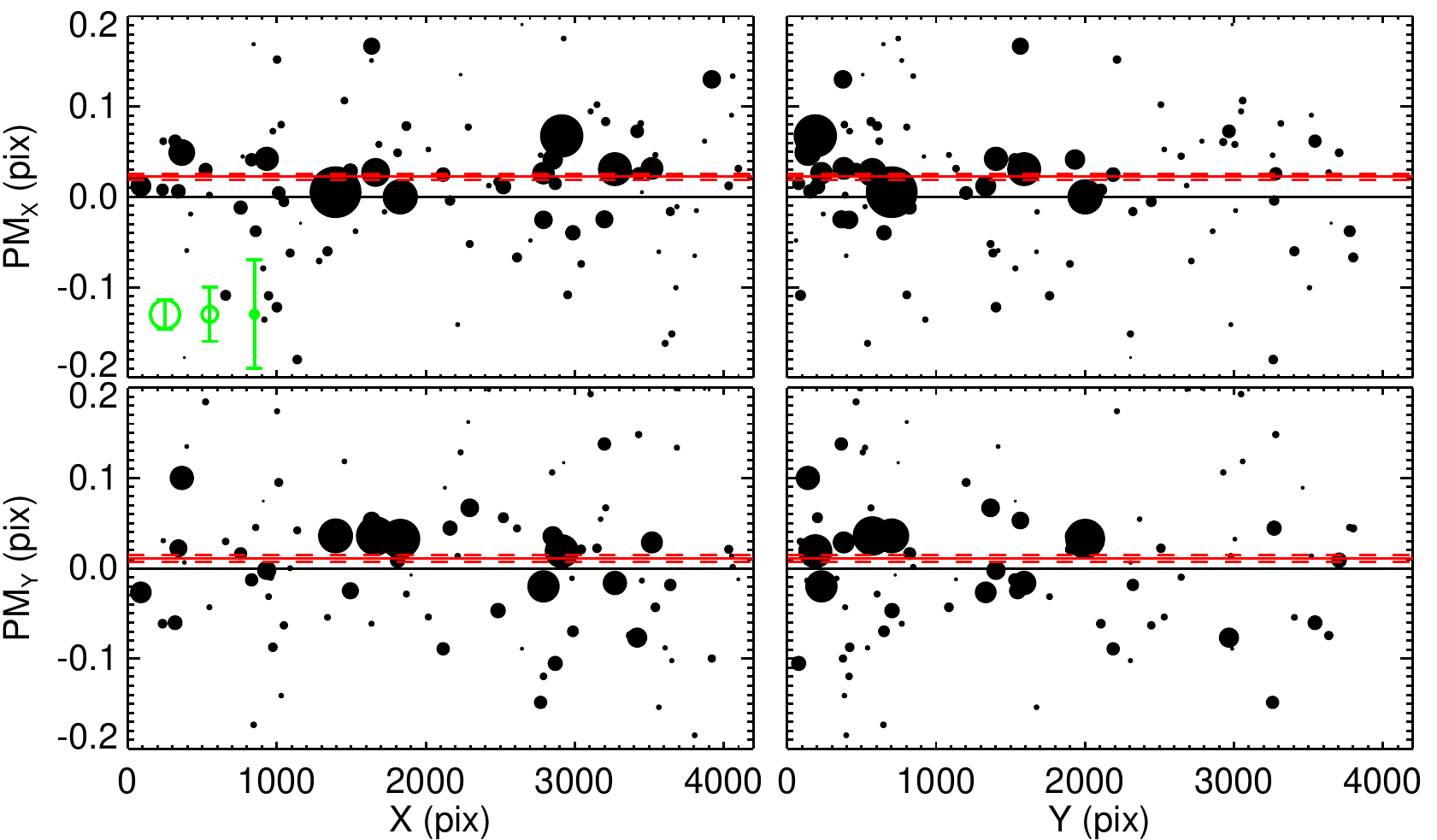}
\caption{Displacements of background galaxies versus detector 
location between one of the 1,700 sec first-epoch images 
({\tt j9gz04tsq}) and the average of the second-epoch images, plotted 
separately for $X$ and $Y$ positions. The black points show the 
relative displacements measured for different background galaxies.
The weighted average for all galaxies is shown as the red line; 
dashed red lines indicate the 1-$\sigma$ error region around the 
average. This region is smaller than the scatter between the points 
by a factor of $\sim\sqrt{N}$, where $N$ is the number of background 
galaxies. The radius of each black point is proportional to $1/\Delta$
, where $\Delta$ is the proper motion measurement uncertainty for the 
particular background galaxy. Hence, the area of each point is 
proportional to the weight a point receives in the final weighted 
average. Symbols in green in the top left panel illustrate how the 
point size relates to the proper motion uncertainty $\Delta$. The 
units are in native ACS/WFC pixels, and $X$ and $Y$ positions are in 
the reference frame.
\label{fig:glxypm1}}
\end{figure*}

Figure~\ref{fig:glxypm1} shows the measured proper motion of each
background galaxy in $X$ and $Y$ as a function of detector
coordinates, for the same 1,700 sec first-epoch image as in
Figure~\ref{fig:starpm1}. The proper motion of each background galaxy
is measured with respect to the average second-epoch position of that
galaxy, including the local correction. There are far more stars than
background galaxies in our images, and star positions are generally
determined more accurately than galaxy positions. Hence, the final
proper motion uncertainty is dominated by the astrometric accuracy for
the background galaxies. For each individual first-epoch exposure we
take the weighted average over all background galaxies (larger symbols
in Figure~\ref{fig:glxypm1} denote background galaxies with more
accurate measurements, which receive more weight) to obtain a 
individual Leo~I proper motion estimate (red line). The 1-$\sigma$ 
confidence region around the weighted average (dashed lines) was 
computed using the bootstrap method \citep{efr93}, with 10,000 
bootstrap samples.


\subsection{Inferred Proper Motion}
\label{sec:ObservedPM}

\begin{figure}
\epsscale{1.0}
\plotone{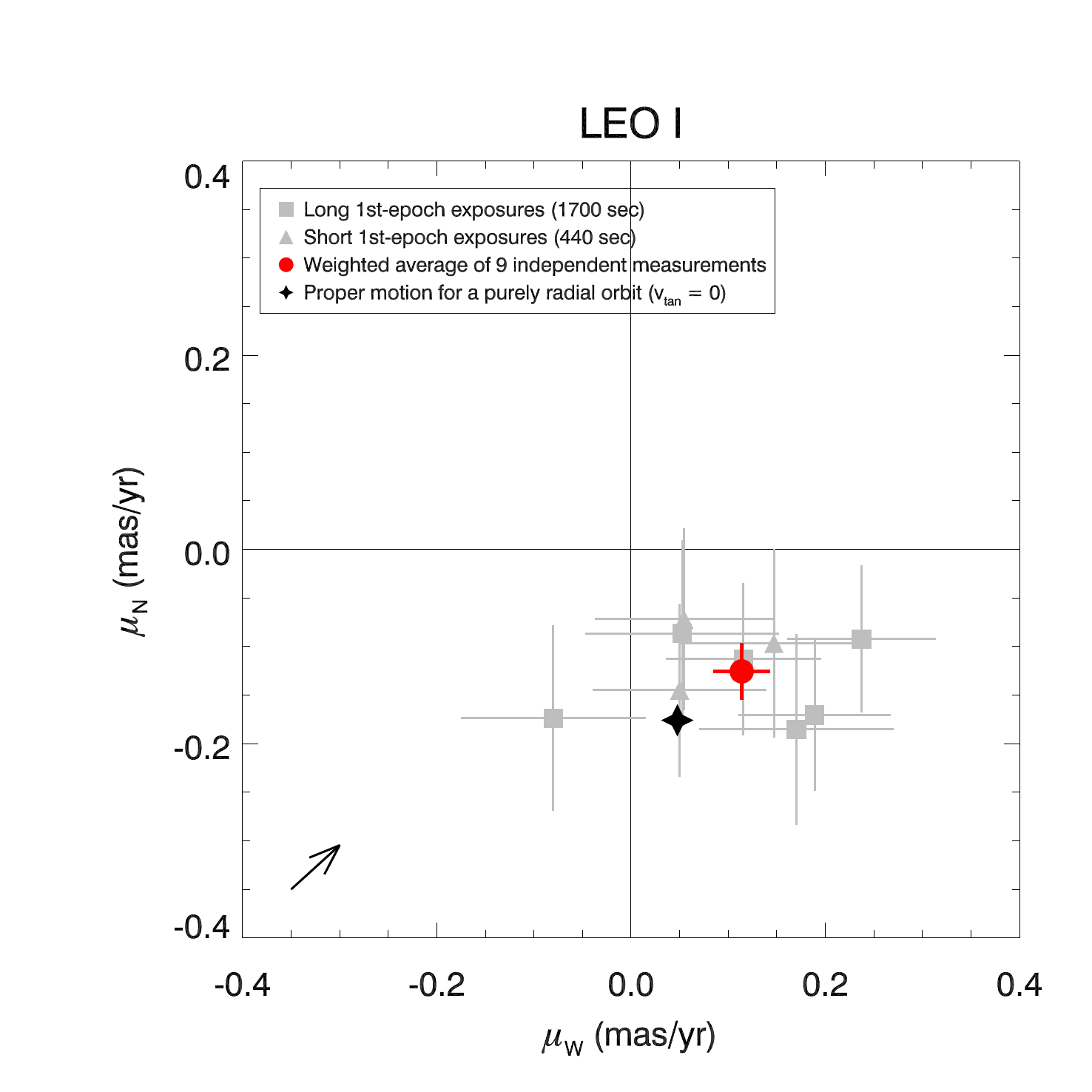}
\caption{Leo~I proper motion results. Each gray dot with an error bar
  indicates the average proper motion of Leo~I stars inferred from a single
  first-epoch exposure. Measurements using images with longer (closed
  square) and shorter (closed triangle) exposure times are indicated
  with different symbols. The solid red data point is the weighted
  average of the nine separate measurements, which is the final result
  of our analysis. The origin corresponds to the velocity such that
  Leo~I has no transverse motion in the heliocentric rest frame. The
  star symbol corresponds to the velocity such that Leo~I has no
  tangential velocity in the Galactocentric rest frame (i.e., a radial
  orbit with respect to the MW). The line segment in the bottom left 
  corner indicates the CCD $Y$-direction. The two ACS/WFC CCDs are 
  read out in the $\pm Y$-directions, respectively.
\label{fig:PMDiagram}}
\end{figure}

The proper motion diagram for the 9 independent first-epoch
measurements is shown in Figure~\ref{fig:PMDiagram}. We plot the data
points for the longer (1,700 sec; open squares) and shorter (440 sec;
open triangles) exposures in the same diagram. We transformed the
proper motions and their associated errors along the detector axes to
the directions West and North using the orientation of the reference
image with respect to the sky ($-47$\fdg 7). Table~\ref{tab:PMresults}
lists the proper motion for each first-epoch image and the
corresponding error, along with the number of background galaxies used
for the proper motion derivation. 

The final average proper motion of Leo~I is calculated by taking the
error-weighted mean of the 9 independent measurements listed in
Table~\ref{tab:PMresults}. This yields 
\begin{equation}
\label{PMleo}
  (\mu_{W}, \mu_{N}) =
  (0.1140 \pm 0.0295, -0.1256 \pm 0.0293)\ {\rm mas\ yr}^{-1} .
\end{equation}
This result differs from zero at approximately 4-$\sigma$ confidence 
in each coordinate direction, so the detected motion of Leo~I is very 
statistically significant.

The quantity 
\begin{equation}
  \chi^2 = \sum_{i} 
     \left[
     \left ( \frac{ \mu_{W,i} - {\overline \mu_{W}} }
               { \Delta \mu_{W,i} } \right )^2 
     +
     \left ( \frac{ \mu_{N,i} - {\overline \mu_{N}} }
               { \Delta \mu_{N,i} } \right )^2
     \right]
\end{equation}
provides a measure of the extent to which different measurements agree
to within the random errors. In absence of systematic errors, one
expects that this quantity follows a $\chi^2$ probability distribution
with $N_{DF} = 18 - 2 = 16$ degrees of freedom.  The expectation value
for such a distribution is $N_{DF}$, and the dispersion is 
$\sim\sqrt{2N_{DF}} = 5.7$. We find $\chi^{2} = 11.2$ for our 
measurements. This indicates that the measurements from the different 
exposures are consistent, and that the errors may actually be slight 
overestimates.\footnote{We could have rescaled the final random error 
downward by a factor $\sqrt{11.2/16} = 0.84$ to force the $\chi^2$ to 
match the number of degrees of freedom. However, we chose not to do so, 
in part because there is another tendency in our technique to slightly 
{\it underestimate} the final random error. We treat the measurements 
in Figure 5 and Table 1 as independent, because they are based on
different first-epoch exposures. However, the proper motions are all
based on comparison to the same average of second-epoch
exposures. This causes a small amount of covariance between the
estimates (so they are not truly independent). Accounting for this
would increase the final Leo~I proper motion random error by
10\%. This roughly balances the effect above, so we decided not apply
any corrections. Either way, differences at the 10-20\% level in the
quoted random error have negligible impact on any of the conclusions
of our paper.}

The final Leo~I proper motion uncertainties correspond to $\sim 29\ 
\mu$as yr$^{-1}$. This is a factor of $>2$ larger than what we 
achieved for M31 \citep{soh12}, due mostly to the fact that for M31 
deeper exposures were available for three different fields. Our M31 
\hst\ measurements approached the accuracy achieved using VLBA water 
maser observations for the M31 satellites M33 and IC10. Whereas this 
is not the case here, our Leo~I measurements are more accurate than 
what has been achieved with HST for other MW satellites, using one or 
more fields centered on background quasars \citep[see the compilation
in Table~4 of][]{wat10}.

\begin{deluxetable}{lccc}
\tablecaption{Proper Motion Results\label{tab:PMresults}}
\tablehead{
   \colhead{} & \colhead{$\mu_{W}$} & \colhead{$\mu_{N}$} &  \\
   \colhead{Data Set} & \colhead{(mas yr$^{-1}$)} & \colhead{(mas yr$^{-1}$)} & \colhead{$N_{\rm used}$\tablenotemark{a}}
   }
\startdata
{\tt j9gz04tsq} & \phs0.2374 $\pm$ 0.0763 & $-$0.0922 $\pm$ 0.0759 &    100 \\
{\tt j9gz04ttq} & \phs0.1156 $\pm$ 0.0801 & $-$0.1129 $\pm$ 0.0784 &    101 \\
{\tt j9gz04tvq} & \phs0.1892 $\pm$ 0.0785 & $-$0.1704 $\pm$ 0.0779 & \phn90 \\
{\tt j9gz05tyq} & \phs0.1702 $\pm$ 0.1003 & $-$0.1853 $\pm$ 0.0984 & \phn96 \\
{\tt j9gz05tzq} &  $-$0.0800 $\pm$ 0.0950 & $-$0.1737 $\pm$ 0.0957 & \phn86 \\
{\tt j9gz05u1q} & \phs0.0528 $\pm$ 0.0996 & $-$0.0865 $\pm$ 0.0962 & \phn98 \\
{\tt j9gz06krq} & \phs0.1471 $\pm$ 0.0977 & $-$0.0966 $\pm$ 0.0970 & \phn56 \\
{\tt j9gz06ksq} & \phs0.0545 $\pm$ 0.0917 & $-$0.0718 $\pm$ 0.0939 & \phn53 \\
{\tt j9gz06kuq} & \phs0.0501 $\pm$ 0.0894 & $-$0.1449 $\pm$ 0.0891 & \phn54 \\
\tableline
weighted av.\tablenotemark{b} & \phs0.1140 $\pm$ 0.0295 & $-$0.1256 $\pm$ 0.0293 & 
\enddata
\tablenotetext{a}{Number of background galaxies used for deriving the average proper motion of each field.}
\tablenotetext{b}{Weighted average of the results for the 9 independent measurements.}
\end{deluxetable}


\subsection{Control of Systematic Errors}
\label{sec:Systematic}

The technique used here for measuring proper motions is identical to
that used in our study of M31, as described in \citet{soh12}. As
discussed in detail in section 4.3 of that paper, the technique has
many built-in features to minimize the impact of systematic errors on
the measurement. For example, we explicitly correct for Y-CTE using the
technique of \citet{and10}, we use different geometric distortion 
solutions for the two epochs, and we model PSF variations between 
epochs. Moreover, any remaining astrometric residuals from
these instrumental effects will have very limited impact on our proper
motion measurement. This is because our measurement is a differential
one between stars and background galaxies, observed at the same time
on the same detector. Through our local correction, we restrict the
differential comparison to sources of similar magnitude observed on
the same part of the detector. This effectively minimizes both geometric
distortion and PSF residuals (which depend on detector position) and
CTE-residuals (which depend on detector position and source
magnitude). Also, our final random proper motion errors in Table 1 are
calculated through bootstrapping, which means that they reflect the
scatter between results from different background galaxies. Any
systematic proper motion residuals that vary with position on the
detector (e.g., from CTE) are therefore accounted for in the random
errors.

Despite our best efforts, there will always be some remaining level of
systematic error in the final proper motion measurement. This could
be, e.g., from the fact that the astrometric CTE impact is different
for point sources and extended sources, from the fact that we have
not explicitly corrected for X-CTE (which is only $\sim 1$\% of the
size of Y-CTE), from color-difference effects, or from other
higher-order effects. In the context of our M31 proper motion study,
we therefore set up the observational experiment such that we would
have several independent limits on the size of any remaining
systematic errors: (1) we observed an object for which an entirely
independent estimate exists of the transverse velocity \citep{vdm08}; 
(2) we observed with two different instruments (ACS/WFC and WFC3/UVIS) 
for the second epoch; and (3) we chose to observe, at different times 
and with different telescope orientation, three well-separated fields, 
which have different types and numbers of background galaxies, and 
different level of stellar densities. We found that different methods, 
different instruments, and different fields, all yielded proper motion 
answers with our \hst\ technique that are statistically consistent. 
Fig. 13 of \citet{soh12} and Fig. 3 \citet{vdm12a} show that there is 
agreement to better than $\sim 100$ km/s, which at the distance of 
M31 corresponds to $0.027$ mas/yr. This scatter can be attributed 
entirely to known random errors, and this therefore sets a rigorous 
and conservative upper limit to any possible systematic errors in our 
technique.

Since systematic errors in our technique had been previously
validated, we did not set up our Leo~I experiment to provide a similar
level of independent systematic error validation. For efficiency of
\hst\ usage, we observed only one field, with one instrument. However,
our setup is otherwise very similar to in our M31 study: we again
compare two epochs of ACS/WFC data, one taken before the Service
Mission 4 (SM4) and the other after the SM4, with a 5 year time
baseline. Therefore, any remaining systematic proper errors in our 
Leo~I result should be similar to what was present in the M31 result, 
and that was rigorously and conservatively bounded by $0.027$ mas/yr. 
This is below the random error in our Leo I result. Therefore, any
systematic errors in our result should be below the quoted random
error.

We also performed a variety of additional tests on our Leo~I data to
confirm that indeed no unidentified systematics are present. We did
not find any dependence of the proper motions of the stars and
background galaxies in the field on $I$-magnitude, $B-I$ color, or
source extent (FWHM). Using a different magnitude range in the local
corrections didn't change the final proper motion result by more than
the random errors. Comparing results for background galaxies near to
and far from the read-out amplifiers yielded consistent results given 
the the uncertainties: when using only galaxies close to the amplifiers 
(i.e., limited CTE losses), our results change to $(\mu_{W}, \mu_{N}) = 
(0.0875 \pm 0.0392, -0.1703 \pm 0.0390)\ {\rm mas\ yr}^{-1}$, while 
using only galaxies far from the amplifiers (i.e., higher CTE losses), 
our results changed to $(\mu_{W}, \mu_{N}) = (0.1467 \pm 0.0403, 
-0.0504 \pm 0.0396)\ {\rm mas\ yr}^{-1}$. These results differ from 
our final PM results (Equation~1) by (0.7$\sigma$, 
1.1$\sigma$) and (0.8$\sigma$, 1.9$\sigma$), respectively. As a test, 
we also reduced the data without inclusion of the \citet{and10} 
pixel-space CTE correction. This is clearly wrong, since we know that
CTE is present and well-corrected by this correction, but even this
only changed the final proper motion result by an average of $1.5$
times the random error per coordinate. So overall, we have no reason
to believe that any systematic errors are present in our final proper
motion result at a level that exceeds the quoted random errors.


\section{The Orbit of Leo~I}
\label{sec:Orbit}


\subsection{Velocity in the Galactocentric Rest Frame}
\label{sec:GRF}

We adopt a Cartesian Galactocentric coordinate system ($X, Y, Z$),
with the origin at the Galactic Center, the $X$-axis pointing in the
direction from the Sun to the Galactic Center, the $Y$-axis pointing
in the direction of the Sun's Galactic rotation, and the $Z$-axis
pointing towards the Galactic North Pole. The position and velocity 
of an object in this frame can be determined from the observed sky 
position, distance, line-of-sight velocity, and proper motion, as in,
e.g., \citet{vdm02}.

To determine the Galactocentric position of an object, it is necessary
to also know the distance $R_0$ of the Sun from the Galactic Center.
Moreover, it is necessary to know the velocity of the Sun inside the
MW to turn observed heliocentric rest-frame velocities into 
Galactocentric rest-frame velocities. Following \citet{vdm12a}, we
adopt the recent values of \citet{mcm11} for the distance of the Sun
from the Galactic center and the circular velocity of the local
standard of rest (LSR): $R_{0} = 8.29 \pm 0.16$ kpc and $V_{0} = 239
\pm 5$ km s$^{-1}$. For the solar peculiar velocity with respect to
the LSR we adopt the estimates of \citet{sch10}: $(U_{\rm pec}, 
V_{\rm pec}, W_{\rm pec}) = (11.10, 12.24, 7.25)$ km s$^{-1}$ with 
uncertainties of $(1.23, 2.05, 0.62)$ km s$^{-1}$.

To obtain the distance of Leo~I, we average the distances measured 
via the tip of the red-giant branch (TRGB) method in the last decade
\citep{men02,bel04,hel10}, which yields $256.7\pm 13.3$ kpc. This
implies a Galactocentric $(X, Y, Z)$ position 
\begin{equation}
  {\vec r} = (-125.0, -120.8, 194.1) \kpc ,
\end{equation}
with an uncertainty of 13.3 kpc along the line-of-sight direction.

The most recent measurement of the systemic heliocentric line-of-sight
velocity of Leo~I is $v_{\rm LOS} = 282.9 \pm 0.5$ km s$^{-1}$
\citep{mat08}. The measured proper motion $\mu$ from 
equation~(\ref{PMleo}) corresponds to a heliocentric transverse 
velocity in km/s equal to $4.7404 \times D({\rm kpc}) \times \mu 
({\rm mas\ yr}^{-1})$. 
This implies
\begin{equation}
  (v_W,v_N) = (138.7 \pm 36.6, -152.8 \pm 36.5) \kms
\end{equation}  
with proper motion errors dominating over distance errors in the
determination of the velocity errors. The internal velocity dispersion
of Leo~I is $\sigma = 9.2 \pm 0.4 \kms$, with little evidence for
rotation \citep{mat08}. This is well below our observational velocity
errors. Hence, there is no need to correct the observed values for the
internal kinematics of Leo~I, even though our field was offset 
from its photometric center (see Figure~\ref{fig:ACSPointing}). 

The velocity for which Leo~I would be on a radial orbit with respect
to the MW (i.e., the velocity for which there is zero tangential 
velocity in the Galactocentric rest frame) is
\begin{equation}
  (v_W,v_N)_{\rm rad} = (58.0, -214.0) \kms .
\end{equation}
This differs from the measured proper motion at almost 3-$\sigma$ 
significance. Therefore, our measurements imply that Leo~I is not 
on a radial orbit about the MW.

Several authors have argued previously that the Galactocentric 
tangential velocity of Leo~I is probably small, given its significant
radial velocity \citep{byr94,soh07,mat08}. Indeed, 
Figure~\ref{fig:PMDiagram} shows that the observed proper motion does 
fall in the same quadrant of proper motion space as a radial orbit. 
This can be interpreted as a consistency/plausibility check on the 
proper motion measurement. The same was found for the case of M31, 
for which we also presented several other successful consistency 
checks on our proper motion analysis methodology \citep{soh12}.

Conversion into the Galactocentric rest frame yields for the velocity
vector of Leo~I
\begin{equation}
  {\vec v} = (-167.7 \pm 31.9, -37.0 \pm 33.0, 94.4 \pm 24.2) \kms .
\end{equation}
The listed uncertainties here and hereafter were obtained from a
Monte-Carlo scheme that propagates all observational distance and
velocity uncertainties and their correlations, including those for 
the Sun. Note that the Galactocentric velocity uncertainties are 
highly correlated because the observational velocity uncertainty is 
much larger in the transverse direction than in the line-of-sight
direction. 

The corresponding Galactocentric radial and tangential velocities are
\begin{equation}
  (V_{\rm rad}, V_{\rm tan}) = 
   (167.9 \pm 2.8, 101.0 \pm 34.4) \kms .
\label{vradtaneq}
\end{equation}
Although the tangential velocity is significantly non-zero, it is 
less than the radial velocity. So while Leo~I is not on a radial 
orbit about the MW, the orbit must be fairly elliptical. The observed 
total Leo~I velocity with respect to the MW is
\begin{equation}
  v \equiv |{\vec v}| = 196.0 \pm 19.4 \kms 
\label{vtoteq}
\end{equation}
with the listed numbers corresponding to the peak and the symmetrized
1$\sigma$ of the $v$ probability distribution.\footnote{The error
  distribution of the total velocity $v$ is somewhat asymmetric, but
  this is more pronounced at the $2\sigma$ than at the $1\sigma$
  level. The median of the $v$ distribution and the surrounding 68\%
  (95\%) confidence intervals are $v = 199.8^{+21.8 (+47.0)}_{-17.1
  (-29.3)} \kms$, as used in Paper~II.}

These inferred Leo~I velocities use a solar velocity inside the MW
based on \citet{sch10} and \citet{mcm11}, which yields an azimuthal 
velocity component $v_{\phi,\odot} = 251.2 \kms$. However, alternative 
values for the solar velocity continue to be in common use. These 
differ from the values used here primarily in the azimuthal direction. 
For example, with the old IAU recommended circular velocity 
$V_0 = 220 \kms$ and the peculiar velocities from \citet{deh98}, 
$v_{\phi,\odot} = 225.2 \kms$. Based on these latter values,
the Galactocentric Leo~I velocities would be $V_{\rm rad} = 180.5
\kms$, $V_{\rm tan} = 93.8 \kms$, and $v = 203.4 \kms$, which can be
compared to the values in equations~(\ref{vradtaneq})
and~(\ref{vtoteq}). While the change in $V_{\rm rad}$ is significant
compared to the small uncertainties, $V_{\rm tan}$ and $v$ change by
much less than the observational uncertainties. The conclusions of the
present paper and Paper~II are therefore not very sensitive to the
adopted solar velocity. The value of $v_{\phi,\odot}$ assumed here is
consistent with the recent determination of $v_{\phi,\odot} =
242_{-3}^{+10} \kms$ by \citet{bov12}, while the alternative
$v_{\phi,\odot} = 225.2 \kms$ discussed above is not.\footnote{
  \citet{bov12} advocate for a lower circular velocity than
  \citet{mcm11}, but the only quantity that matters for the
  calculations presented here is $v_{\phi,\odot}$.}


\subsection{Keplerian Orbit Calculations}
\label{sec:kep}

To assess the implications of the new measurements, we start with the
assumption that the MW can be approximated as a point mass, and that
Leo~I orbits in its potential as a test particle on a Keplerian orbit.
The assumption of a Keplerian potential for the MW is not as 
unreasonable as it may seem at first. The large Galactocentric 
distance of Leo~I, $r \equiv |{\vec r}| = 260.6$ kpc, combined with 
its significant tangential velocity, implies that much of the MW's 
mass is inside the Leo~I orbit at all times. We calculate models with 
more realistic MW potentials in Section~\ref{sec:orbits}.

The escape velocity for a point mass $M_{\rm MW}$ is 
\begin{equation}
  v_{e} = \sqrt{2GM_{\rm MW}/r} .
\end{equation}
Hence, given the observed total Leo~I velocity with respect to the MW 
given by equation~(\ref{vtoteq}), Leo~I is bound to the MW if 
$M_{\rm MW} \geq (1.16 \pm 0.24) \times 10^{12} \Msun$. Cosmological 
simulations imply that it is very unlikely to find an unbound satellite 
at the present epoch near a MW-type galaxy 
\citep[e.g.,][and Paper~II]{dea11}. So if we assume that Leo~I must be 
bound to the MW, then this can be interpreted as a new crude lower 
limit on the MW mass.

Alternatively, one may assume that the mass of the MW is already 
constrained from other arguments. In that case, one can use the new 
measurement to assess the probability that Leo~I is in fact bound. 
Studies of the MW mass have advocated many different values, roughly 
covering the range $0.75$--$2.25 \times 10^{12} \Msun$ 
\citep[see][for a compilation of recent mass estimates of the MW]{boy12}.
\citet{vdm12a} assumed a flat prior probability over this range, and 
then used a Bayesian scheme to include the latest measurements of the 
Local Group timing mass, based on our M31 HST proper motion work. 
The Local Group timing mass is relatively high, which increases the 
likelihood of high MW masses compared to low MW masses. We combined 
the probability distribution for the MW mass from Figure~4 of 
\citet{vdm12a} with the measured $v$ for Leo~I. This implies that 
there is a 77\% probability that Leo~I is bound to the MW, and a 23\% 
probability that it is not bound. The preference for a bound state 
is consistent with expectations from cosmological simulations 
\citep[e.g.,][]{ben05,wet11}.

For any assumed point mass $M_{\rm MW}$, and given
Galactocentric Leo~I phase-space vectors ${\vec r}$ and ${\vec v}$,
the shape of the Keplerian orbit is determined analytically. We
calculated these orbits in a Monte-Carlo sense. At each Monte-Carlo
step we draw a mass $M_{\rm MW}$ from the previously discussed
probability distribution derived by \citet{vdm12a}, and we draw
Leo~I phase-space vectors from the observationally determined values 
and uncertainties. We then determine the statistics of the orbital 
characteristics over the Monte-Carlo sample. 

The Monte-Carlo analysis yields an average ratio of pericenter to 
apocenter distance $r_{\rm peri}/r_{\rm apo} = 0.06 \pm 0.03$ for the 
bound elliptical Keplerian orbits (i.e., orbits with eccentricity 
of less than 1). Leo~I has a positive radial velocity, and is therefore 
past pericenter. The pericentric passage occurred at $t_{\rm peri} = 
1.18 \pm 0.45$ Gyr ago at a Galactocentric distance of $r_{\rm peri} 
= 67 \pm 39$ kpc. The velocity at pericenter was $v_{\rm peri} = 
561 \pm 475 \kms$. The uncertainties in these quantities are determined 
largely by the uncertainties in the Leo~I phase-space vectors, and much 
less so by uncertainties in MW mass. For example, if $M_{\rm MW}$ is kept 
fixed at $1.5 \times 10^{12} \Msun$ for all Monte Carlo drawings, 
then the orbital characteristics become: $r_{\rm peri}/r_{\rm apo} = 
0.04 \pm 0.02$, $t_{\rm peri} = 1.08 \pm 0.13$ Gyr, $r_{\rm peri} = 
69 \pm 38$ kpc, $v_{\rm peri} = 518 \pm 417 \kms$. In 
Section~\ref{sec:orbits} we compare these simple Keplerian results to 
the results from orbit calculations in more detailed 
cosmologically-motivated halo models.

Another useful application of Keplerian orbits is through the timing
argument \citep{kah59}. This argument assumes that bound galaxy pairs
follow a Newtonian Keplerian trajectory starting soon after the Big
Bang, which corresponds to the ``first pericenter.'' The galaxies
initially move away from each other due to the expansion of the
Universe, but then fall back towards each other due to gravity.  In
this picture, Leo~I is just passed its second pericenter.  In general,
there are four observables (the time since Big Bang $t$, relative
distance $r$, radial velocity $V_{\rm rad}$, and tangential velocity
$V_{\rm tan}$) and four independent orbital parameters (eccentric
anomaly $\eta$, semi-major axis length $a$, eccentricity $e$, and the
total mass $M$). Hence, the Keplerian orbit can be solved for
analytically, as described in e.g., \citet{vdm08}. One may call this
the ``complete timing argument'' (cta). In many applications however,
the transverse velocity $V_{\rm tan}$ is not known and it is then
often assumed that $V_{\rm tan} = 0$ and $e=1$. This yields the
so-called ``radial-orbit timing argument'' (rta).

The timing argument has traditionally been applied to the MW--M31
system \citep[see][and references therein]{vdm12a}, but it can also be
applied to the MW--Leo~I system \citep{zar89}. The radial-orbit timing
argument as applied to Leo~I with the previously derived
Galactocentric position and velocity implies a mass $M_{\rm MW,rta} =
(1.50 \pm 0.12) \times 10^{12} \Msun$ \citep[consistent with the value
previously inferred by][using a slightly different assumed solar 
velocity]{li08}. Any tangential velocity increases the timing mass. 
Since we have now measured the tangential velocity of Leo~I, we can 
use instead the complete timing argument without assuming a radial 
orbit. This implies a mass $M_{\rm MW,cta} = (1.93 \pm 0.42) \times
10^{12} \Msun$.

The error bars in the listed timing masses reflects only the
propagation of errors in the observational quantities. However, it is
important to also quantify any inherent biases and cosmic scatter, and
to calibrate the timing mass to more traditional measures of mass,
such as the virial mass. \citet{li08} addressed these issues for the
radial-orbit timing argument using the cosmological Millennium
simulation.  They identified a set of host galaxy - satellite pairs
with properties similar to the MW-Leo~I pair. For these pairs
they studied the statistics of the ratio $M_{\rm 200}/M_{\rm rta}$. 
In the following we find it more convenient to use the virial mass
$M_{\rm vir}$ rather than the quantity $M_{\rm 200}$, so where
necessary we transform the latter to the former using $M_{\rm
vir}/M_{200} = 1.19$ \citep[as appropriate for an NFW halo of
concentration $c=9.5$; see the Appendix of] [for the relevant
equations and mass definitions]{vdm12a}.

Yang-Shyang Li kindly made available the catalog of galaxy pairs used
in the analysis of \citet{li08} (the sample defined by the bottom row
of their Table~3). This allowed us to perform an analysis similar to
theirs, but now for the complete timing argument. We find that the
bias $M_{\rm vir}/M_{\rm cta} = 1.46_{-0.62}^{+0.89}$. This estimate
was obtained, as in the \citet{li08} analysis, by averaging over all
satellites in the simulation sample, independent of tangential
velocity. However, we have now measured the tangential velocity of
Leo~I. So we can get a more appropriate measure of the bias by
including only the satellites with tangential velocities similar to
that of Leo~I.  If we require agreement in $V_{\rm tan}$ to within 
$25 \kms$, we find that $M_{\rm vir}/M_{\rm cta} = 1.63_{-0.61}^{+0.74}$. 
For comparison, this same selection yields for the radial-orbit timing 
argument that $M_{\rm vir}/M_{\rm rta} = 2.10_{-0.69}^{+0.95}$. 
So the complete timing argument yields estimates of $M_{\rm vir}$ that 
are biased low, but not by as much as the radial orbit timing argument. 
This is because part of the bias is due to the fact that satellite 
galaxies generally have non-zero tangential velocities, and this is 
explicitly taken into account in the complete timing argument.

These results for cosmic bias and scatter can be combined with the
previously inferred values of $M_{\rm MW,cta}$ and $M_{\rm MW,rta}$.
This yields $M_{\rm MW,vir} = 3.15_{-1.36}^{+1.58} \times 10^{12}
\Msun$ from the complete timing argument, and $M_{\rm MW,vir} =
3.14_{-1.06}^{+1.45} \times 10^{12} \Msun$ from the radial-orbit
timing argument, respectively. So the two timing arguments give
similar results and uncertainties. The mass estimates are higher than
most MW mass estimates based on other methods \citep[consistent with
the results of][]{li08}, but they are probably not inconsistent with
other MW mass estimates given the significant cosmic scatter. This
situation is similar to what was found for MW mass estimates based on
the timing of the MW-M31 system \citep{vdm12a}.


\subsection{Detailed Orbit Integrations}
\label{sec:orbits}


\subsubsection{Methodology \& Overview}
\label{subsec:methodology}

To get a better understanding of the past orbital history of Leo~I we
need to use more detailed models for the MW's gravitational potential 
$\Phi_{\rm MW}$. Following \citet{bes07}, we describe this 
potential as a static, axisymmetric, three-component model consisting 
of dark matter (DM) halo, disk \citep{miy75} and stellar 
bulge \citep{her90}:
\begin{eqnarray}
\Phi_{\rm MW} = \Phi_{\rm DM~halo}  + \Phi_{\rm disk} + \Phi_{\rm bulge} .
\end{eqnarray}
The DM halo is initially modeled as an NFW halo \citep{nav97} with a 
virial concentration parameter ($c_{\rm vir}$) defined as in 
\citet{kly11} from the Bolshoi Simulation \citep[see also,][]{vdm12b}. 
We apply the adiabatic contraction of the NFW halo in response to 
the slow growth of an exponential disk using CONTRA code \citep{gne04}.
The density profile of the MW is then truncated at the virial radius
\footnote{The virial radius is defined as the radius such that 
$\rho_{\rm vir} = \Delta_{\rm vir}\Omega_{m}\rho_{\rm crit}$, where 
the average overdensity $\Delta_{\rm vir}$ = 360 and the mean matter 
density parameter $\Omega_{m} = 0.27$ 
\citep[see equation A1 in][]{vdm12a}.}. The bulge mass of the MW is 
kept fixed at $10^{10} \Msun$ with a Hernquist scale radius of 0.7 
kpc. The exponential disk scale length is also kept fixed at 3.5 kpc. 

We adopt three different mass models for the MW with total virial
masses of $1.0 \times 10^{12}$, $1.5\times 10^{12}$, and 
$2.0 \times 10^{12} \Msun$. In all cases, the bulge is modeled with 
a scale radius of 0.7 kpc and a total mass of $10^{10} \Msun$. 
The disk scale radius is also kept fixed at 3.5 kpc, but the disk mass 
is allowed to vary to reproduce the circular velocity at the Solar 
circle. We adopt a MW circular velocity of 239 km s$^{-1}$ at the 
solar radius of 8.29 kpc \citep{mcm11}. Model parameters are listed in 
Table~\ref{tab:ModelParamsMW}.

\begin{deluxetable}{cccc}
\tablecaption{Model Parameters for the Milky Way\label{tab:ModelParamsMW}}
\tablehead{
\colhead{$M_{\rm vir}$\tablenotemark{a}} & \colhead{} & \colhead{$R_{\rm vir}$\tablenotemark{c}} & \colhead{$M_{\rm disk}$\tablenotemark{d}} \\
\colhead{($\Msun$)} & \colhead{$c_{\rm vir}$\tablenotemark{b}} & \colhead{(kpc)} & \colhead{($\Msun$)}
}
\startdata
$1.0\times10^{12}$ & 9.86 & 261 & $6.5 \times 10^{10}$ \\
$1.5\times10^{12}$ & 9.56 & 299 & $5.5 \times 10^{10}$ \\
$2.0\times10^{12}$ & 9.36 & 329 & $5.0 \times 10^{10}$
\enddata
\tablenotetext{a}{Mass contained within the virial radius.}
\tablenotetext{b}{The virial concentration parameter \citep{kly11}.}
\tablenotetext{c}{The virial radius. See text for definition.}
\tablenotetext{d}{Mass of the disks.}
\end{deluxetable}

The escape velocities at the distance $r = 260.6$ kpc of Leo~I are
$182$, $222$, and $256 \kms$ for the models with masses of $1.0 \times
10^{12}$, $1.5\times 10^{12}$, and $2.0 \times 10^{12} \Msun$,
respectively. \footnote{In a halo that is not truncated at the virial
radius, the escape velocities are larger by 45--$55 \kms$. At fixed
$M_{\rm MW,vir}$, this increases the probability that Leo~I is
bound; or at fixed probability, this means that Leo~I is bound even
at lower $M_{\rm MW,vir}$. While the escape velocity of an NFW halo
is finite, its mass is not. So truncation at some large radius is
always physically motivated.} For the latter two models, this
exceeds our best estimate $v = 196.0 \pm 19.4 \kms$ for the total
Galactocentric velocity of Leo~I. So Leo~I is most likely bound to the
MW if $M_{\rm MW,vir} \gtrsim 1.5 \times 10^{12} \Msun$, as already
suggested in Section~\ref{sec:kep}. For the lowest-mass MW model
studied here, with $M_{\rm MW,vir} = 1.0 \times 10^{12} \Msun$,
Leo~I is on an unbound hyperbolic orbit. This has repercussions
for the viability of such low mass MW models in a cosmological
context, since satellites are rarely found on hyperbolic
orbits. We explore this in more detail in Paper~II.

Using the current Galactocentric position and velocity vectors of
Leo~I as initial conditions, we can solve the differential equations
of motion numerically to follow the velocity and position of Leo~I
backward in time. If we consider only the gravitational influence of
the MW, the equation of motion has the form:
\begin{eqnarray}
  \ddot{\boldsymbol r} = \frac{\partial}{\partial\boldsymbol r} \Phi_{\rm MW}(|\boldsymbol r|),
  \label{eq:eqmotion}
\end{eqnarray}
 
In Section~\ref{sec:LeoMW} this equation of motion is solved using
well-established numerical methods in order to constrain Leo~I's
interaction history with the MW (i.e., pericentric distance and epoch
of accretion) over a Hubble time. Leo~I is modeled as a Plummer
potential with a softening length of 0.5 kpc, and total mass of
$1.3\times 10^{8} \Msun$ (see Table~\ref{tab:Sat}). With these
parameters, the dynamical mass of Leo~I within 0.93 kpc is $8.9 \times
10^{7} \Msun$, as expected from Table~2 of \citet{wal09} (i.e. the
inferred mass within the outermost data point of the empirical
velocity dispersion profile, referred to as $r_{\rm last}$ in
\citealt{wal09}). We note that dynamical friction is expected to be
negligible for such a low mass satellite (even when its extended dark
mass outside the outermost data point is included) and it is thus not
computed in the equation of motion.

\setlength{\tabcolsep}{2pt} 
\begin{deluxetable*}{lccccccccccccc}
\tablecaption{Properties of Local Group Galaxies\label{tab:Sat}}
\tablehead{ \colhead{} & \colhead{$M_{\rm total}$} & \colhead{$K$\tablenotemark{a}} & \colhead{$r_{\rm last}$\tablenotemark{b}} & \colhead{$M(<r_{\rm last})$\tablenotemark{c}} & \colhead{$X$\tablenotemark{d}} & \colhead{$Y$\tablenotemark{d}} & \colhead{$Z$\tablenotemark{d}} & \colhead{$V_X$\tablenotemark{e}} & \colhead{$V_Y$\tablenotemark{e}}        & \colhead{$V_Z$\tablenotemark{e}} & \multicolumn{3}{c}{Ref.\tablenotemark{f}}\\
\cline{12-14}
 \colhead{Galaxy} & \colhead{($\Msun$)} & \colhead{(kpc)} & \colhead{(kpc)} & \colhead{($\Msun$)} & \colhead{(kpc)} & \colhead{(kpc)} & \colhead{(kpc)} & \colhead{(km s$^{-1}$)} & \colhead{(km s$^{-1}$)} & \colhead{(km s$^{-1}$)} & \colhead{Dist.} & \colhead{RV} & \colhead{PM}
}
\startdata
Leo~I      & $1.30 \times 10^8$\phantom{$^{0}$}   & \phn0.5 & 0.93 & $8.9 \times 10^7$\phantom{$^{0}$} & $-$125.0       & $-$120.8       & \phs194.1 & $-$167.7 & \phn$-$37.0 & \phs\phn94.4 & 1,2,3            & 4  & This study\\
Carina     & $5.70 \times 10^7$\phantom{$^{0}$}   & \phn0.5 & 0.87 & $3.7 \times 10^7$\phantom{$^{0}$} & \phn$-$24.8    & \phn$-$94.8  & \phn$-$39.3 & \phn$-$72.9 & \phs\phn\phn6.9 & \phs\phn38.0 & 2,5              & 6  & 7 \\
Draco      & $3.90 \times 10^8$\phantom{$^{0}$}   & \phn0.5 & 0.92 & $2.6 \times 10^8$\phantom{$^{0}$} & \phn\phn$-$3.5 & \phn\phs76.3 & \phs\phn53.0 &   \phs\phn17.1 & \phs\phn56.0 & $-$227.9 & 8                & 9  & 10 \\
Fornax     & $1.45 \times 10^8$\phantom{$^{0}$}   & \phn0.5 & 1.70 & $1.3 \times 10^8$\phantom{$^{0}$} & \phn$-$40.0    & \phn$-$49.2      & $-$129.4 & \phn$-$24.5 & $-$140.7 & \phs106.2 & 5,11,12,13,14,15 & 16 & 17 \\
Leo~II     & $6.40 \times 10^7$\phantom{$^{0}$}   & \phn0.5 & 0.42 & $1.7 \times 10^7$\phantom{$^{0}$} & \phn$-$77.3    & \phn$-$58.3      & \phs215.3 &  \phs102.2 & \phs237.0 & \phs118.4 & 18,19            & 20 & 21 \\ 
Sculptor   & $1.35 \times 10^8$\phantom{$^{0}$}   & \phn0.5 & 1.10 & $1.0 \times 10^8$\phantom{$^{0}$} & \phn\phn$-$5.3 & \phn\phn$-$9.6       & \phn$-$84.1 &  \phn$-$19.4 & \phs224.6 & $-$101.6 & 13,22            & 23 & 24 \\   
Sextans    & $2.90 \times 10^7$\phantom{$^{0}$}   & \phn0.5 & 1.00 & $2.0 \times 10^7$\phantom{$^{0}$} & \phn$-$40.0    & \phn$-$63.6      & \phs\phn64.6 & $-$181.1 & \phs113.6 & \phs113.6 & 25               & 26 & 27 \\		
Ursa Minor   & $7.70 \times 10^7$\phantom{$^{0}$} & \phn0.5 & 0.74 & $4.4 \times 10^7$\phantom{$^{0}$} & \phn$-$22.2    & \phs\phn52.1 & \phs\phn53.6 & $-$107.5 & \phn$-$15.2 & $-$116.1 & 8                & 9  & 28 \\
LMC        & $5.00 \times 10^{10}$              &    11.0 & 9.00 & $1.3 \times 10^{10}$              & \phn\phn$-$1.1 & \phn$-$41.0      & \phn$-$27.8 & \phn$-$57.4 & $-$225.6 & \phs220.7 & 29               & 30 & 31 \\
M31        & See text  & \nodata & \nodata & \nodata                        & $-$378.9 & \phs612.7  & $-$283.1 & \phs\phn66.1 &  \phn$-$76.3 & \phs\phn45.1 & 32 & 33 & 34
\enddata
\tablenotetext{a}{Softening parameter for the Plummer profile.}
\tablenotetext{b}{Radius of the outermost data point of the 
empirical velocity dispersion profile for each satellite, as 
defined in \citet{wal09}.}  
\tablenotetext{c}{Mass inferred within $r_{\rm last}$. Note that 
for the LMC, this is the mass within the radius of the last data 
point in the carbon star analysis of \citet{vdm02}.}
\tablenotetext{d}{Galactocentric positions with the origin at the 
Galactic center, the $Z$-axis pointing toward the Galactic north 
pole, the $X$-axis pointing in the direction from the Sun to the 
Galactic center, and the $Y$-axis pointing in the direction of 
the Sun's Galactic rotation.}
\tablenotetext{e}{Galactocentric velocities with vectors pointing 
toward $X$, $Y$, and $Z$ as defined above.}
\tablenotetext{f}{Data for distance (Dist.), heliocentric radial 
velocities (RV), and proper motions (PM) were taken from the 
following references. 
 (1) \citealt{men02};   (2) \citealt{bel04};   (3) \citealt{hel10};   
 (4) \citealt{mat08};   (5) \citealt{pie09};   (6) \citealt{mat93}; 
 (7) \citealt{pia03};   (8) \citealt{bel02};   (9) \citealt{arm95}; 
(10) \citealt{sch94};  (11) \citealt{ber00};  (12) \citealt{sav00}; 
(13) \citealt{riz07a}; (14) \citealt{riz07b}; (15) \citealt{gul07}; 
(16) \citealt{wal06};  (17) \citealt{pia07};  (18) \citealt{bel05}; 
(19) \citealt{gul08};  (20) \citealt{koc07};  (21) \citealt{lep11}; 
(22) \citealt{pie08};  (23) \citealt{que95};  (24) \citealt{pia06}; 
(25) \citealt{lee03};  (26) \citealt{har94};  (27) \citealt{wal08}; 
(28) \citealt{pia05};  (29) \citealt{fre01};  (30) \citealt{vdm02}; 
(31) \citealt{kal13};  (32) \citealt{fre90};  (33) \citealt{cou99};
(34) \citealt{soh12}.
}
\end{deluxetable*}
\setlength{\tabcolsep}{5pt} 

Other members in the Local Group with significant masses may exert 
dynamical influence on the orbital history of Leo~I. Relevant to our 
analysis are the LMC and M31. It has been theorized that a number of 
satellite galaxies of the MW lie in a similar orbital plane as the 
Magellanic Clouds, referred to as the ``Magellanic Plane" of galaxies 
\citep{kun76,lyn76}. The ubiquity of this statement is directly 
relevant to cosmological studies of how satellite galaxies are 
accreted by MW mass halos \citep[e.g.,][]{don08,met09,sal11,wan13}. 
The Sagittarius dSph is known to be orbiting in a plane that is 
perpendicular to that of the Magellanic Clouds (which orbit 
approximately in the $Y-Z$ Galactocentric plane). As such, it is 
clear that at least one of the MW satellites is an outlier. Without 
accurate proper motions, it has been unclear to what extent Leo~I's 
orbit lies in a different plane or has a different rotational sense. 
Furthermore, the Keplerian orbit analysis of Section~\ref{sec:kep} 
demonstrates that Leo~I is likely on a fairly eccentric orbit. 
On such a high eccentricity orbit, it is possible that M31 may exert 
an important gravitational influence at early times.

In Section~\ref{sec:Plane} we compute the orbital histories of the 
LMC and M31 in addition to that of Leo~I to define its orbital 
plane relative to that of the LMC and also to assess the 
dynamical significance of M31 to its orbital history and origin. 
We modify the equations of motion (Eq.~\ref{eq:eqmotion}) to include 
the gravitational influence of both M31 and the LMC while computing 
the orbital history of Leo~I over the past 8 Gyr. We simultaneously 
compute the equations of motion for each of the MW, M31 and LMC, 
accounting for the gravitational acceleration from the other bodies. 
The orbits are computed in the Galactocentric frame, using the 
current velocities and positions of the LMC \citep{kal13} and M31 
\citep{vdm12a}, as listed in Table~\ref{tab:Sat}. The masses of the 
MW and M31 are assumed to be static over time, which limits the 
accuracy of this analysis past 6 Gyr \citep[i.e., when the mass of 
these galaxies is expected to be about half of their current value, 
e.g.][]{fak10}. 

M31 is modeled using an NFW halo, where its virial mass is determined
by preserving the total mass of the Local Group, given the mass of 
the MW in each model. The density profile of M31 is also modeled to 
be truncated at the virial radius. Using the recent proper motions of 
M31 by \citet{soh12} and other mass arguments in the literature,
\citet{vdm12a} estimate the Local Group mass to be $M_{\rm LG} = 
(3.17 \pm 0.57) \times 10^{12} \Msun$. In this analysis we thus 
require that the combined $M_{\rm MW} + M_{\rm M31} = 3\times 10^{12}
\Msun$. Given the large distances involved, the contribution of 
M31's disk/bulge component to the gravitational influence on Leo~I 
is irrelevant. 

The LMC is modeled as a Plummer sphere with a softening parameter of 
11 kpc and total mass of $5\times 10^{10} \Msun$. With these 
parameters, the mass contained within 9 kpc is $\sim1.3\times10^{10} 
\Msun$, as observed \citep{vdm02}. To model the orbital evolution 
of a massive satellite such as the LMC accurately, dynamical friction 
effects owing to its motion through the dark matter halo of the MW 
must be accounted for. Dynamical friction is included using the 
Chandrasekhar formula with an approximation to the Coulomb logarithm 
as in \citet{bes07}. Meanwhile, dynamical friction is irrelevant to 
the motion of the MW and M31 because their halos do not overlap over 
the past 8 Gyr. 

The star formation history of Leo~I has been the topic of many
studies, especially with {\hst} \citep{cap99,gal99,dol02,sme10}. 
From the most recent {\hst} ACS/WFC observations \citep{sme10}, 
Leo~I is known to have formed stars continuously since $> 12$ Gyr 
ago, with two pronounced star formation activities at $\sim 4.5$ and 
$\sim 2$ Gyrs ago. After this last activity, star formation abruptly 
dropped until a complete cessation at $\sim 0.5$ Gyr ago. Some of 
the inferred increases and decreases in star formation activity may 
be related to features in Leo~I's orbit about the MW, including the 
time of accretion and the pericenter time. We determine these times 
in Section~\ref{sec:LeoMW}.

The origin of enhanced star formation activities in Leo~I's past may
also be related to interaction with other satellites.  Furthermore,
three-body encounters may also alter the orbital trajectory of Leo~I
as proposed by \citet{mat08}, potentially explaining its high speed
today \citep[e.g.,][]{sal07}. To test for possible interactions with
other MW satellites, we extend in Section~\ref{sec:Sat} the analysis
of Section~\ref{sec:LeoMW} such that the equations of motion now
account for gravitational interaction terms with the following
satellites for which proper motions are available from other studies
(see Table~\ref{tab:Sat}): Carina, Draco, Fornax, Leo~II, Sculptor, 
Sextans, Ursa Minor, and the LMC. The Sagittarius dSph is not 
included in this analysis because its orbit is too close to the MW 
disk plane to dynamically influence that of Leo~I. The SMC is also 
not included because its orbit is likely closely matched to that of 
its binary companion, the LMC. Since the LMC is the more massive of 
the pair, it is likely to be the more significant perturber. Each 
satellite is modeled as a Plummer potential with total mass and 
softening parameters as listed in Table~\ref{tab:Sat}; model 
parameters are chosen to match the observed masses within 
$r_{\rm last}$, as defined in \citet{wal09}.


\subsubsection{Leo~I Orbital Properties}
\label{sec:LeoMW}

\begin{figure*}
\centering
\mbox{{\includegraphics[width=3in]{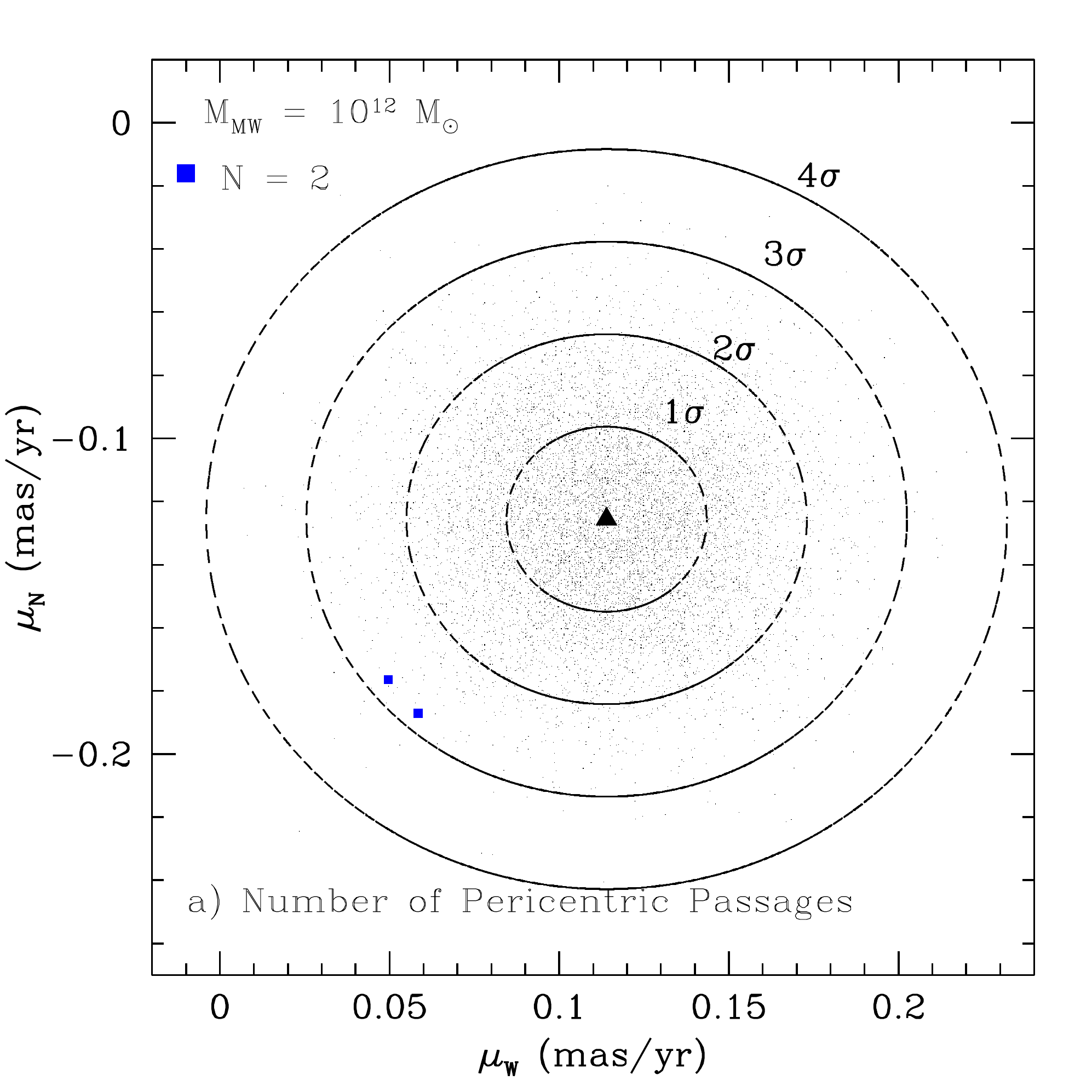}}
{\includegraphics[width=3in]{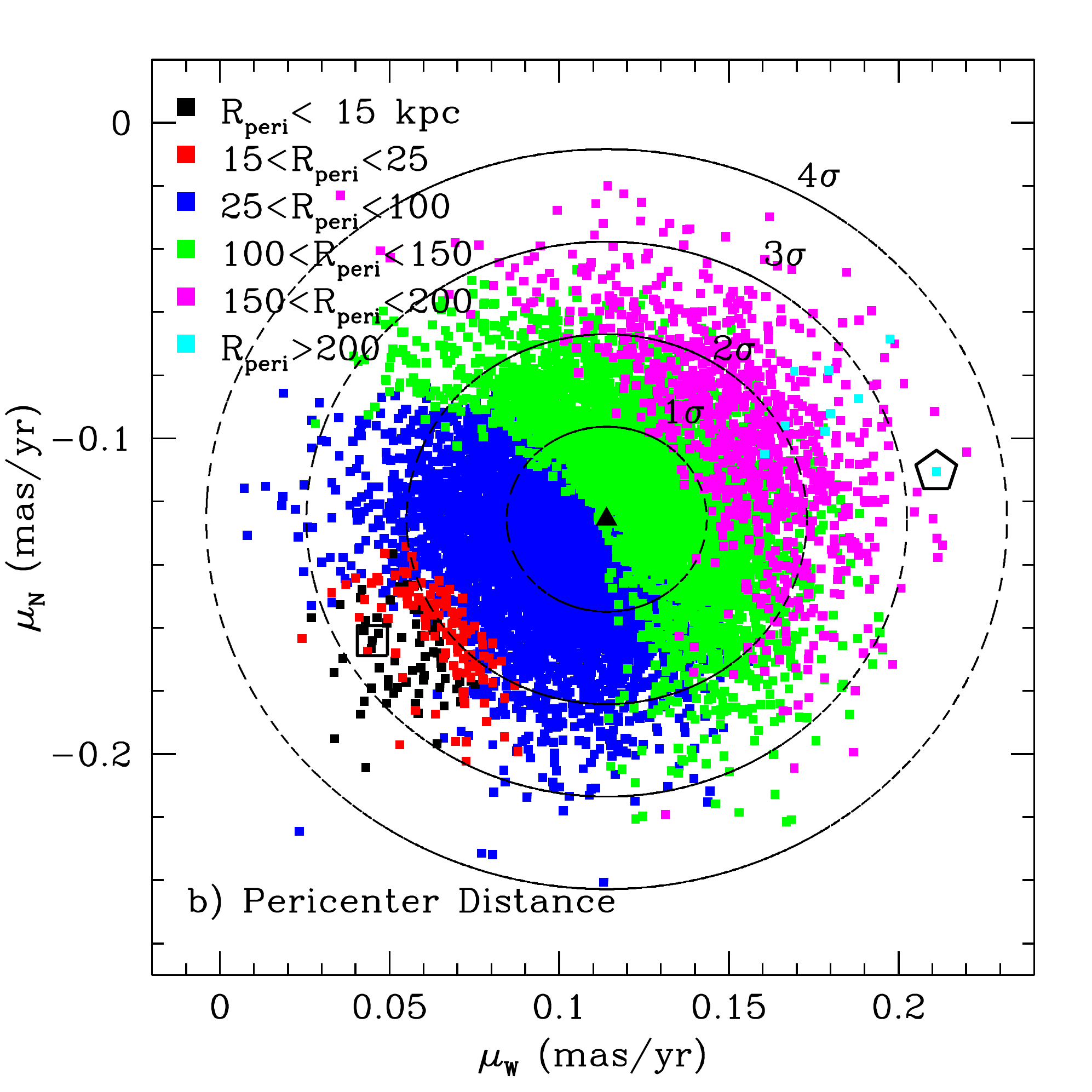}}}\\
\mbox{
{\includegraphics[width=3in]{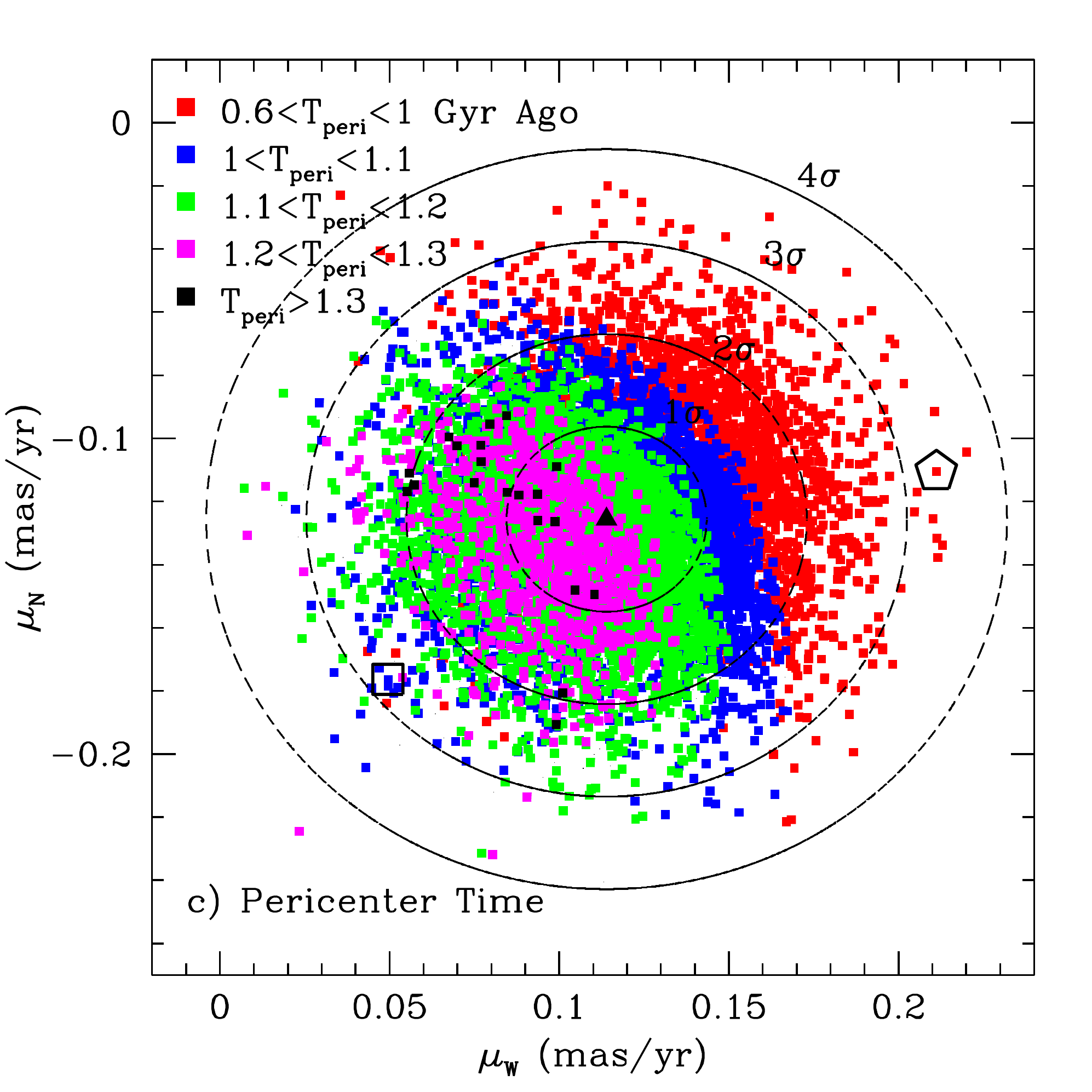}}
{\includegraphics[width=3in]{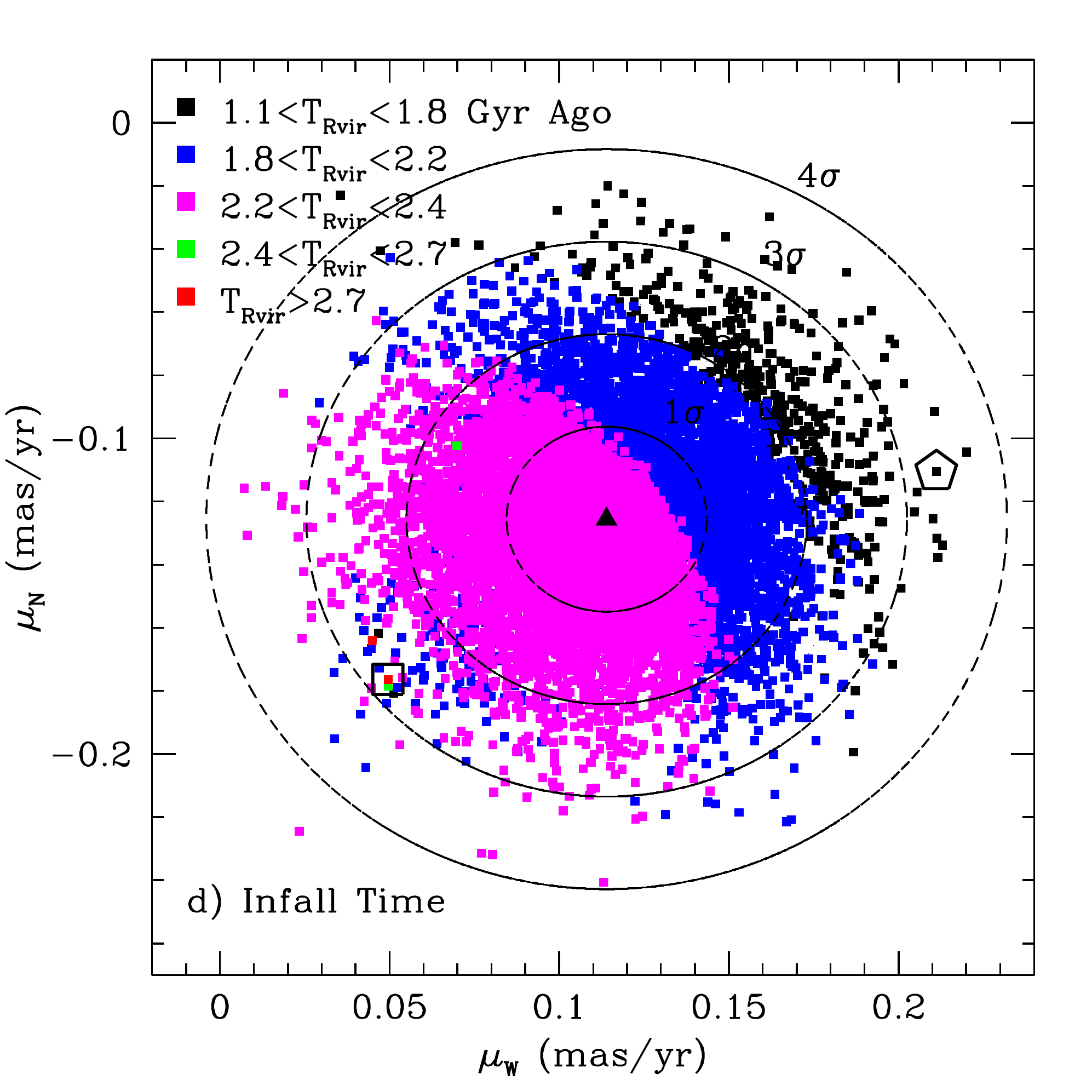}}
}
\caption{Ten thousand points randomly sampled from 
the 4$\sigma$ proper-motion error space of Leo~I. The dashed 
ellipses indicate the standard deviation of the enclosed points 
from the mean (filled triangle). For each point, the orbital 
history of Leo~I was computed by integrating the equations of 
motion [equation~(\ref{eq:eqmotion})] backward in time for the 
lowest mass MW model ($M_{\rm vir} = 1.0\times10^{12} \Msun$). 
In panel (a), solutions with more than one previous pericentric 
approach are highlighted. All cases have at least one recent 
pericentric approach with the MW. In panel (b), points are color 
coded based on the distance of their most recent pericentric 
approach, as indicated in the legend. The minimum (maximum) 
pericentric approach is indicated by the black open square 
(pentagon). In panel (c), the color coding indicates the time 
of the pericentric approach. Finally, panel (d) indicates the 
infall time, i.e. when Leo~I last entered the virial radius of 
the MW.
\label{fig:MC1e12}}
\end{figure*}  

\begin{figure*}
\centering
\mbox{{\includegraphics[width=3in]{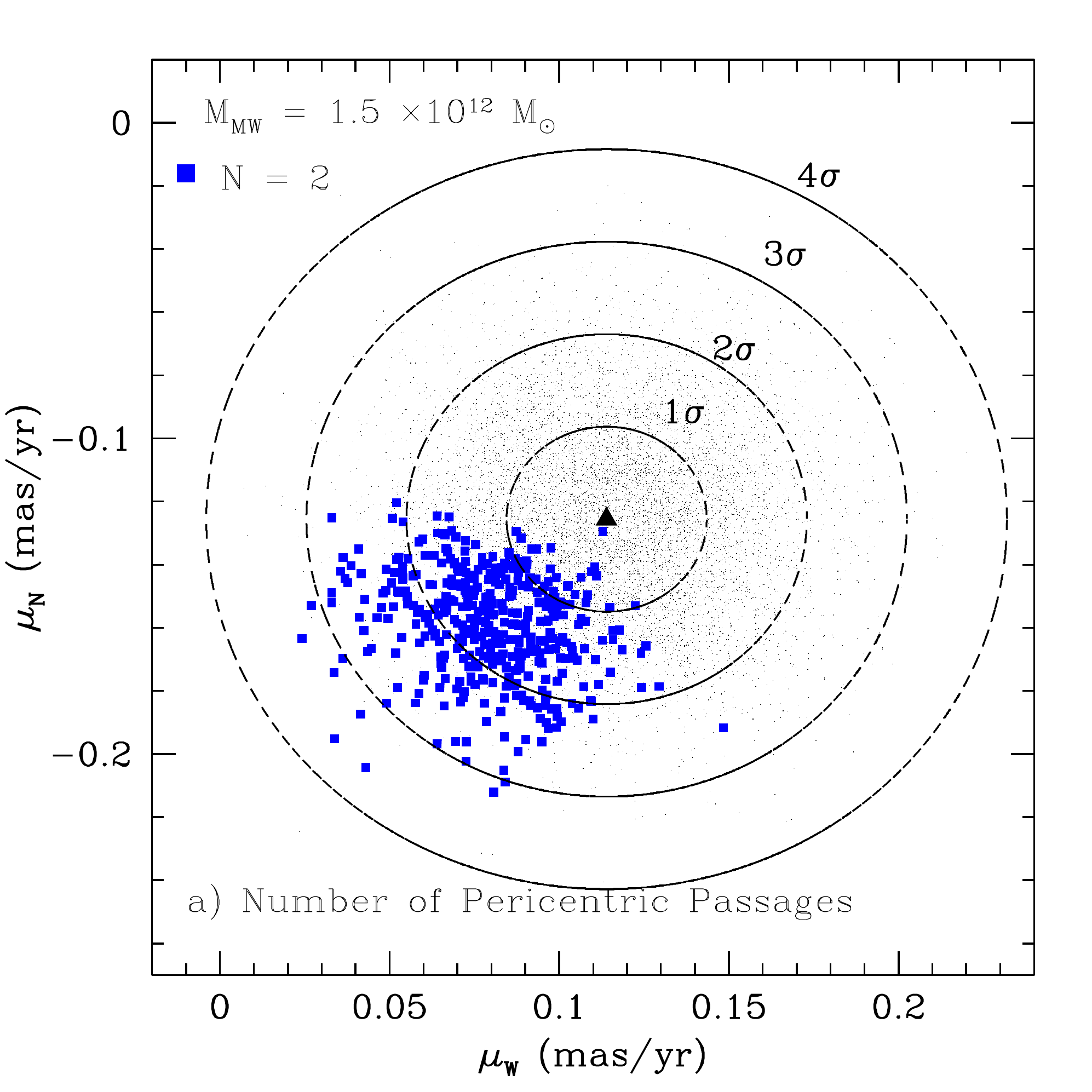}}
{\includegraphics[width=3in]{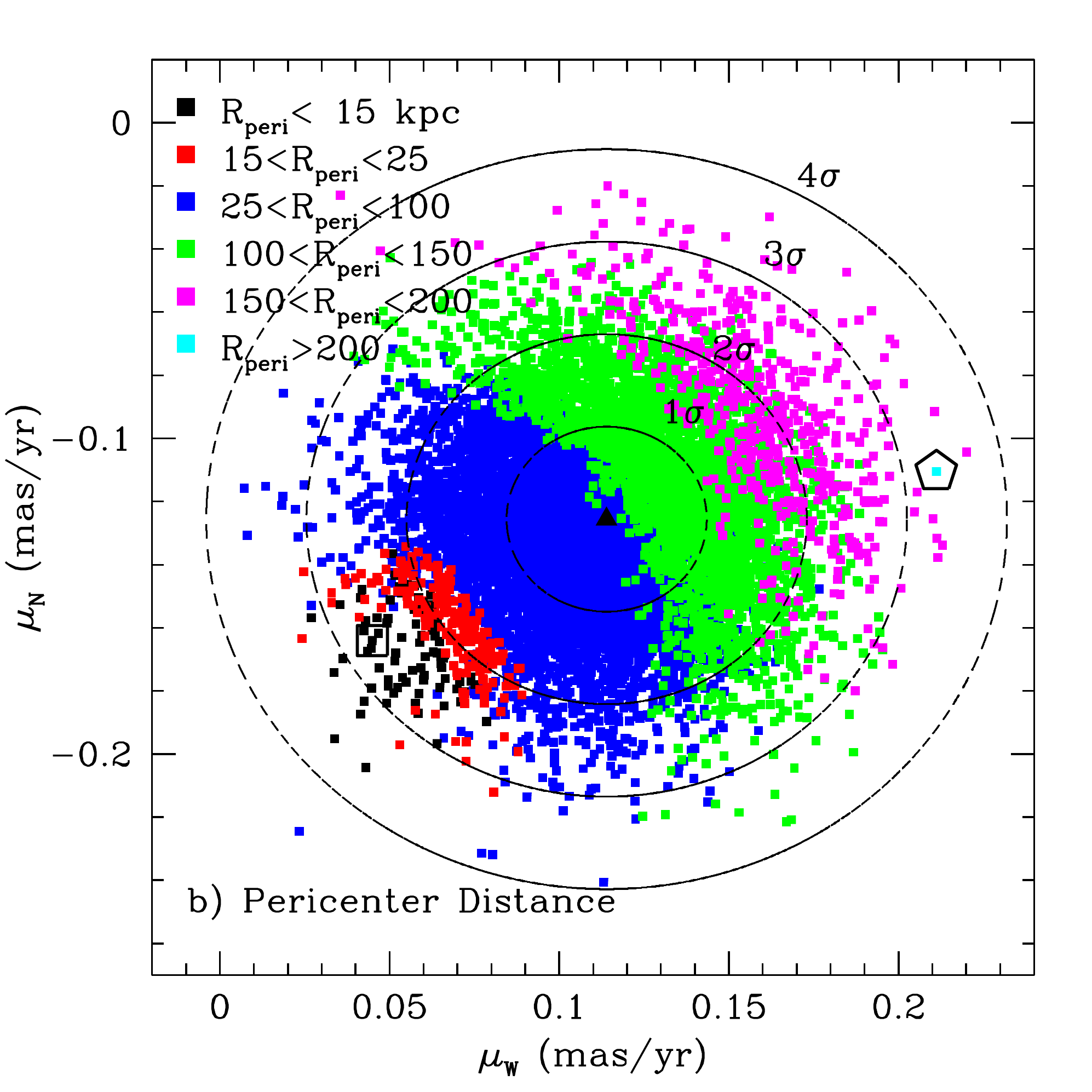}}}\\
\mbox{
{\includegraphics[width=3in]{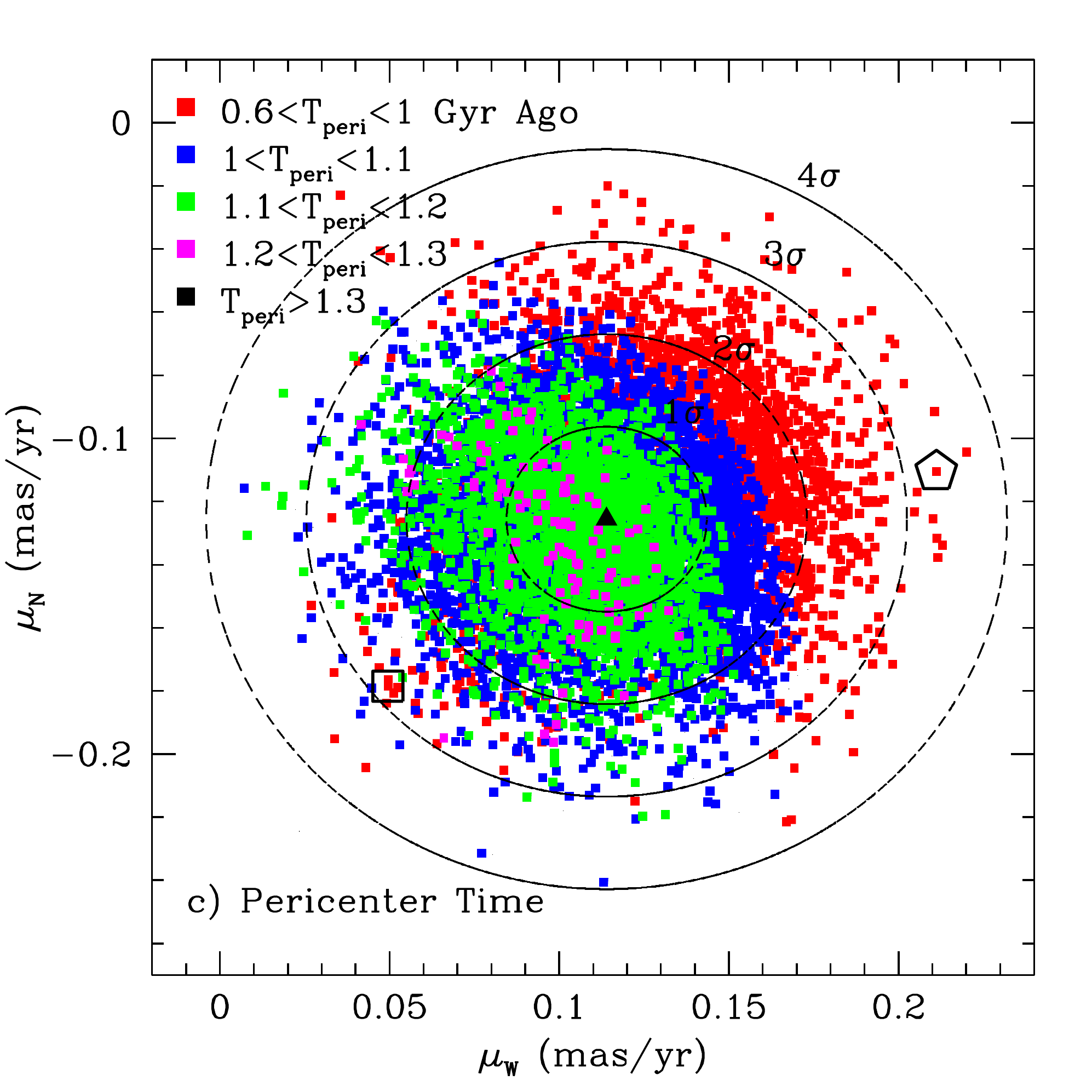}}
{\includegraphics[width=3in]{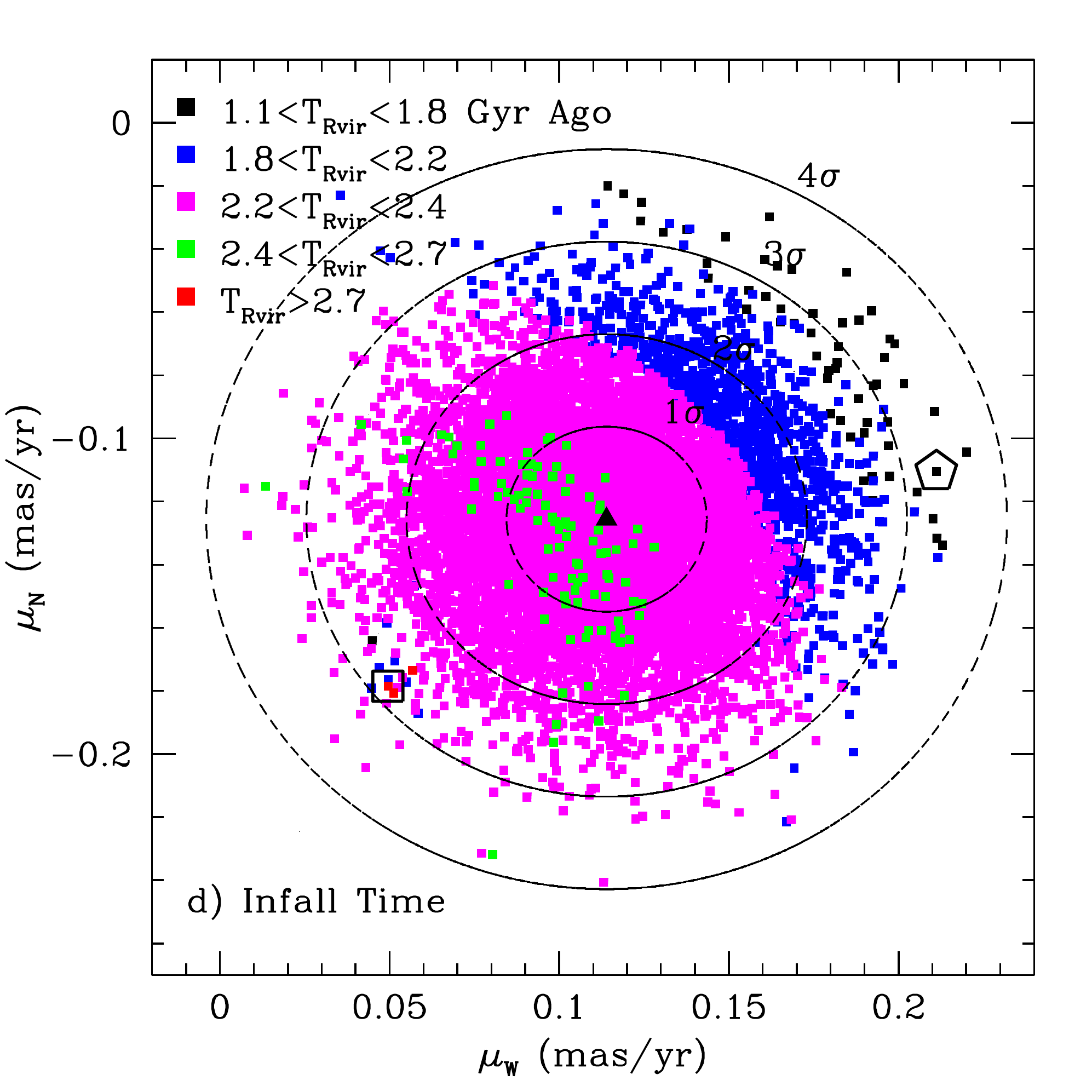}}
}
\caption{\label{fig:MC1.5e12} Similar to Figure~\ref{fig:MC1e12}, 
but now for the intermediate mass MW model 
($M_{\rm vir} = 1.5\times10^{12}$).
}
\end{figure*}  

\begin{figure*}
\centering
\mbox{{\includegraphics[width=3in]{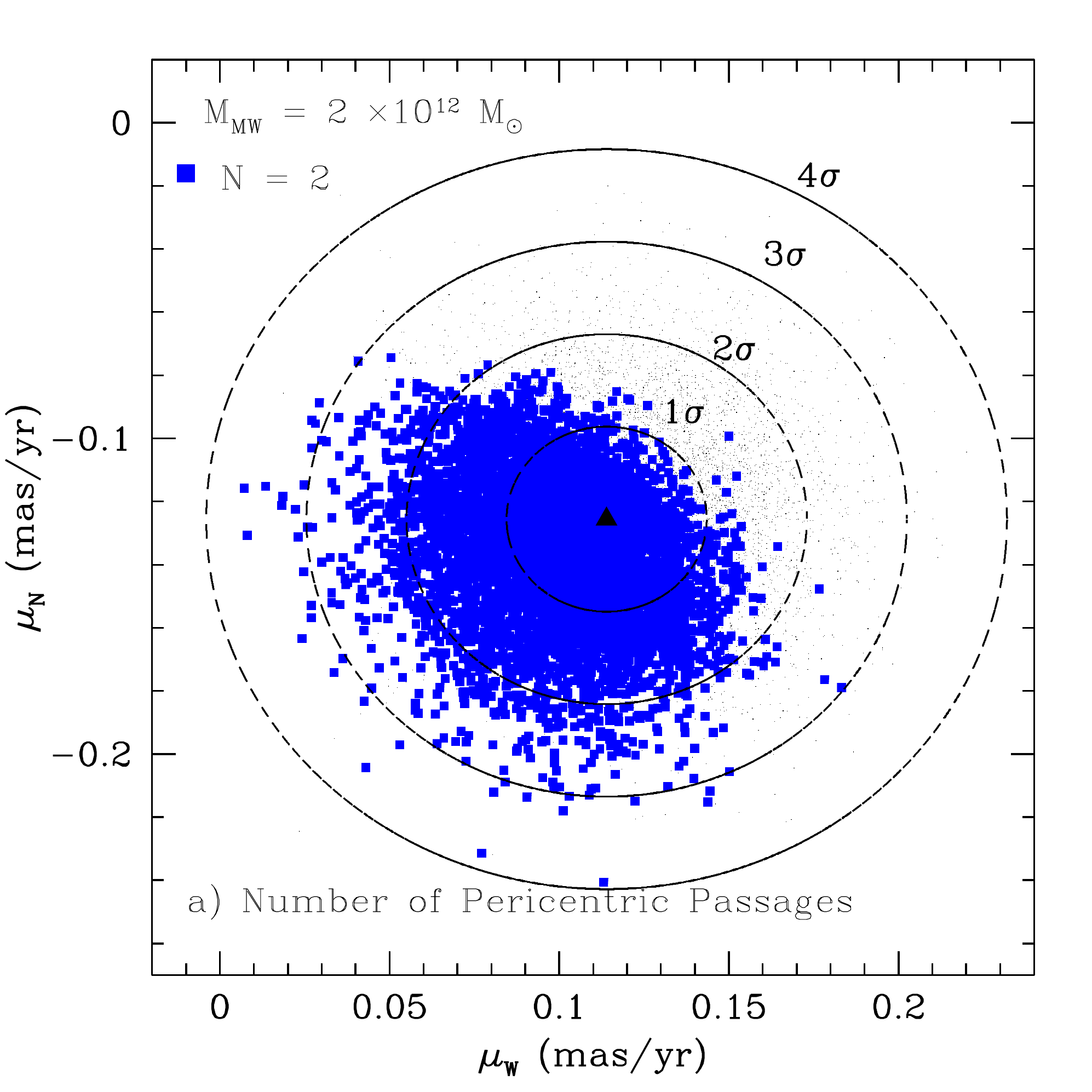}}
{\includegraphics[width=3in]{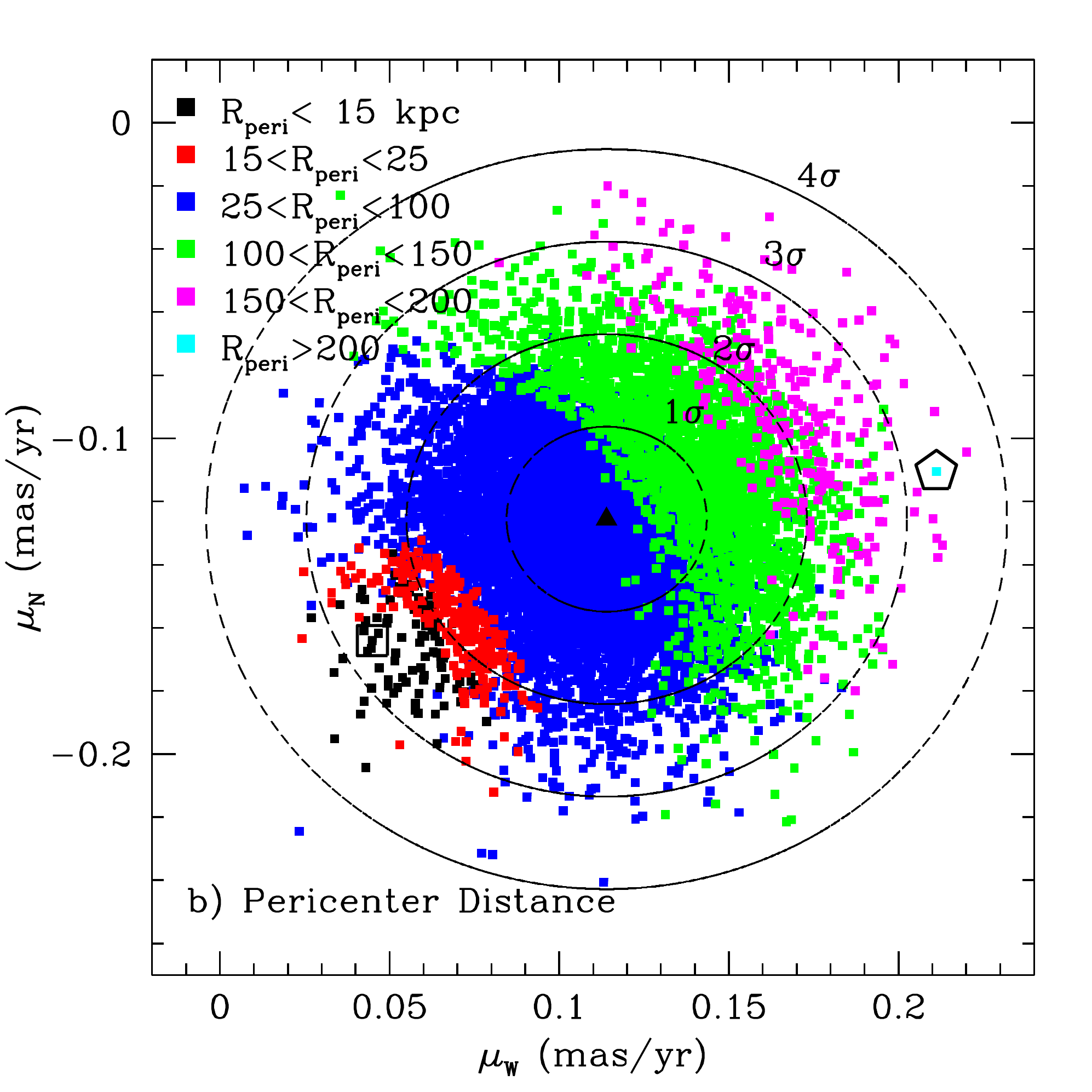}}}\\
 \mbox{
 {\includegraphics[width=3in]{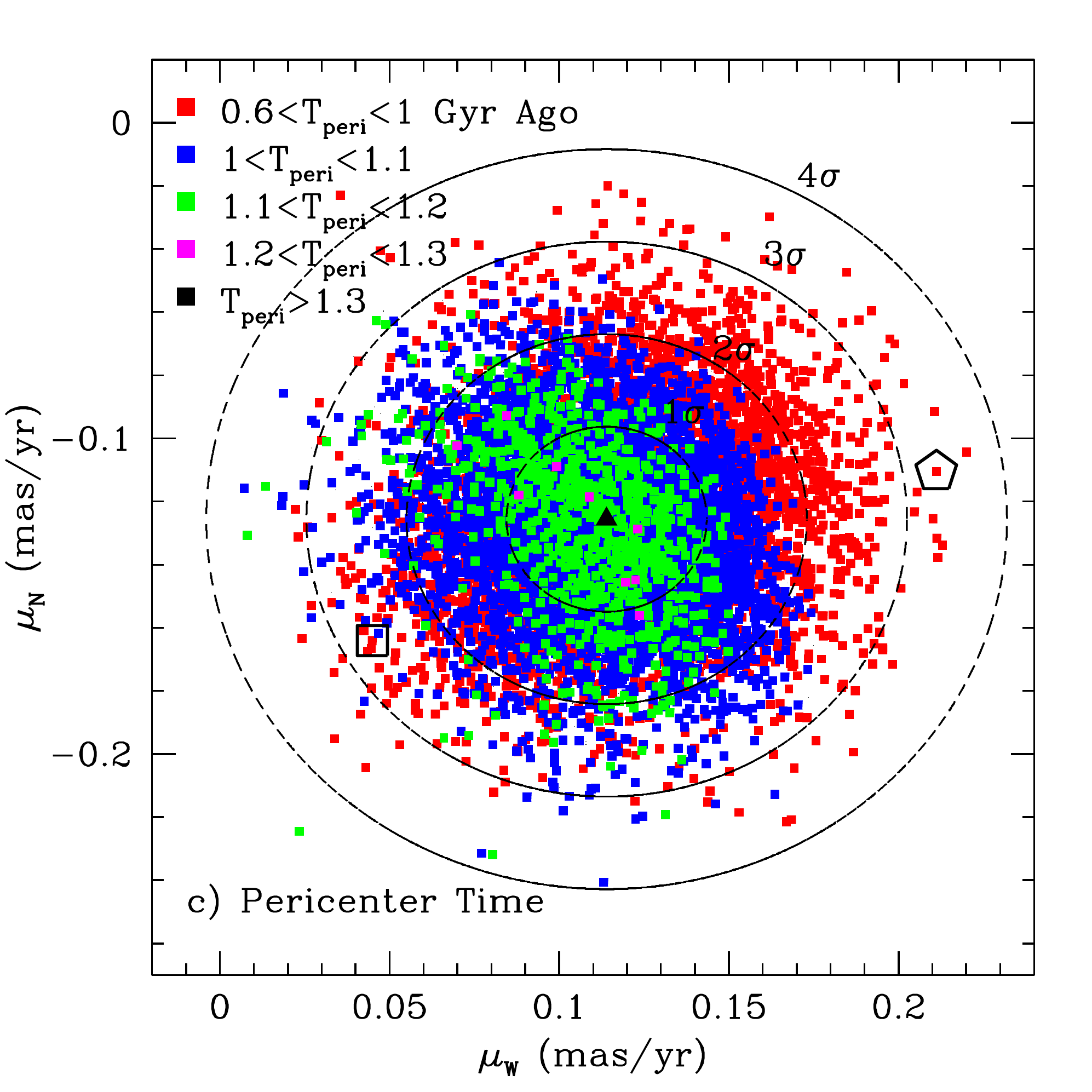}}
 {\includegraphics[width=3in]{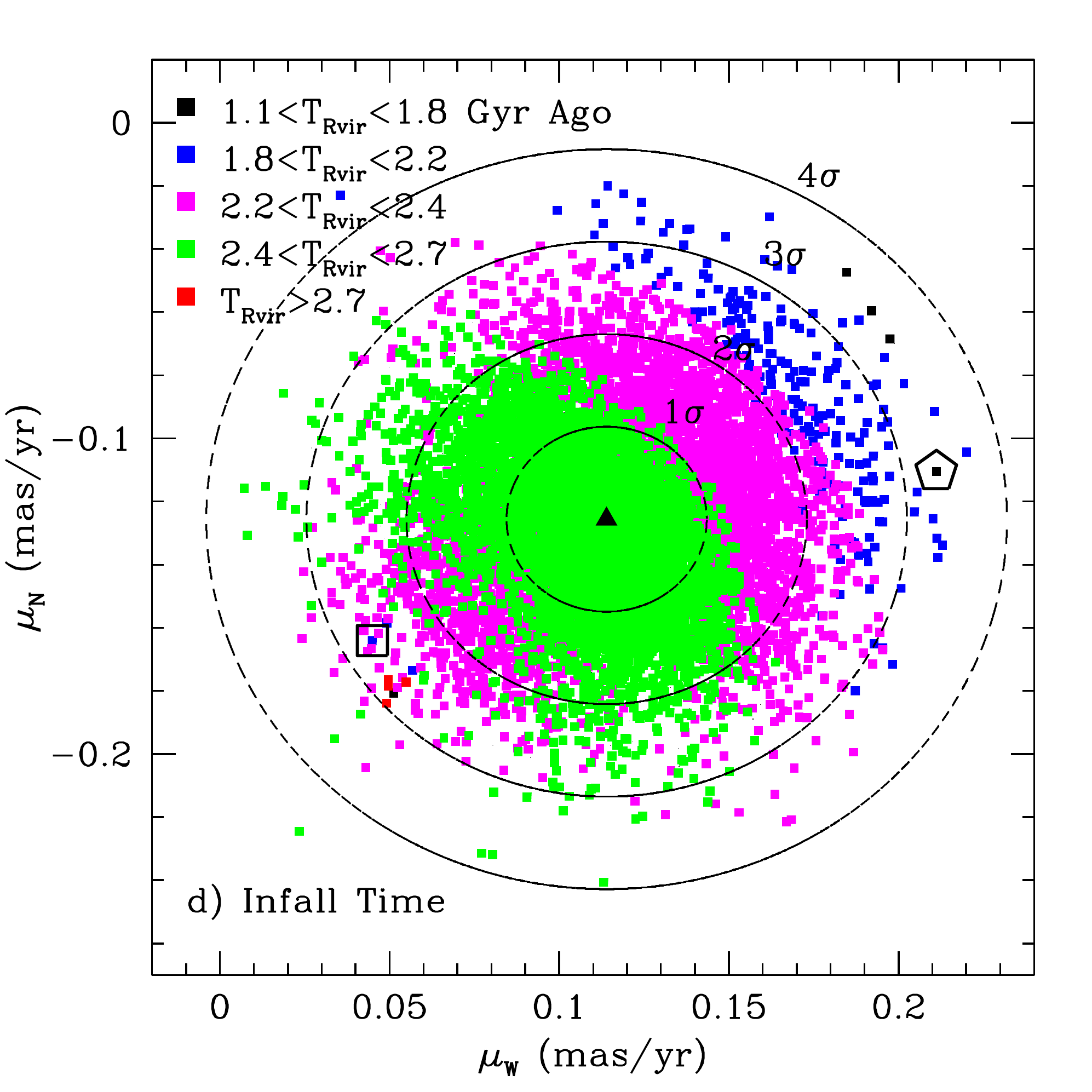}}}
 \caption{\label{fig:MC2e12} Similar to Figure~\ref{fig:MC1e12}, but 
 now for the high mass MW model ($M_{\rm vir} = 2.0\times10^{12}$). 
 }
\end{figure*}

Here we present the plausible orbital histories of Leo~I following 
the methodology outlined above. To propagate the observed errors, we 
randomly sampled 10,000 combinations of the west and north components 
of the observed Leo~I proper motion within the error space provided in 
Section~\ref{sec:ObservedPM}. For each of these combinations, the 3D 
velocity was derived and the orbit of Leo~I was followed backward in 
time for a Hubble time in each of our three MW models. The analysis 
presented in this section only considers the influence of the MW's 
gravitational field on the orbit of Leo~I, as described in 
equation~(\ref{eq:eqmotion}).

We seek to define Leo~I's interaction history with the MW by 
constraining the time and distance of Leo~I's recent pericentric 
approach to our MW ($r_{\rm peri}$, $t_{\rm peri}$), the number of 
such encounters it might have had in the past ($N_{\rm peri}$) and the 
time of infall to our system ($t_{\rm infall}$). We define infall time 
as the time at which Leo~I last crossed the virial radius of the MW. 
The allowed range in pericenter distance will inform us about the 
maximal tidal influence the MW may have exerted over Leo~I, which 
may have caused its transformation into a dSph or triggered enhanced 
star formation activities. The pericenter distance also 
informs us about the importance of ram pressure stripping by the MW's 
gaseous halo in removing gas from this system \citep[e.g.,][]{grc09}; 
the deeper that Leo~I travels into the MW's halo, the higher the 
background gas density and the more likely gas gets stripped. 
The infall time is similarly relevant; the longer ago that Leo~I 
was accreted, the longer the time scale for ram pressure stripping 
to operate. Ultimately, the number of previous pericentric passages 
will tell us whether Leo~I is in fact a recent interloper in our 
system. These properties can be further used to constrain the halo 
mass of our own MW statistically by identifying analogs of the 
Leo~I satellite (in terms of mass and orbital properties) about 
MW-type hosts in large-scale cosmological simulations (see Paper~II).

\begin{deluxetable*}{ccccccccc}
\setlength{\tabcolsep}{5pt} 
\tablecaption{Mean Orbital Properties of Leo~I\label{tab:MC}}
\tablehead{
\colhead{$M_{\rm MW}$} & \colhead{} & \colhead{$\langle r_{\rm peri} \rangle$} & \colhead{$\langle t_{\rm peri} \rangle$} & \colhead{} & \colhead{$\langle d_{\rm apo} \rangle$} & \colhead{$\langle t_{\rm peri,2} \rangle$} & \colhead{$\langle t_{\rm infall} \rangle$} & \colhead{} \\
\colhead{($\times 10^{12} \Msun$)} & \colhead{$\langle N_{\rm peri} \rangle$} & \colhead{(kpc)} & \colhead{(Gyr)} & \colhead{$\langle v_{\rm peri}/v_{\rm esc} \rangle$} & \colhead{(kpc)} & \colhead{(Gyr)} & \colhead{(Gyr)} & \colhead{$\langle v_{\rm infall}/v_{\rm Rvir} \rangle$}
}
\startdata
1.0 & $1.0 \pm 0.2$ &    $100 \pm 37$   & $1.1 \pm 0.1$ & $1.04 \pm 0.07$ &    $1177 \pm    640$ & \phn$7.7 \pm 5.6$ & $2.2 \pm 0.2$ & $1.6 \pm 0.2$ \\
1.5 & $1.0 \pm 0.2$ & \phn$90 \pm 35$	& $1.1 \pm 0.1$ & $0.96 \pm 0.04$ & \phn$558 \pm \phn61$ &    $12.6 \pm 1.3$ & $2.3 \pm 0.1$ & $1.3 \pm 0.2$ \\
2.0 & $1.5 \pm 0.5$ & \phn$82 \pm 33$	& $1.0 \pm 0.1$ & $0.91 \pm 0.02$ & \phn$530 \pm \phn55$ &    $11.0 \pm 1.7$ & $2.5 \pm 0.2$ & $1.0 \pm 0.1$ 
\enddata
\tablecomments{Orbital properties as defined in the text. Standard
deviations for the mean values are marked. Apocenter distances are
computed only for the cases that have a second pericentric
approach.}
\end{deluxetable*}

Figures~\ref{fig:MC1e12},~\ref{fig:MC1.5e12} and ~\ref{fig:MC2e12} 
show the 4$\sigma$ proper motion error space that is sampled to 
determine the orbital properties of Leo~I in the three different 
MW mass models. Points are color coded to reflect the range of 
allowed values for the quantity of interest ($N_{\rm peri}$,
$r_{\rm peri}$, $t_{\rm peri}$ and $t_{\rm infall}$). The mean 
values for these quantities are listed in Table~\ref{tab:MC}. 
In addition, we also list the mean velocity at pericenter relative 
to the escape speed at $r_{\rm peri}$, ($v_{\rm peri}$/$v_{\rm esc}$), 
and the velocity at infall relative to the circular velocity of the 
MW at the virial radius ($v_{\rm infall}/v_{\rm Rvir}$). In the 
cases where there is a second pericentric approach some time in the 
past, we list the time this occurs ($t_{\rm peri,2}$) and the mean 
apocenter distance ($r_{\rm apo}$). We mark the minimum and maximum 
$r_{\rm peri}$ allowed within the 4$\sigma$ proper motion error 
space and the times at which they occur in the second and third 
panels in each of Figures~\ref{fig:MC1e12}, \ref{fig:MC1.5e12} and 
\ref{fig:MC2e12} by the black open square (min) and pentagon (max). 

In all cases, Leo~I has recently had a pericentric passage with
respect to the MW, and so panels (a) in Figures~\ref{fig:MC1e12}, 
\ref{fig:MC1.5e12} and \ref{fig:MC2e12} only note if a second 
pericentric approach occurs. As the MW mass increases, the 
likelihood of a second close passage also increases; however there 
are no cases where a third pericentric approach occurs. In the lowest 
MW mass model, there are solutions for a second pericentric passage 
only outside the 2$\sigma$ error ellipse. However, there are only two 
such solutions out of our 10,000 realizations, one with a second 
pericenter time of 13 Gyr and the other with 2 Gyr (which is likely 
a slingshot scenario where Leo~I got too close to the MW center).
This reflects the fact that for the low-mass MW model, Leo~I is 
generally on a hyperbolic or near-hyperbolic orbit. 
Solutions for a second pericenter are readily obtainable within the 
1$\sigma$ error ellipse for the higher MW mass models. However, the 
Monte-Carlo statistics (see Table~\ref{tab:MC}) still favor orbits 
with only one previous pericenter. Moreover, in those cases with a 
second pericenter, the time since that pericenter is $\gtrsim 10$ Gyr. 
Also, the apocenter distances are well beyond the virial radius of 
the MW for all models ($d_{\rm apo} > 500$ kpc). Recall that the MW 
mass was assumed to be static in time; with an accurate treatment of 
the mass evolution of the MW it is doubtful that any of these second 
pericentric passages would still occur. It is thus most likely that 
Leo~I has passed its {\it first} infall into the MW. Moreover, if the 
previous pericentric passage occurred $\gtrsim 10$ Gyr ago, this 
implies that Leo~I would be at large distances from the MW exactly 
a Hubble time ago, which is an implausible scenario in the view of 
the timing argument.

The average time of Leo~I's most recent pericentric passage is 
$\sim$1 Gyr ago for all MW models. This is similar to what was found 
from the Keplerian calculations. The average pericentric approach 
ranges from 80--100 kpc, and the average velocity at pericenter ranges 
from 300--$370 \kms$ (Table~\ref{tab:MC}). This distance is somewhat
larger, and the velocity somewhat smaller, than found from the
Keplerian calculations. This is easy to understand from the fact that
a Keplerian model is too concentrated, and therefore overestimates 
the acceleration as Leo~I approaches the MW. The average ratio of 
pericenter to apocenter distance in the orbit calculations ranges 
from 0.08--0.15. This is larger than in the Keplerian calculations, 
in part because those calculations ignore the mass outside of the 
Leo~I distance, and therefore overestimate the apocenter distance.

The tidal radius $r_{\rm t} = r_{\rm peri} [2M_{\rm MW}(r_{\rm peri}) 
/ M_{\rm Leo~I}]^{-1/3}$ of Leo~I, given pericentric distances of 
80--100 kpc, is 3--4 kpc for the three MW models. The present optical 
radius of Leo~I is $\sim 1$ kpc \citep{soh07,wal09}. This means that 
on average, the tidal field of the MW does not appear to be sufficient 
to tidally truncate Leo~I to this radius, implying that Leo~I likely 
has an extended dark matter halo. There is a clear trend in 
$r_{\rm peri}$ with the proper motion as evidenced in panel (b) of 
Figures~\ref{fig:MC1e12},~\ref{fig:MC1.5e12} and ~\ref{fig:MC2e12}; 
$r_{\rm peri}$ increases with increasing $\mu_{W}$ and $\mu_{N}$. 
The minimum pericentric approach determined in the 10,000 Monte-Carlo 
orbits is $\sim$1--2 kpc. However, a pericentric approach of $< 5$ kpc 
is unlikely, because the tidal radius of Leo~I in all MW mass models 
is less that 0.7 kpc. Leo~I is not sufficiently disturbed to have 
approached this close to the MW. On the other hand, pericentric 
approaches of 20 kpc yield tidal radii of $\sim$1 kpc. Cases where 
$15< r_{\rm peri}< 25$ kpc are found within the 1.5--2$\sigma$ error 
ellipse in all MW mass models. So such close approaches are not ruled 
out by our data. However, they are not so likely given our data, with 
only 2--3\% of the Monte-Carlo calculated orbits yielding 
$r_{\rm peri} < 25$ kpc.

The time of Leo~I's last pericentric passage, $\sim 1$ Gyr ago,
corresponds roughly to the time when star formation stopped in Leo~I
\citep{cap99,gal99,dol02,sme10}. The pericentric approach is the point
in time where Leo~I would experience maximal ram pressure stripping,
which could lead to quenching. All satellites of the MW within 300 kpc
(apart from the Magellanic Clouds) are devoid of gas \citep{grc09},
consistent with this picture. As the MW mass increases, $t_{\rm peri}$
decreases; for the most massive MW model the maximal $t_{\rm peri}$ is
1.3 Gyr and the minimal value is 0.6 Gyr.  These values are remarkably
consistent regardless of MW mass, when searching the full 4$\sigma$
proper motion error space [panels (c)].  Hence, it is likely that star
formation stopped in Leo~I owing to ram pressure effects at
pericentric approach, implying a quenching time scale of $t_{\rm
  infall} - t_{\rm peri} \approx 1.3 \Gyr$ (see Table~\ref{tab:MC}). 

Of course, in general, star formation can cease in galaxies for many
other reasons, e.g., exhaustion or blowout of the gas supply. However, 
if this were the cause of star formation ceasing in Leo I, then there 
would be no natural explanation for why this would coincide with a 
pericenter passage. On the other hand, this coincidence could 
certainly happen by chance, especially since the uncertainties in 
both the SFH and the orbital analysis are significant.

The average infall time ranges from 2.2--2.5 Gyr with little
variation, regardless of MW mass. The infall time is similar to the
time scale of the most recent enhanced star formation observed in 
Leo~I \citep{cap99,gal99,dol02,sme10}, suggesting that this enhanced 
star formation activity was triggered by either ram pressure 
compression as Leo~I entered a higher gas density environment relative 
to the Local Group, or as it began to feel gravitational torques 
exerted by the MW. Note that $t_{\rm infall}$ refers to the most recent 
time at which Leo~I entered the virial radius; there are cases where 
Leo~I has made an earlier pericentric approach about the MW. But, as 
discussed earlier, such orbits may not be physical or plausible. There 
are a few cases where Leo~I remains within the virial radius of the MW 
for approximately a Hubble Time [indicated by red squares in panels
(d)]. However, such a scenario has low likelihood, since these cases
are all 4$\sigma$ outliers that only occur in the high mass MW model.

The ratio between the infall velocity and the circular velocity at the
virial radius ($v_{\rm Rvir}$) ranges from 1.0--1.6
(Table~\ref{tab:MC}). For the low mass MW model, the average Leo~I
infall velocities are higher than expected based on cosmological
simulations of structure formation, where subhalos are typically
accreted with characteristic orbital velocities of $\sim 1.1 v_{\rm
Rvir}$ at the virial radius \citep[1$\sigma$ scatter of 25\%,][]{wet11}. 
This further disfavors masses $M_{\rm MW,vir} \lesssim 10^{12} \Msun$.

Our conclusion that Leo~I has most likely passed its {\it first}
infall into the MW, and our value for $t_{\rm infall}$ are consistent
with the results of \citet{roc12} for Leo~I. They find that there is a
tight correlation between the present day orbital energies of the MW
satellites and their infall times as inferred from cosmological
simulations. We explore the implications of this for our understanding
of Leo~I further in Paper~II.


\subsubsection{Comparison to Previous Orbit Estimates}
\label{subsec:compare}

\begin{figure}
\epsscale{1.0}
\plotone{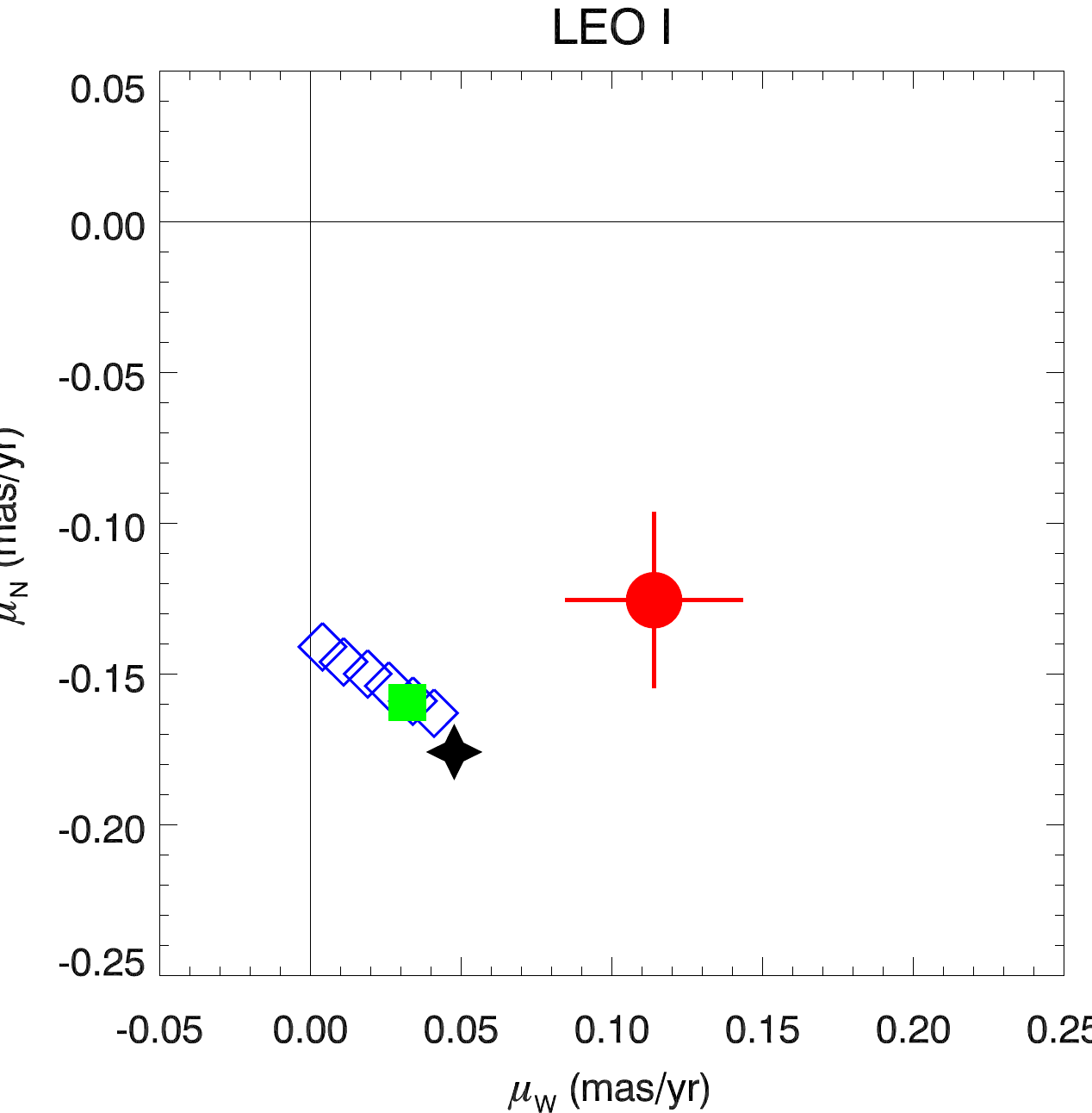}
\caption{Comparison between the average Leo~I proper motion measured
  in this study (red data point with error bars) and the proper
  motions predicted by the models of \citet{soh07} (green closed 
  square) and \citet{mat08} (blue open diamonds). The black cross 
  indicates the proper motion for a radial orbit, i.e., with 
  $V_{\rm tan} = 0$, as in Figure~\ref{fig:PMDiagram}.
  \label{fig:PMcomp}}
\end{figure}

\begin{figure*}
\centering
\mbox{{\includegraphics[width=3in]{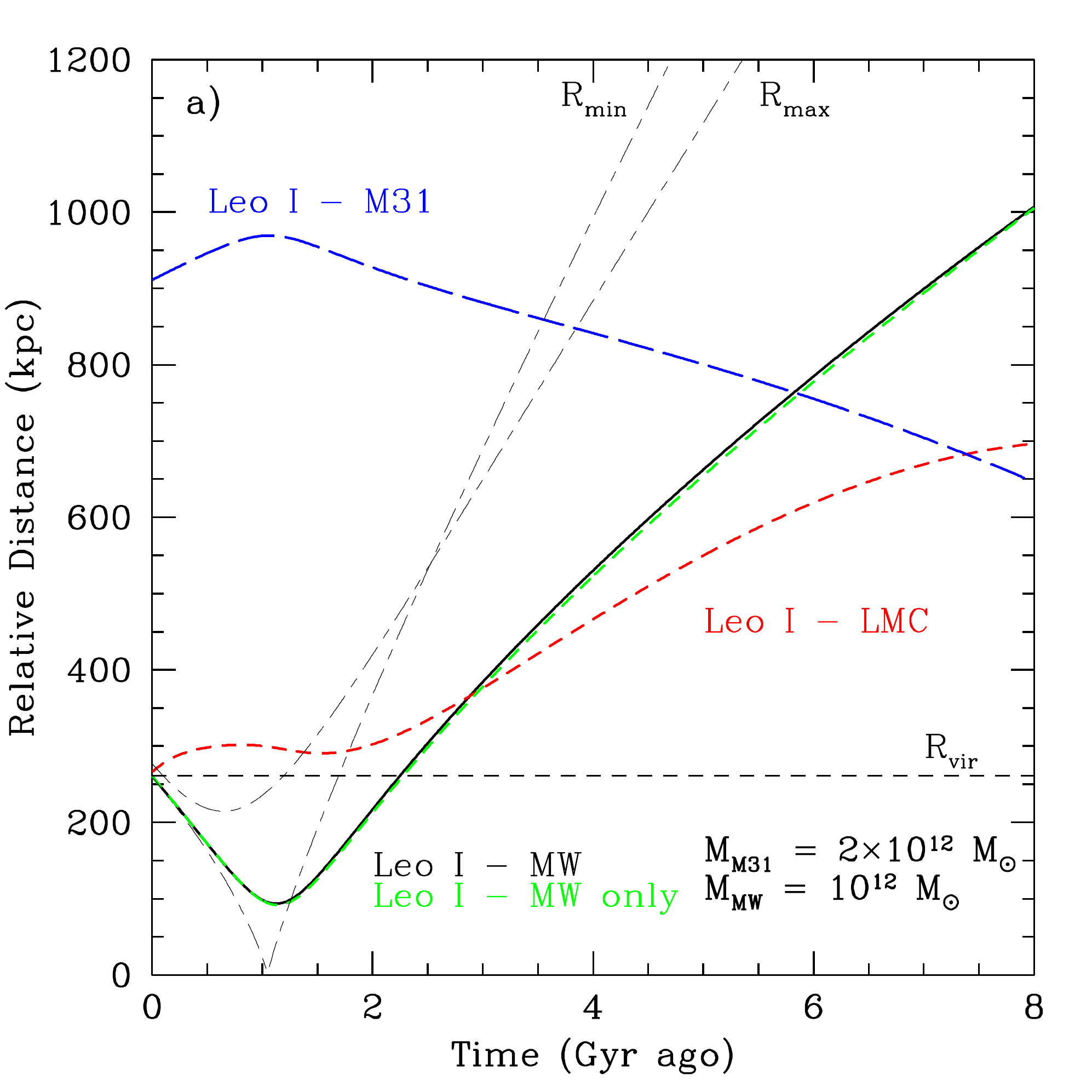}}
{\includegraphics[width=3in]{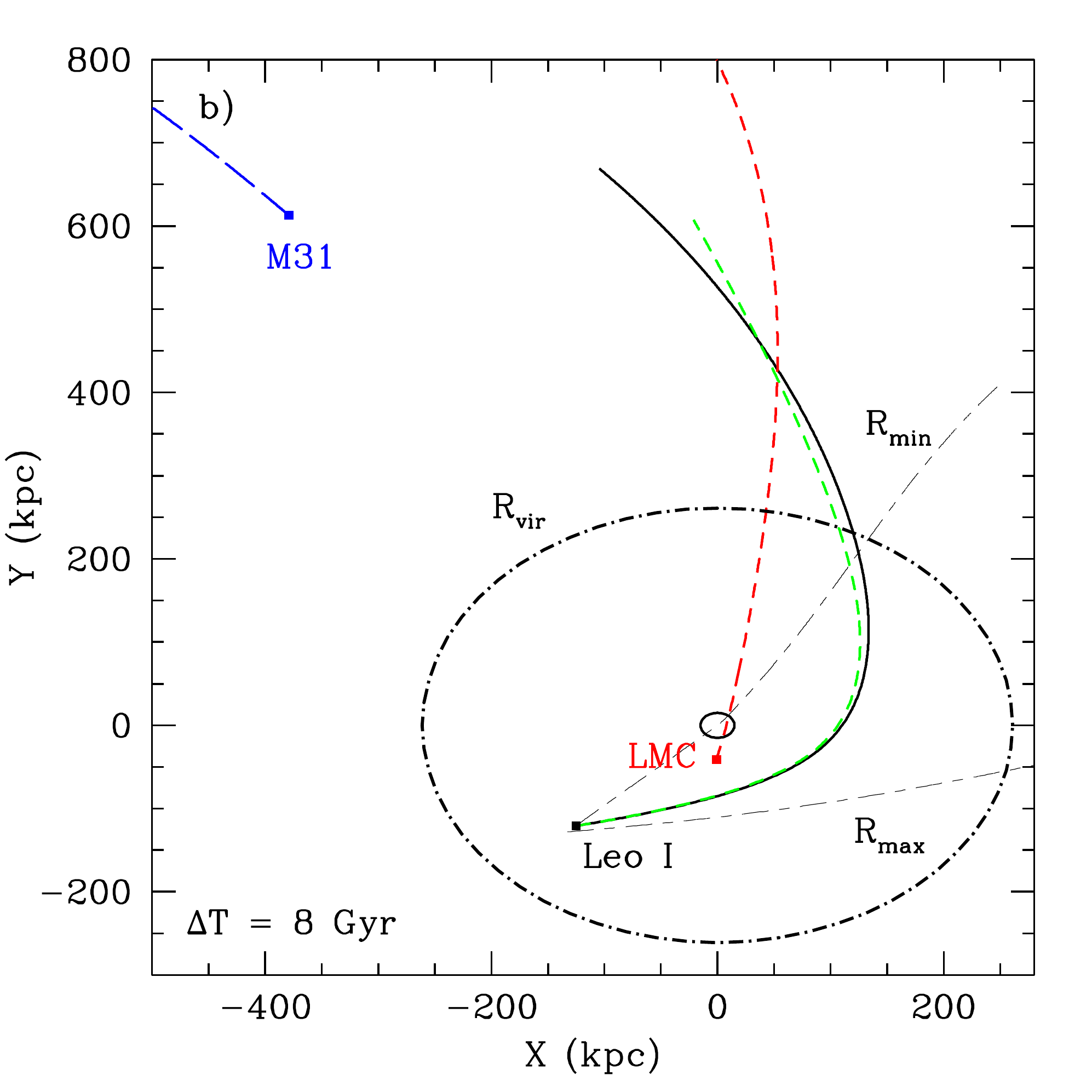}}}\\
 \mbox{
 {\includegraphics[width=3in]{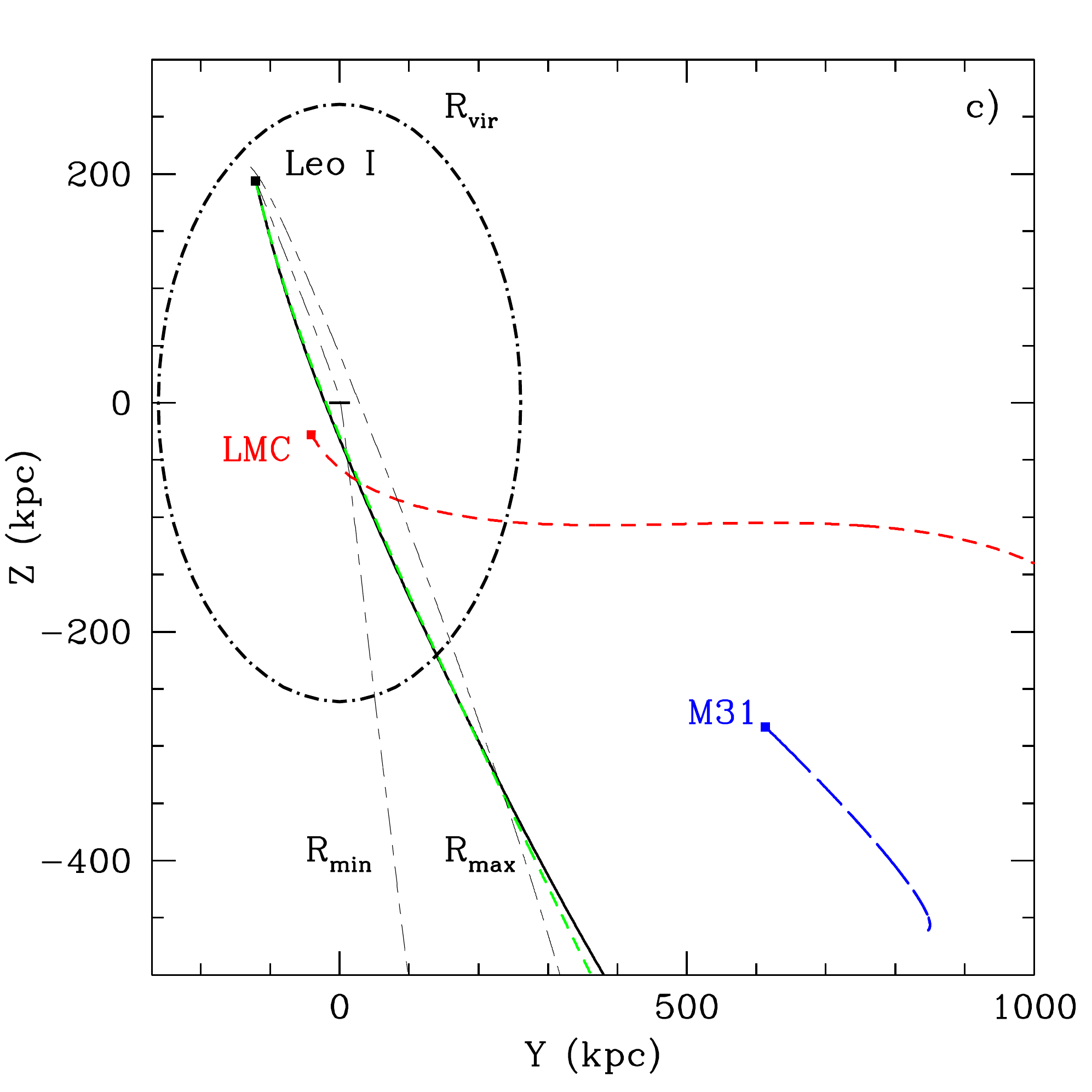}}
 {\includegraphics[width=3in]{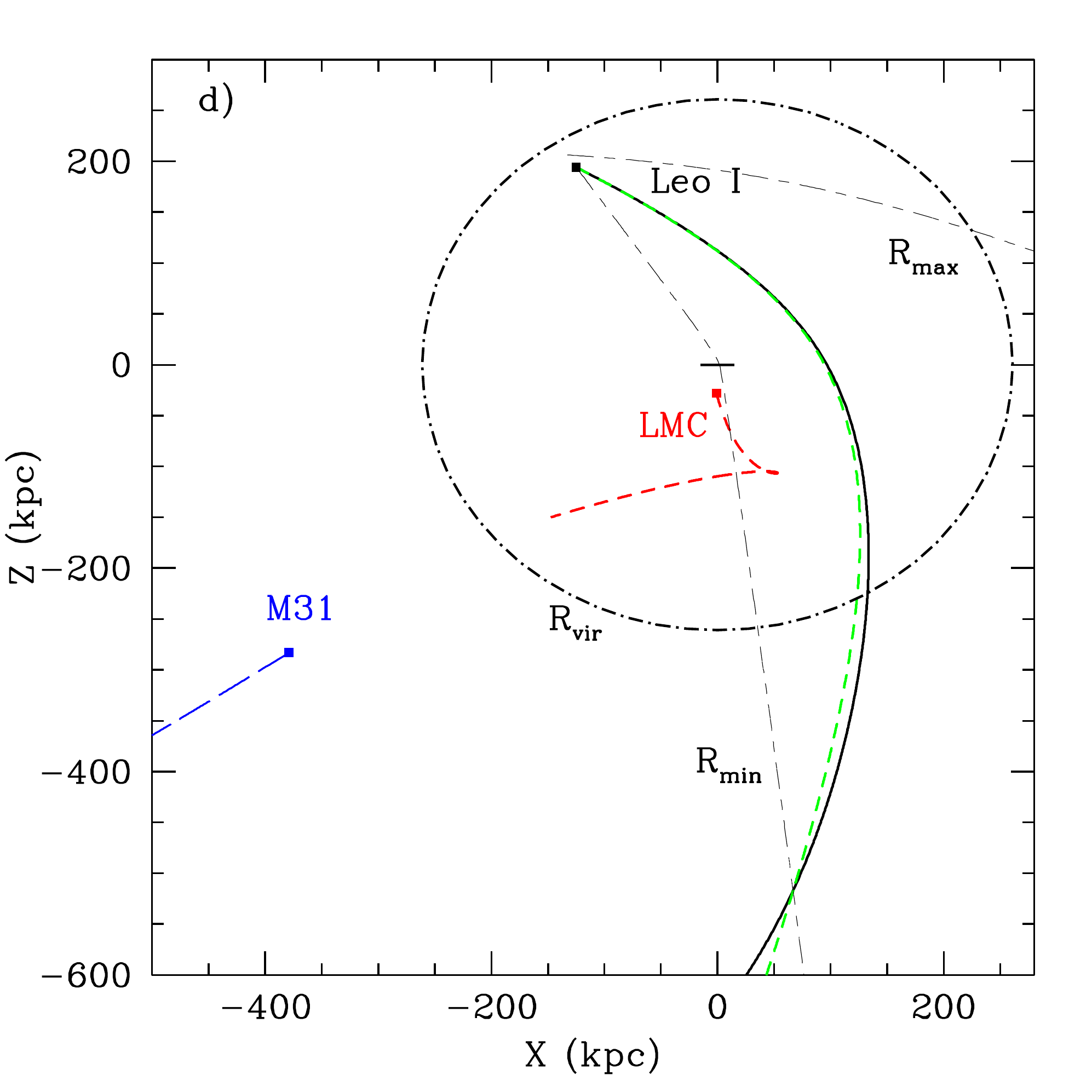}}}
 \caption{\label{fig:Orb1e12} The mean orbital history of Leo~I for
   the low mass MW model ($M_{\rm vir} = 1.0\times10^{12} \Msun$). 
   The mass of M31 is chosen to preserve the total Local Group mass 
   of $3\times10^{12} \Msun$. Panel (a) shows the separation between 
   Leo~I and the LMC, MW and M31 as a function of time. In panels 
   (b), (c), and (d), the orbital plane is presented in the 
   Galactocentric $X-Y$, $Y-Z$, and $X-Z$ planes, respectively.  
   The virial radius (R$_{\rm vir}$) of the MW is indicated by the 
   dashed horizontal line in panel (a), and the dot-dashed ellipses in 
   the other panels. In panels (b)--(d), the MW location is indicated 
   by the thick black line (circle in the case of the $Y-Z$ plane) 
   centered at (0,0). The current location of each galaxy is indicated 
   by the solid squares. Solid black lines indicate results when Leo~I 
   is assumed to be moving with the mean proper motion determined in 
   this study. For comparison, we also plot the orbit of Leo~I 
   accounting only for the influence of the MW in dashed green lines.
   The thin dashed-dotted black lines indicate solutions 
   where the pericenter distance is minimized and maximized [3$\sigma$ 
   outliers; see panel (b) in Figure~\ref{fig:MC1e12}].  }
\end{figure*}  
 
\begin{figure*}
\centering
\mbox{{\includegraphics[width=3in]{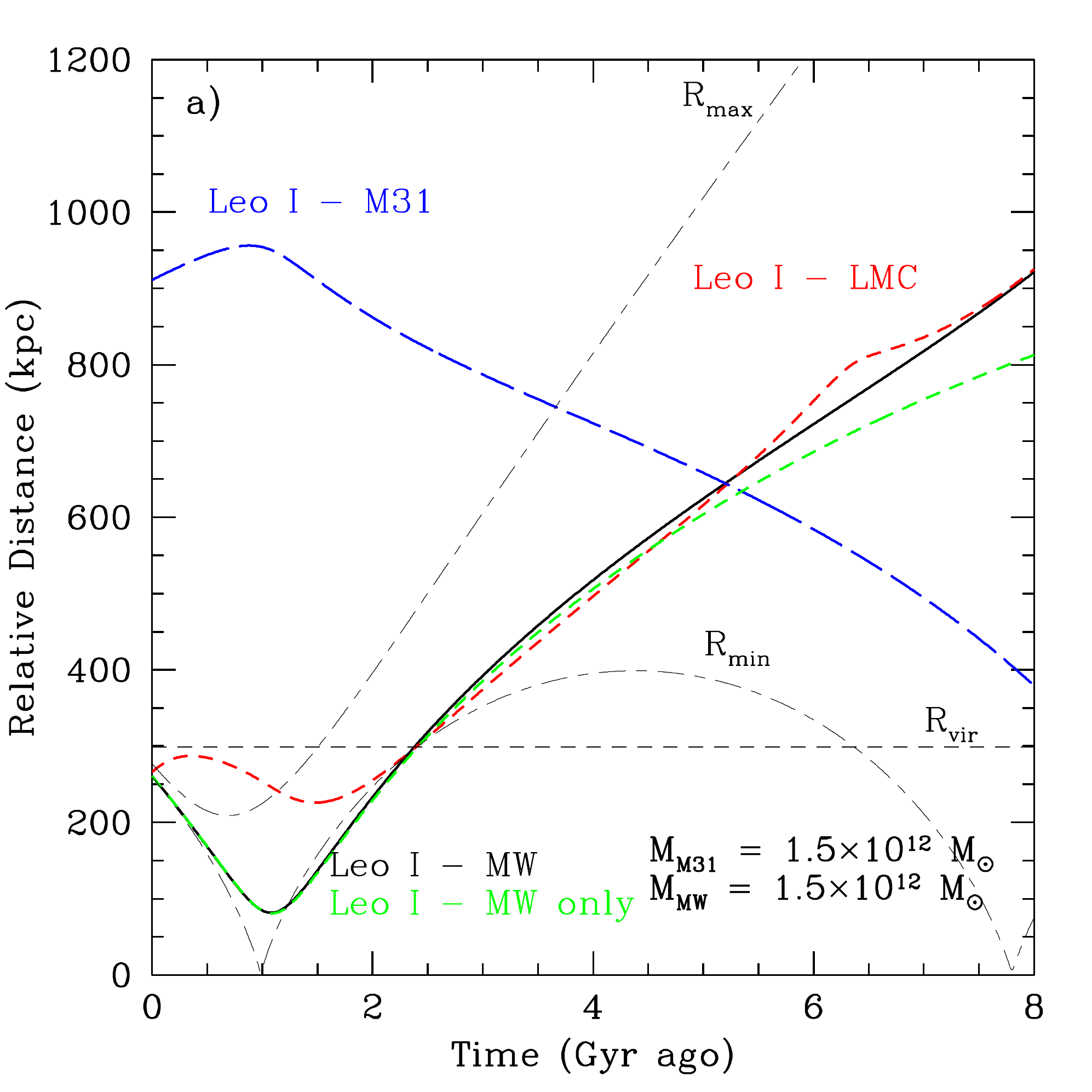}}
{\includegraphics[width=3in]{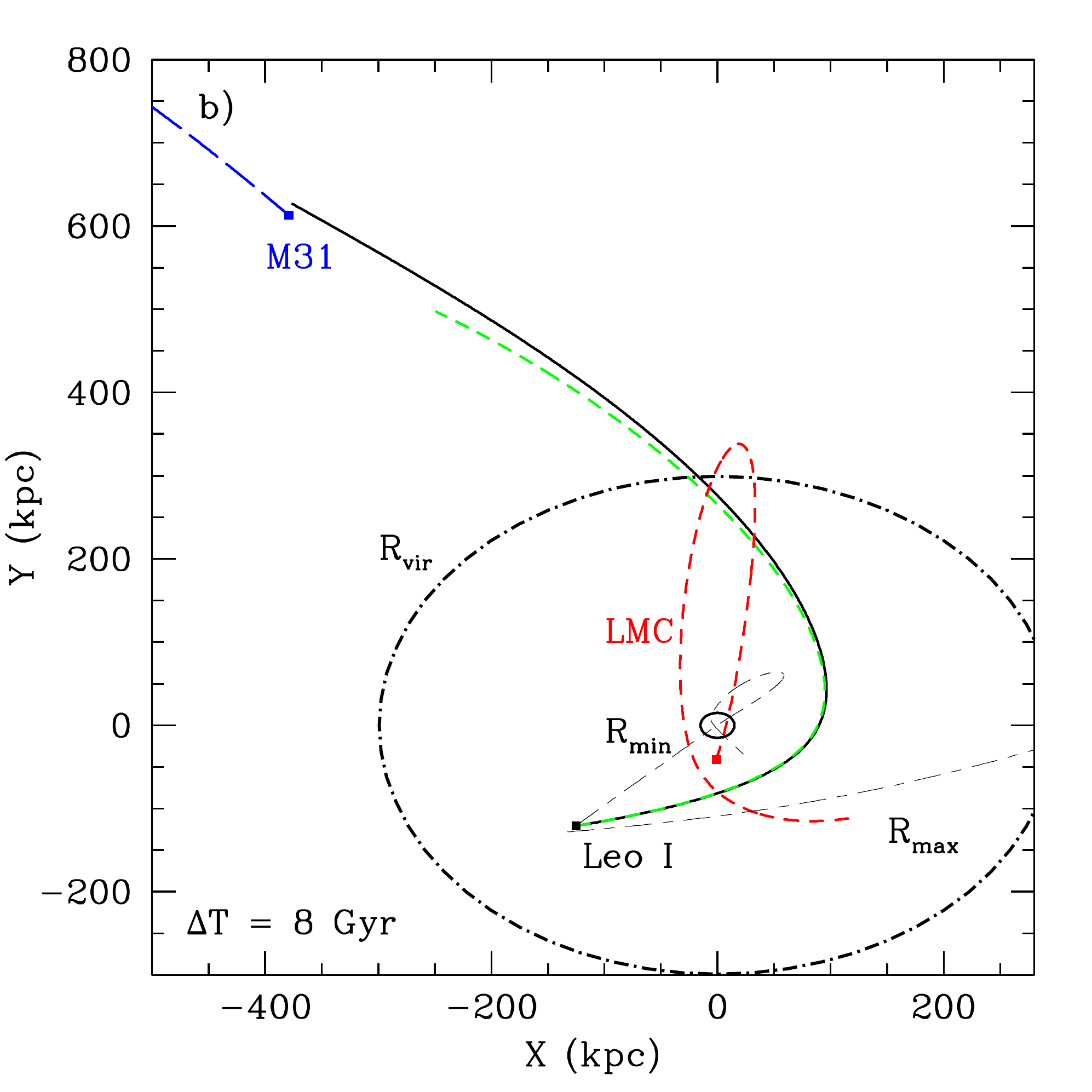}}}\\
 \mbox{
 {\includegraphics[width=3in]{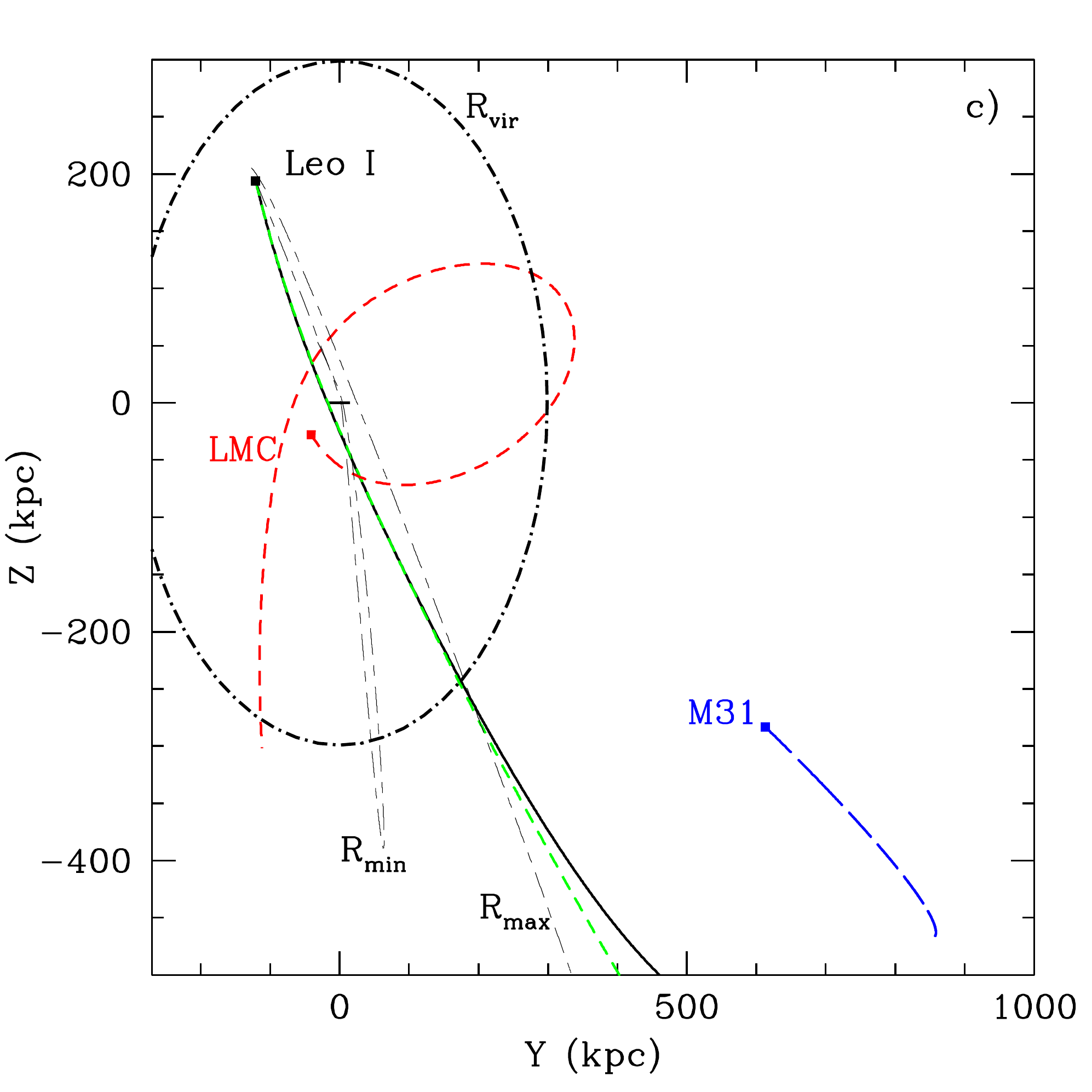}}
 {\includegraphics[width=3in]{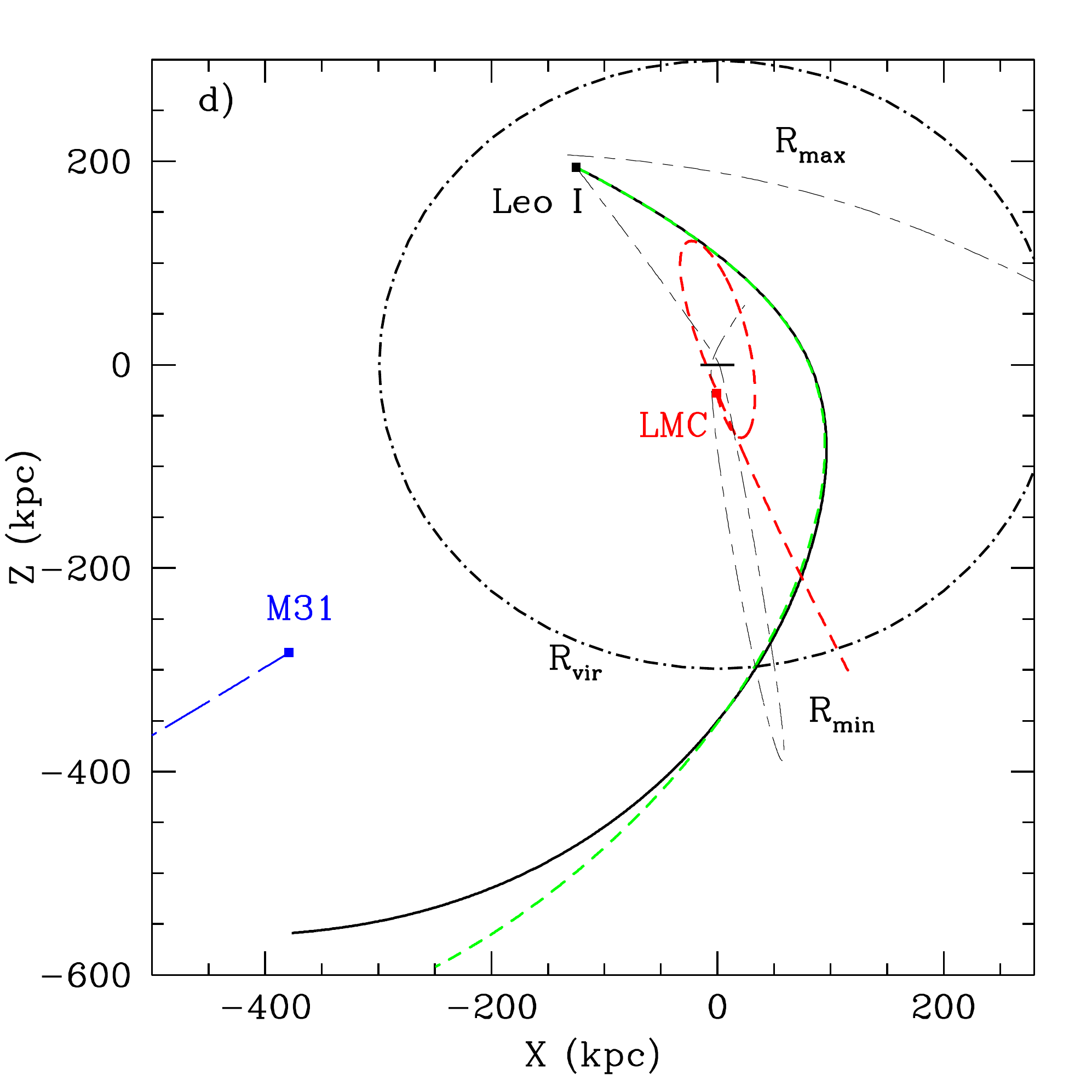}}}
 \caption{\label{fig:Orb1.5e12} Similar to Figure~\ref{fig:Orb1e12}, 
 but now for the intermediate MW mass model 
 ($M_{\rm vir} = 1.5\times10^{12} \Msun$).
  }
\end{figure*}  

\begin{figure*}
\centering
\mbox{{\includegraphics[width=3in]{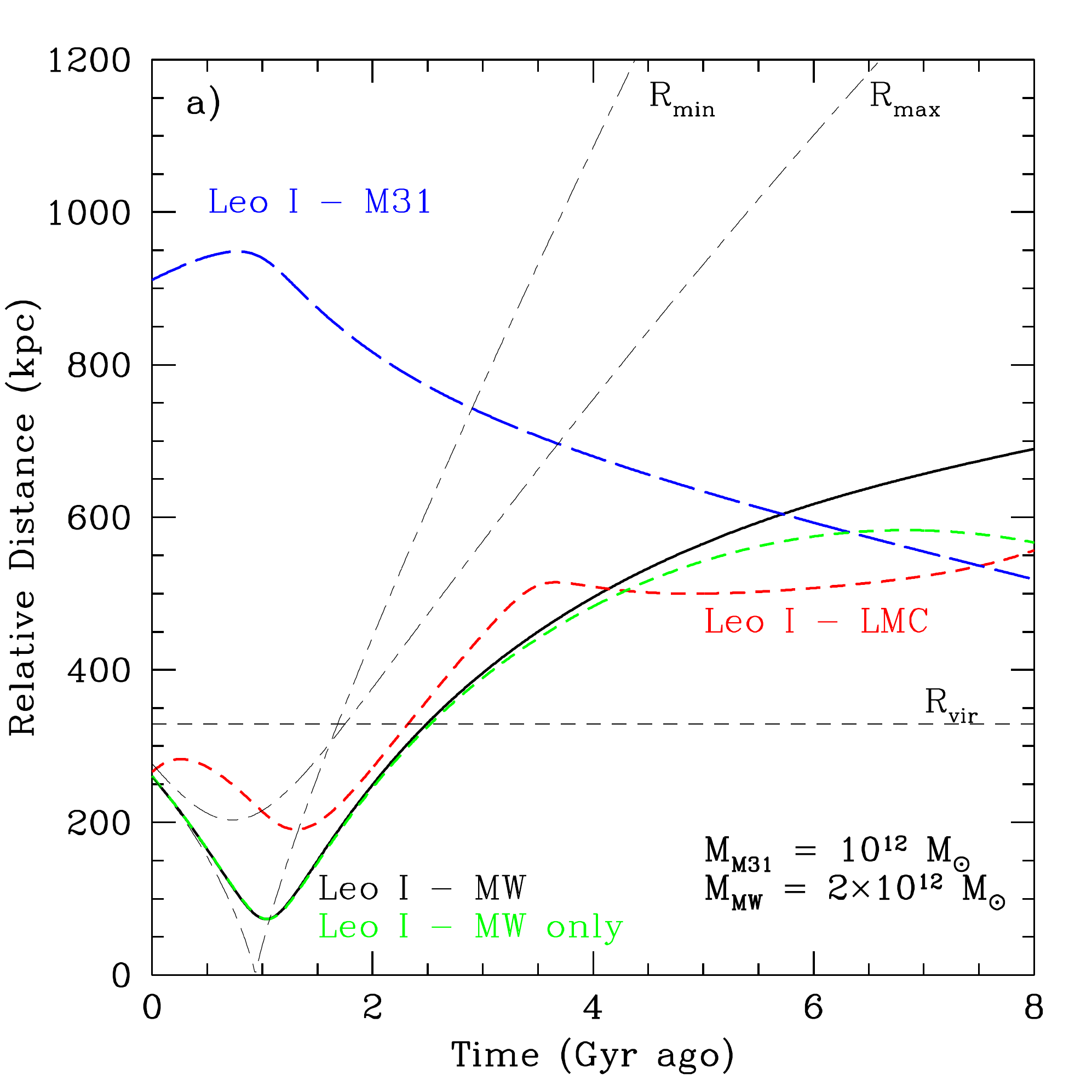}}
{\includegraphics[width=3in]{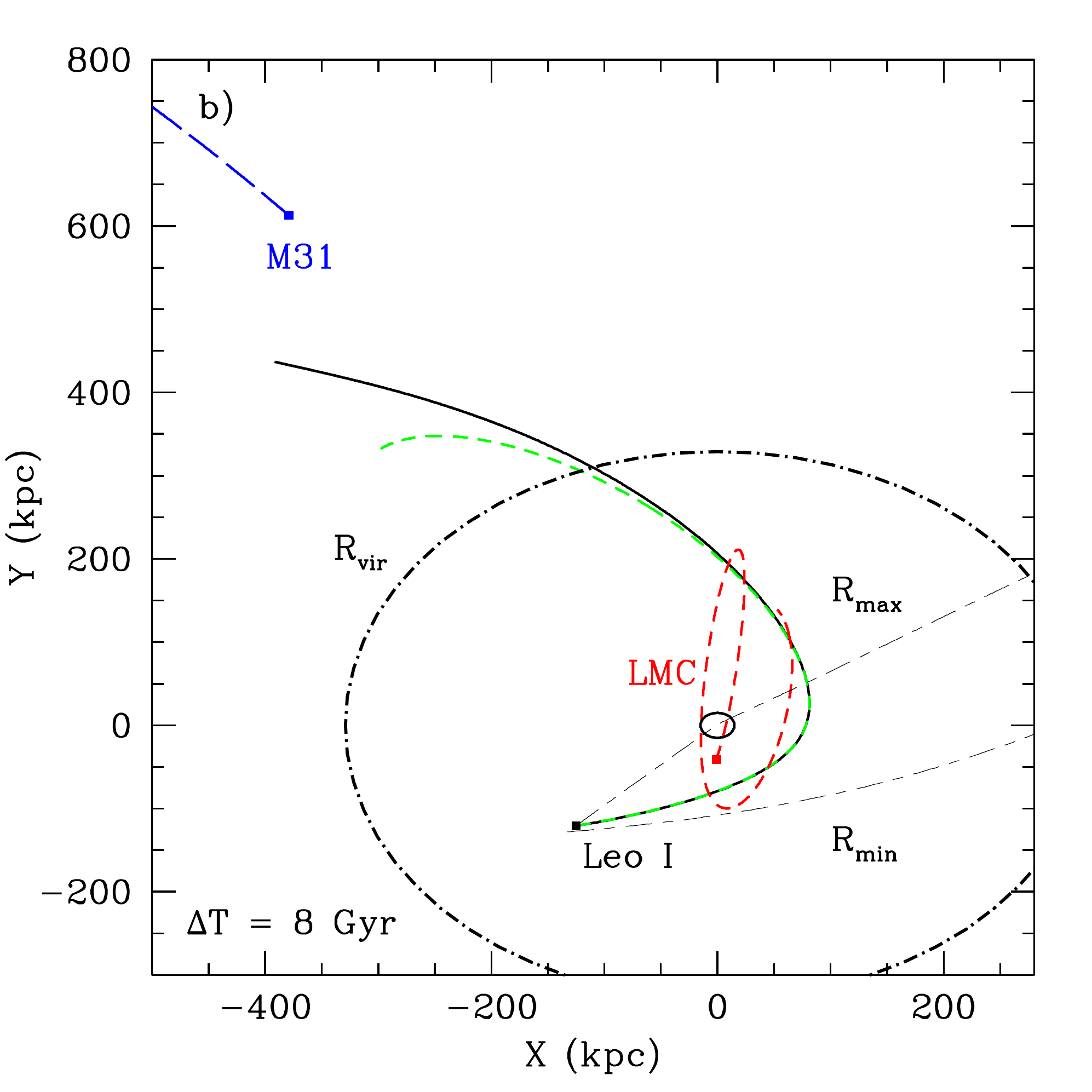}}}\\
 \mbox{
 {\includegraphics[width=3in]{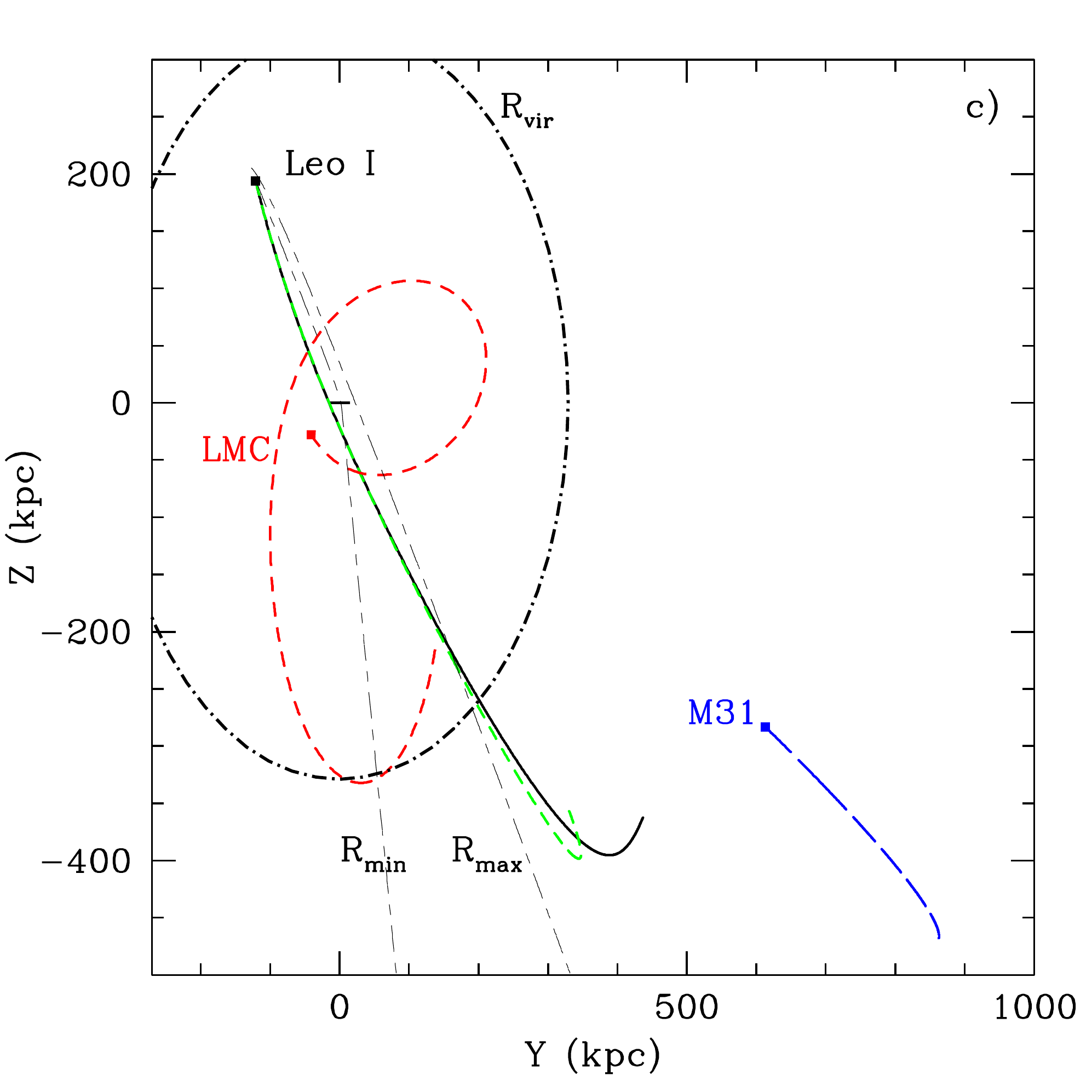}}
 {\includegraphics[width=3in]{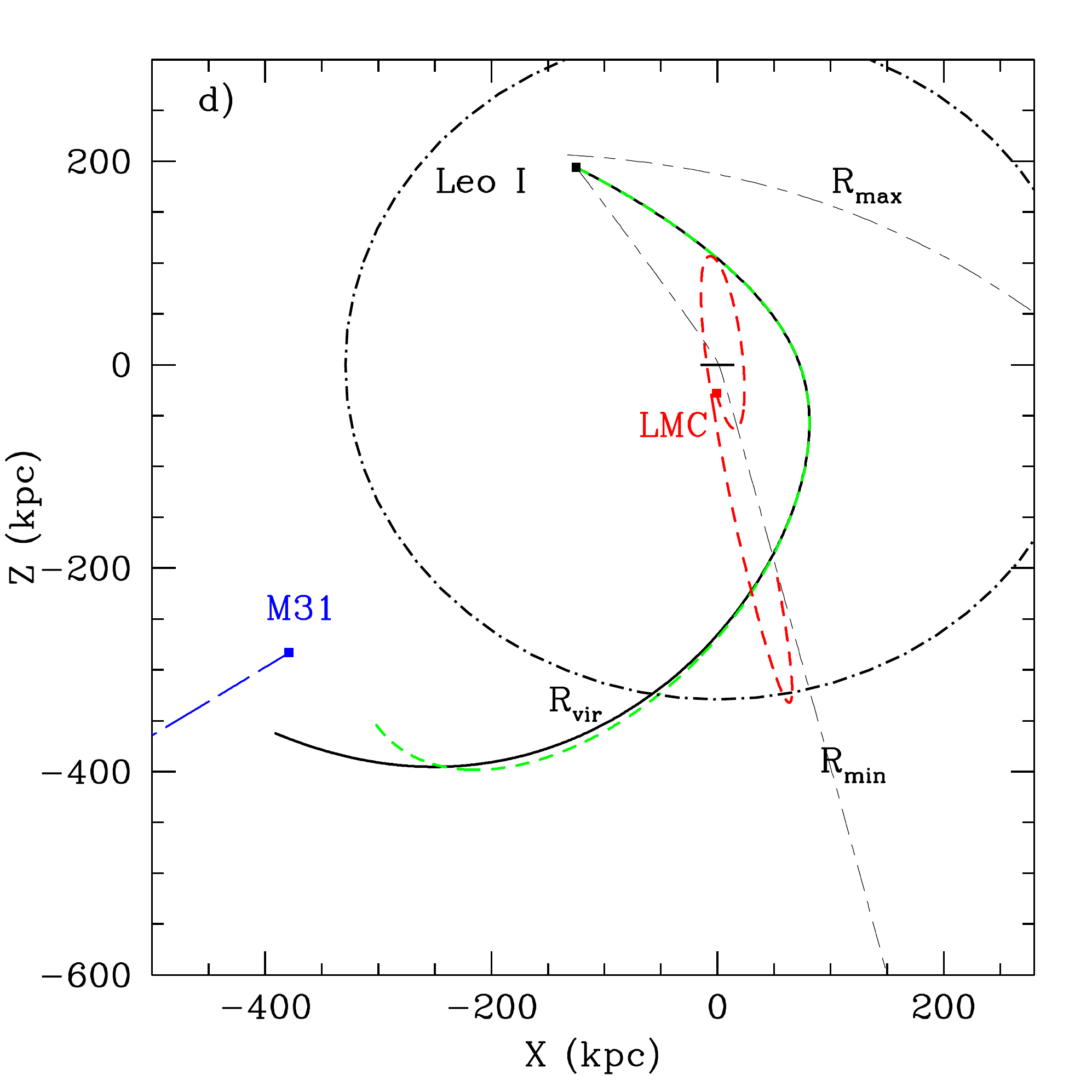}}}
 \caption{\label{fig:Orb2e12} Same to Figure~\ref{fig:Orb1e12}, 
 but now for the high MW mass model 
 ($M_{\rm vir} = 2.0\times10^{12} \Msun$).
  }
\end{figure*}

\citet{soh07} and \citet{mat08} provided estimates of the orbital
history of Leo~I based on indirect arguments, rather than proper
motion measurements. They aimed in particular to explain the
photometric and kinematical data of giants stars in Leo~I. The proper
motions corresponding to the proposed orbits are compared to our new
HST measurements in Figure~\ref{fig:PMcomp}.\footnote{The proper
motion predictions reported in \citet{soh07} are erroneous. We
re-derived the predicted proper motions based on the orbital
positions and velocities of their model 117 at $t=0$ and the result 
is $(\mu_{\rm W}, \mu_{\rm N}) = (0.032, -0.160) \masyr$. This assumed
the same position and velocity of the Sun as the present study.}
The orbits of \citet{soh07} had a pericentric approach of only $\sim
10$ kpc, so the predicted proper motion is very close to the $V_{\rm
  tan} = 0$ point. \citet{mat08} provided proper motion predictions
for a range of assumed Leo~I masses (their Table~8). While their
predicted perigalactic distances reach out to 30 kpc, their proper
motions lie on the opposite side of the $V_{\rm tan} = 0$ point
compared to our measurements. So our measurements do not confirm the
predictions of these studies. More specifically, the previous studies
argued for highly eccentric orbits with smaller perigalactic distances
than what we find here.

\citet{soh07} focused primarily on trying to reproduce the observed
photometric and kinematic features of Leo~I by adopting a tidal
disruption scenario. The orientation of their model orbits was
determined by assuming that the position angle of the Leo~I
ellipticity and the orientation of the break population are caused by
tidal effects and tidal stripping, respectively. They showed that the
observed features can be plausibly produced by the tidal effects of
the MW. However, the tidal effects may have been overestimated, given
that the orbital properties derived here imply that Leo~I is on a less
eccentric orbit than assumed by \citet{soh07}. This discrepancy does
not imply though that the tidal scenarios used by \citet{soh07} and
\citet{mat08} are necessarily wrong. It may just be that some of the
specific assumptions in their models were oversimplified. For
example, if \citet{soh07} had not modeled the progenitor satellite as
a spherical Plummer profile, the best-fit orbits may well have been
more consistent with the observed proper motion. New $N$-body models
based on the observed proper motion should be able to further improve
our understanding of the tidal disruption features observed in Leo~I.
However, such models are beyond the scope of the present paper.


\subsubsection{Leo~I Orbital Plane}
\label{sec:Plane}

Here we compute the orbital history of Leo~I, relative to the other 
major players in the Local Group, namely the MW, the LMC, and M31. 
We aim to define Leo~I's orbital plane and compare it to that of the 
LMC, and to determine whether Leo~I was ever close enough to M31 for 
it to have exerted any dynamical influence in Leo~I's history. 

In Figures~\ref{fig:Orb1e12},~\ref{fig:Orb1.5e12},
and~\ref{fig:Orb2e12} we plot the orbit of Leo~I using the mean proper
motions determined in this study, including the influence 
of the MW, LMC, and M31 (solid black line). For comparison, we also
plot the orbit of Leo~I accounting only for the influence of the MW 
(dashed green line). Orbits corresponding to proper
motions that are $\pm$3$\sigma$ from the mean (identified as having
min/max pericenter distances to the MW from panels (b) in
Figures~\ref{fig:MC1e12},~\ref{fig:MC1.5e12}, and~\ref{fig:MC2e12})
are indicated by the thin dash-dotted black lines (Rmin and Rmax) in
all panels. 

Panels (a) illustrate the separation of Leo~I from the other galaxies
and Panels (b), (c), and (d) respectively show the orbits of the
galaxies in the $X$-$Y$, $Y$-$Z$, and $X$-$Z$ Galactocentric planes.
As the MW mass increases, Leo~I's past orbit becomes less eccentric
and increasingly directed towards M31.  It is clear that Leo~I does
not get closer than 400 kpc from M31 in any model over the past 8 Gyr
and neither the presence of M31 nor the LMC have an impact on the
infall time or pericenter properties\footnote{This is unsurprising
  given that these properties are computed using a backwards
  integration scheme.}.  However, M31 may play a role at early times
in the higher MW mass models; the orbits are more energetic, reaching
larger distances than if the gravity of the MW were considered alone.
We note that we have not accounted for the mass evolution of the MW or
M31, which will diminish the role that M31 plays at early times.  But,
this analysis does suggest that accounting for the local overdensity
(i.e. that there are two roughly equal mass galaxies in our Local
Group) may be a relevant parameter in understanding the origin of the
angular momentum of high speed satellites.  In this study, such
considerations are only relevant for the higher mass MW models; in the
lowest mass MW model, Leo~I is too far from M31 at all times for
torques to be relevant. Our conclusions regarding the hyperbolic
nature of Leo~I's orbit in low-mass MW models is thus robust to the
influence of M31 and can be compared statistically to satellite orbits
found in cosmological simulations of isolated MW analogs (see
Paper~II).

\begin{figure*}
\centering
\mbox{{\includegraphics[width=2in]{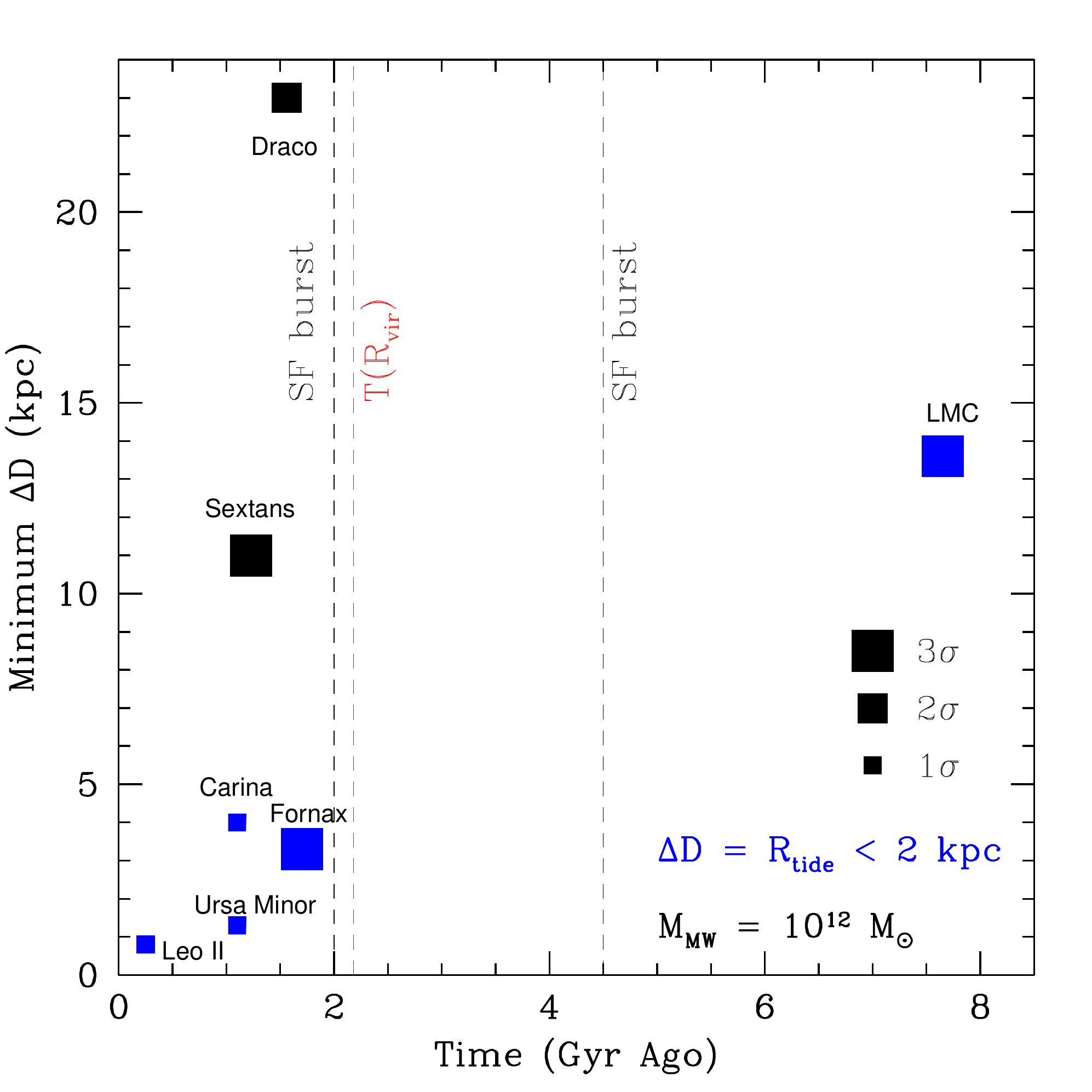}}
{\includegraphics[width=2in]{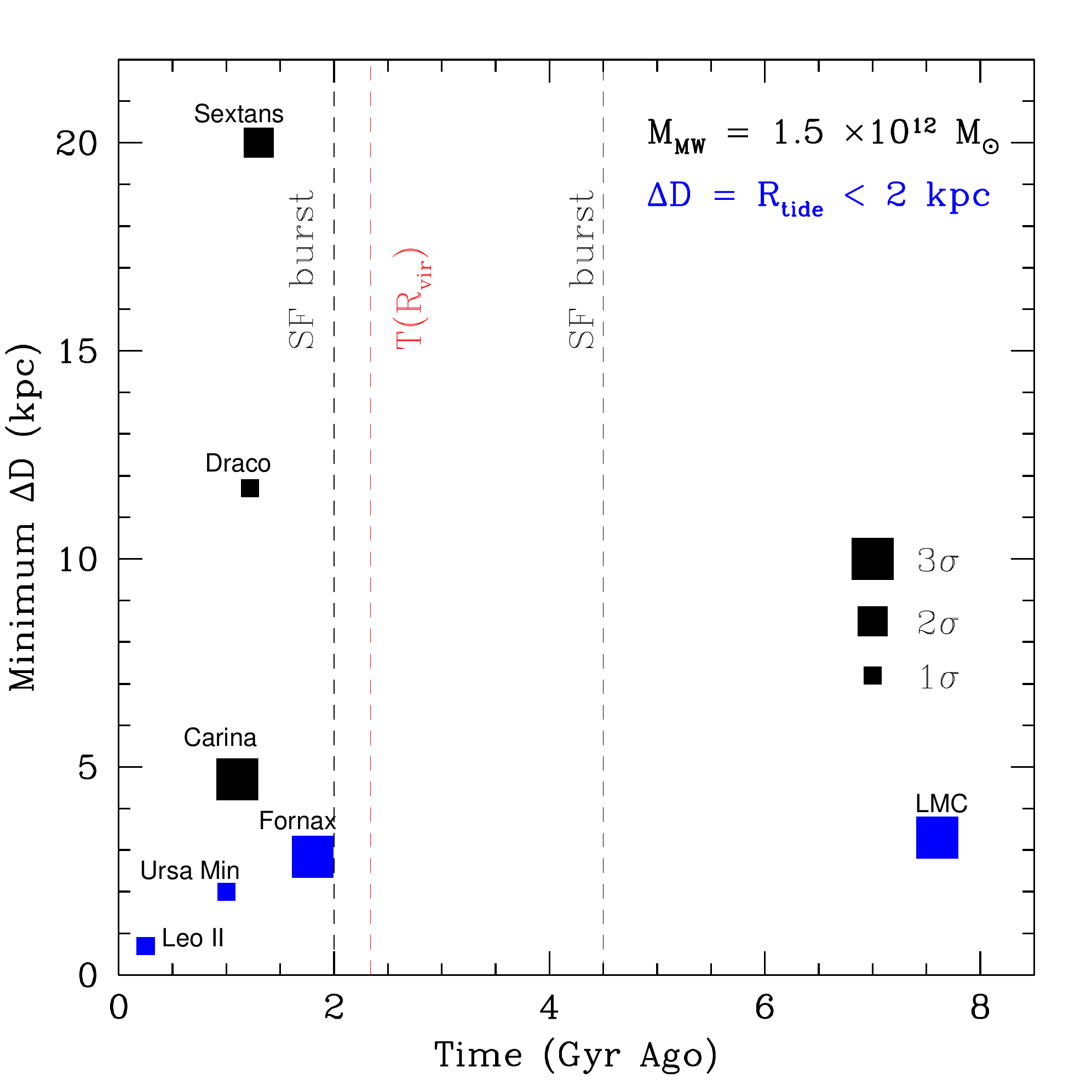}}
{\includegraphics[width=2in]{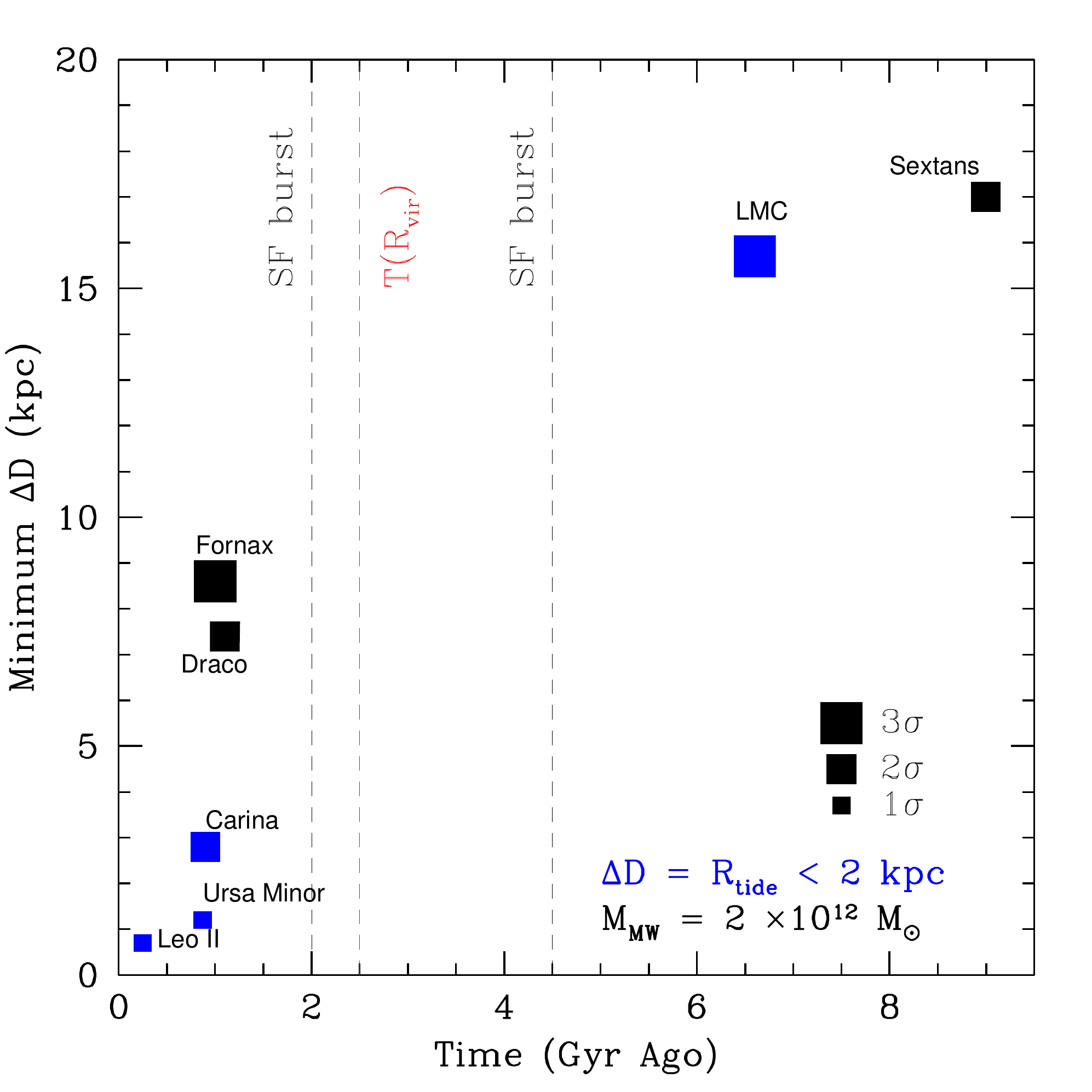}}}
 \caption{\label{fig:SatSep} The minimum separation to Leo~I that is 
 allowed within the proper motion error space  of each satellite is 
 plotted as a function of time for each MW mass model. Cases where 
 the separation is  sufficiently small so as to tidally truncate Leo~I 
 to within at least 2 kpc are marked in blue. Given the low mass 
 ratios, such separations will also allow Leo I to influence that 
 satellite in turn, except for the LMC. Black dashed lines indicate 
 known epochs of enhanced star formation activity in Leo~I, as 
 discussed in the text. The red dashed line indicates its infall time 
 within the virial radius of the MW. The size of each point indicates 
 the distance of the corresponding Leo~I proper motion from the mean 
 observed proper motion, as described in the legend; smaller points 
 are more likely. Although events were found within 1$\sigma$ for some 
 satellites, they typically represent only $\lesssim 4$\% of the Monte-Carlo 
 drawings. This is shown explicitly in Figure~\ref{fig:UrsaLeoII} for 
 the satellites that have the closest encounters, Ursa Minor and 
 Leo~II, to illustrate the low probability of such encounters. We note 
 that Sculptor never gets closer to Leo~I than its current separation 
 and is thus omitted from the plot.
}
\end{figure*}

The minimum pericentric passage orbits appear to be slingshot orbits,
approaching near the MW center, gaining energy and traveling to larger
distances. However, as discussed in Section~\ref{sec:LeoMW}, such
orbits are likely unphysical because Leo~I is not sufficiently
disturbed to have traveled this close to the MW. Since the orbit of
Leo~I is still bound to the MW in the vast majority of models, we do
not need a small pericenter, slingshot orbit to explain the orbital
properties of Leo~I.

Panels (b), (c), and (d) respectively show the orbits of the galaxies 
in the $X$-$Y$, $Y$-$Z$, and $X$-$Z$ Galactocentric planes. The orbital 
angular momentum of Leo~I is not coincident with that of the LMC. 
This is most clearly illustrated when looking at the orbital history 
in the $X$-$Z$ Galactocentric plane, especially in the higher-mass MW 
models.\footnote{In the lower-mass MW models (Figure~\ref{fig:Orb1e12}d) 
the LMC and Leo~I are also clearly unassociated, because they are more 
than 300 kpc from each other for most of their evolution. However, the
LMC orbit has a less clear sense of rotation in the $X$-$Z$ plane, 
owing to the gravitational influence of M31 which causes a kink/twist 
in the orbit. The mass of M31 is the highest in the lowest MW mass model, 
thereby making its gravitational perturbation the strongest.} The LMC is 
moving in a clockwise direction in this plane whereas Leo~I is moving
counterclockwise.


\subsubsection{Interactions with other Satellites} 
\label{sec:Sat}

We have so far established that Leo~I is likely on its first orbit 
around the MW and that its most recent pericentric approach was likely 
at a separation of 80--100 kpc, too large for the MW to have exerted
significant tidal torques. Yet, Leo~I shows signs of a past
interaction. \citet{soh07} found an excess of red giant stars along
the major axis of Leo~I's main body relative to a symmetric King
profile. In addition to this spatial configuration, Leo~I red giant
stars have an asymmetric radial velocity distribution at large radii
\citep[cf., see also][]{mat08}. If the MW is not the culprit for the
distorted structure and kinematics of Leo~I, then what is?

Here we consider the orbital history of not only Leo~I, but also of
the other satellite galaxies of the MW simultaneously. We randomly
sampled 10,000 combinations of the observed west and north components 
of the proper motion within the 4$\sigma$ error space for Leo~I and 
for each satellite. The proper motions, distances, and radial 
velocities were taken from various sources as listed in
Table~\ref{tab:Sat}. As for Leo~I, we took the distance for each
satellite to be the error-weighted average of TRGB measurements in the
last decade. The proper motions and radial velocities were adopted
from the most recent measurements available in the literature. We
computed the orbits of all satellites backwards in time for 10 Gyr
using each of our three MW models, and using the mass model for each
satellite outlined in Table~\ref{tab:Sat}.  Our goal is to determine
the closest separation that Leo~I may have reached to any of these
other satellites within the error space and to assess whether such
separations are sufficient to exert torques on Leo~I and induce star
formation or to significantly modify Leo~I's orbit.

Given the small masses of these satellites, their dynamical influence
is minimal unless the separation between them is small.  Overall,
satellite separations as low as 3--4 kpc are required before one can
significantly influence the other, i.e. such that the tidal radius of
the satellite is less than 2 kpc. Because the LMC is much more massive
than the other satellites, a larger separation of at least 20 kpc will
allow it to distort Leo~I to within 2 kpc. Of course, Leo~I is much
too small to strongly affect the LMC\footnote{Indeed the SMC's tidal 
field has had limited influence over the star formation history of 
the LMC \citep{bes12}.}.

Figure~\ref{fig:SatSep} indicates the minimum separation 
between each satellite and Leo~I as a function of time for each 
respective MW mass model. Satellites that reach separations 
within their 4$\sigma$ error space small enough to influence 
Leo~I within a radius of 2 kpc (and vice versa) are highlighted 
in blue. The size of the point reflects the probability of that 
encounter, based on the sigma deviation of the required Leo~I 
proper motion from the mean. Vertical dashed lines indicate 
relevant events in the history of Leo~I, such as epochs of star 
formation and the epoch of infall into the MW. 

\begin{figure*}
\centering
\mbox{{\includegraphics[width=3in]{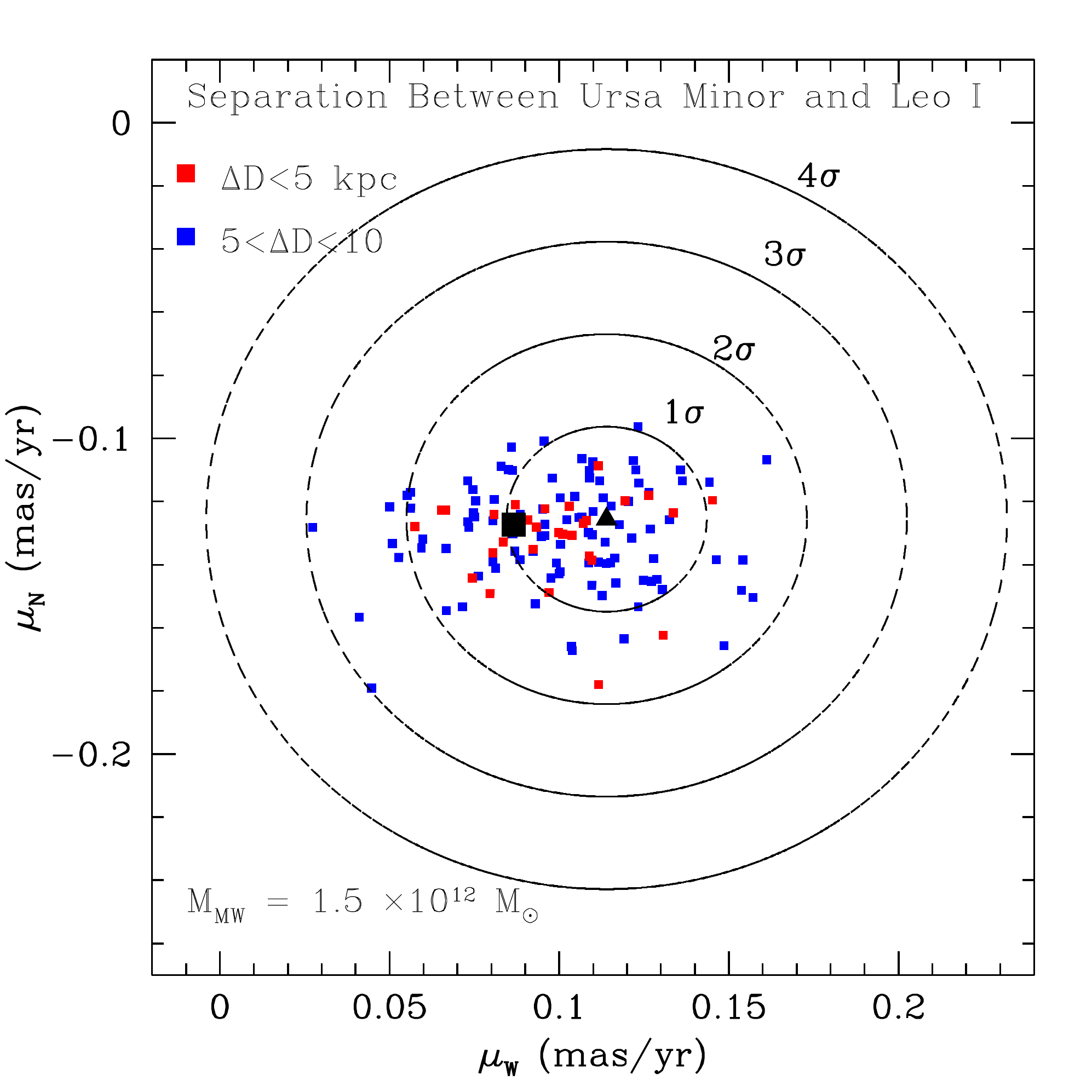}}
{\includegraphics[width=3in]{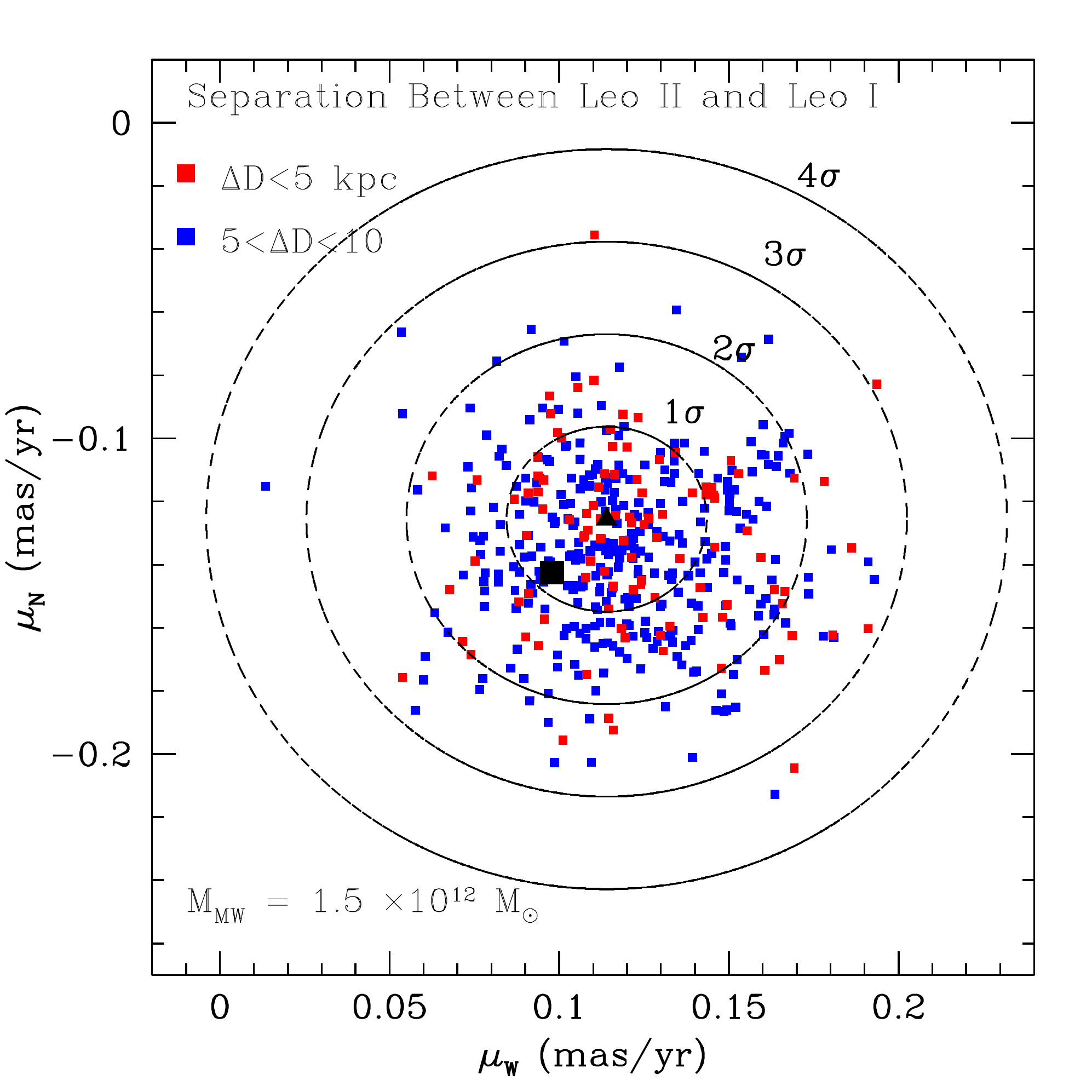}}}
 \caption{\label{fig:UrsaLeoII} Orbits, out of 10,000 randomly
   selected points from the 4$\sigma$ error space of the Leo~I proper
   motions, that produce a passage of Ursa Minor (left panel) or
   Leo~II (right panel) within 10 kpc of Leo~I. Orbits are color
   coded based on the minimum separation, as calculated for the
   intermediate MW mass model ($M_{\rm vir} = 1.5\times10^{12} 
   \Msun$). Separations less than 3 kpc are readily found within 
   the 1$\sigma$ error ellipse, indicating that such past encounters 
   are possible. For Ursa Minor, there are 62 blue points and 
   19 red points that fall within the 1$\sigma$ region (the innermost 
   circle), and for Leo~II, there are 171 blues points and 57 red 
   points within the 1$\sigma$ region.
}
\end{figure*}  

There are no obvious satellite encounters found that can explain 
the enhanced star formation activities at $\sim 2$ and $\sim 4.5$
Gyr ago. It appears much more likely that the enhancement of star 
formation that took place $\sim 2$ Gyr ago is related to the epoch 
of accretion by the MW. While there are some parameter combinations 
that allow for close encounters between Leo~I and Fornax, Carina 
and the LMC, these are likely random events as there are only ever 
2, or 3 points within the 10,000 orbit search that yield such 
encounters. Also, Sculptor never gets close to Leo~I than it is 
today, and is thus omitted from the plot. Among the MW satellites 
we consider, there is a higher probability -- though still 
relatively small -- of an encounter with Ursa Minor and Leo~II for 
all MW models. Specifically, 1.3\% and 4.5\% of the Monte-Carlo 
calculated orbits yield passages within 10 kpc for Ursa Minor and 
Leo~II, respectively. These encounters are expected to have occurred 
within the past 1 Gyr, well after star formation has ceased in all 
of these galaxies \citep{car02,koc07,kir11}. The encounter is thus 
unlikely to have signatures in the star formation histories of these 
galaxies, as they should already have been devoid of gas at that point. 
However, there may be kinematical signatures instead. In particular 
the distortions in Leo~I noted by \citet{soh07} could be explained by 
collisions with Ursa Minor or Leo~II, rather than interactions with 
the MW. Ursa Minor is also known to be kinematically disturbed 
\citep{kle98,wil04}. It has an inner bar that has been suggested to 
be tidally induced \citep{lok12} and has an extended stellar halo 
\citep{pal03}. To date, however, no sign of tidal disturbance is 
found for Leo~II.

To illustrate the non-negligible probability of such encounters, 
we focus in Figure~\ref{fig:UrsaLeoII} on orbit calculations for the
intermediate mass MW mass model ($M_{\rm vir} = 1.5 \times 10^{12} 
\Msun$). We plot from the 10,000 randomly sampled points within 
the 4$\sigma$ proper motion error space, those points that produce a
passage of Ursa Minor (left panel) or Leo~II (right panel) within 
10 kpc of Leo~I. There are several orbital solutions within 1$\sigma$ 
with such small separations to Leo~I. Since the error space of the 
Leo~II and Ursa Minor proper motions were also searched, we note here 
that solutions are also found within 1$\sigma$ of their respective 
means. So close encounters between these galaxies and Leo~I are not 
ruled out by the data, although the probability of such encounters is low.
The encounter time for Ursa Minor occurs 
within a range of 0.8-1.1 Gyr ago, while the encounter time for 
Leo~II is relatively recent, with a range of 0.15--0.3 Gyr ago. 

To determine whether the presence of the satellites of the MW 
can modify the orbit of Leo~I, we repeated the analysis of
Section~\ref{sec:LeoMW}, but now also accounting for the tidal torques
exerted by the other satellites. We followed the method outlined above
to test the proper motion error space for each satellite in
Monte-Carlo sense.\footnote{We decided not to explore the full proper
motion error space of M31. The preceding analysis already
established that M31 is unlikely to have come close enough to 
modify exert sufficient torques to modify the star formation history of
Leo~I.} This showed that any multi-body tidal effects by the other
MW satellites are insufficient to modify the average orbital
properties listed in Table~\ref{tab:MC}. In particular, the average
velocity and time at infall are unaffected by the presence of the
other MW satellites. We find that the Leo~I orbit is significantly
affected by the presence of the other satellites in only
$0.1$--$0.2$\% of Monte-Carlo orbits. So Leo~I's high velocity is
probably not a product of multi-body tidal torques. This makes it
important to address whether such a high velocity can arise naturally
in $\Lambda$CDM galaxy assembly scenarios without assistance from
multi-body interactions. We explore this topic in detail in Paper~II.


\section{Conclusions}
\label{sec:Conclusions}

We have presented the first absolute proper motion measurement of
Leo~I, based on \hst\ ACS/WFC images taken in two different epochs
separated by $\sim 5$ years. We used the method of \citet{soh12} to
measure the average bulk motion of Leo~I stars with respect to
stationary distant galaxies in the background. We detect motion of
Leo~I at $4\sigma$ confidence, and find its proper motion to be
($\mu_{W}, \mu_{N}$) = ($0.1140 \pm 0.0295, -0.1256 \pm 0.0293$) mas
yr$^{-1}$. The uncertainties are smaller than those obtained in
previous \hst\ studies of other MW satellites that used a background
quasar as stationary reference. To derive the velocity of Leo~I with
respect to the MW, we combined the proper motion with the known
line-of-sight velocity and corrected for the solar reflex motion. The
resulting Galactocentric radial and tangential velocities are $V_{\rm
  rad} = 167.9 \pm 2.8$ and $V_{\rm tan} = 101.0 \pm 34.4 \kms$,
respectively. Hence, Leo~I has a significant transverse velocity, but
it is less than the radial velocity.

Combined with its current position, the new knowledge of the
three-dimensional velocity of Leo~I has allowed us to study its
orbital history in detail. To evaluate the past orbit we employed a
range of mass models of increasing complexity. Starting from Keplerian 
models for the MW, we progressed first to cosmologically motivated MW
models of $M_{vir} = 1.0\times 10^{12}, 1.5\times 10^{12}$, and 
$2.0\times 10^{12} \Msun$, and then to models in which other MW 
satellites and Local Group galaxies are included as well. In each of 
these models we solved the equations of motion to follow the Leo~I 
velocity and position backward in time. We used a Monte-Carlo analysis 
to explore the impact of the observational measurement uncertainties.

Allowing for both observational uncertainties and uncertainties in the
MW mass model (Table~\ref{tab:MC}, with a flat prior in mass), we find
that Leo~I's most recent perigalactic passage was $1.05 \pm 0.09$ Gyr
ago at a Galactocentric distance of $91 \pm 36$ kpc. The ratio of the
orbital pericenter to apocenter distance is $0.17 \pm 0.07$, so the
orbit extends well outside of the MW's virial radius. Leo~I entered
the virial radius $2.33 \pm 0.21$ Gyr ago. This was most likely
Leo~I's first infall into the MW. A previous pericenter, which would
have occurred almost a Hubble time ago, becomes slightly plausible
only in the highest-mass MW models.

Stellar population studies of Leo~I have inferred that it experienced 
an enhancement in star formation $\sim 2$ Gyr ago. This may have been 
the result of Leo~I entering the virial radius of the MW for the first 
time, leading to gas compression through ram pressure or MW tidal 
torques. Stellar population studies have also shown that star formation 
in Leo~I was quenched $\sim 1$ Gyr ago. This may have been due to the 
pericentric approach of Leo~I with respect to the MW, at which point 
ram pressure stripping of gas was maximal.

A previously inferred enhanced star formation activity in Leo~I that 
occurred $\sim 4.5$ Gyr ago is not obviously associated with any orbital 
time scale, or interaction with any other galaxy. The orbital plane of 
Leo~I is not coincident spatially or in rotational sense to that of 
the LMC. The separation between M31 and Leo~I remains large at 
early times, but it may have been possible that M31 applied sufficient 
torques at early times to have modified Leo~I's orbit. Leo~I may have 
closely approached (within $\sim 10$ kpc) other MW satellites, but only 
in the last Gyr, and with probabilities of at most a few percent. 
The probabilities are highest for past encounters with Ursa Minor or 
Leo~II, which may have left marks in the kinematical properties of 
Leo~I or these other galaxies.

Given the observed velocity of Leo~I and prior constraints on the MW
virial mass, Leo~I is most likely bound to the MW. However, this is
not true in MW models with masses $M_{vir} \lesssim 10^{12}$, at the
low end of the allowed range. Leo~I has just passed pericenter,
probably from its first infall into the MW. So its kinematics are
probably not virialized. So even though Leo~I is probably bound, it is
not necessarily appropriate to include it in equilibrium models of the
MW satellite population, such as those that are often used to estimate
the MW virial mass \citep[e.g.,][]{wat10}.

The velocity of Leo~I can also be used to estimate the MW mass through
the timing argument. Previous studies of this kind have used the
assumption of a radial orbit, but with a statistical correction for
any possible transverse velocity \citep[e.g.,][]{li08}. Now that the
transverse velocity has actually been measured, it is possible to
solve the complete timing equations for a non-radial orbit. This
yields $M_{\rm {MW,vir}} = 3.15_{-1.36}^{+1.58} \times 10^{12} \Msun$,
with the large uncertainty dominated by cosmic scatter. This is higher
than, but not inconsistent with, the range of MW mass estimates
obtained from other methods.

This is the first paper in a series of two. In Paper~II, we compare
the new observations to the properties of Leo~I subhalo analogs
extracted from state-of-the-art cosmological simulations. We show
there that Leo~I is most likely bound to the MW, since unbound
subhalos are extremely rare at the present epoch. We also show there
that the observed kinematics of Leo~I are more consistent with
high-mass than with low-mass MW models. Both these conclusions are
consistent with what we have inferred in the present paper through
different arguments.

In this paper and in \citet{soh12}, we have used our new methodologies
to successfully measure with {\hst} the absolute proper motions of
Leo~I and M31, respectively. The new measurements have allowed us to
derive new constraints on the past and future orbital evolution of the
target galaxies, and on the mass of the Local Group's dominant
galaxies, the MW and M31. Motivated by these results, we are
continuing to apply our proper motion measurement techniques to a
range of other topics in Local Group galaxy research. For example, our
group has ongoing \hst\ observing programs to measure the proper
motions of: (1) dwarf galaxies near the edge of the Local Group
(GO-12273, PI: R. P. van der Marel); (2) stars at different locations
in the Sagittarius Stream (GO-12564, PI: R. P. van der Marel); and (3)
the dwarf galaxy Leo~T, which is likely a galaxy falling into the MW
for the first time (GO-12914, PI: T. Do). We expect that the results
from these ongoing programs will further constrain the dark matter
distribution in the Local Group and its dominant galaxies.


\acknowledgments

We thank the anonymous referee for constructive feedback that helped
improve the presentation of our results.
Support for this work was provided by NASA through a grant for program
GO-12270 from the Space Telescope Science Institute (STScI), which is
operated by the Association of Universities for Research in Astronomy
(AURA), Inc., under NASA contract NAS5-26555. GB acknowledges support
from NASA through Hubble Fellowship grant HST-HF-51284.01-A.  The
authors wish to thank Jay Anderson for his pioneering efforts on
\hst\ proper motion analyses, for providing the dither pattern for the
second epoch \hst\ observations, and for helpful comments throughout
the proper motion derivation process. Yang-Shyang Li kindly provided
the Millennium simulation sample used in \citet{li08}, and reanalyzed
in Section~\ref{sec:kep}. MB-K acknowledges support from the Southern
California Center for Galaxy Evolution, a multi-campus research
program funded by the University of California Office of
Research. This research has made use of the NASA/IPAC Extragalactic
Database (NED) which is operated by the Jet Propulsion Laboratory,
California Institute of Technology, under contract with the National
Aeronautics and Space Administration.

{\it Facilities:} \facility{HST (ACS/WFC)}.


\end{document}